\documentstyle[12pt,epsfig]{article}
\begin{document}
\def\e{\epsilon}
\def\d{{\rm d}}
\def\Li{{\rm Li}}
\def\S12{{\rm S}_{12}}
\def\zcut{z_{\rm cut}}
\def\ycut{y_{\rm cut}}
\def\ymin{y_{\rm min}}
\def\scut{y_{\rm cut}}
\def\smin{s_{\rm min}}
\def\yqqb{y_{q\bar{q}}}
\def\yqg{y_{q\gamma}}
\def\yqbg{y_{\bar{q}\gamma}}
\def\Dg{D_{g\to \gamma}}
\def\DBqp{D^B_{q\to \gamma}}
\def\DBq{D^B_{q\to \gamma}}
\def\DBqpp{D^B_{q'\to \gamma}}
\def\DBgp{D^B_{g\to \gamma}}
\def\Dq{D_{q\to \gamma}}
\def\Dqp{D_{q'\to \gamma}}
\def\DBa{D^B_{a\to \gamma}}
\def\Pqqzero{P^{(0)}_{q\to q}}
\def\Pqgzero{P^{(0)}_{q\to g}}
\def\Pgqzero{P^{(0)}_{g\to q}}
\def\Pqpzero{P^{(0)}_{q\to \gamma}}
\def\Pqpone{P^{(1)}_{q\to \gamma}}
\def\Pgpone{P^{(1)}_{g\to \gamma}}
\def\Pqg{P_{q\to g}}
\def\Pqp{P_{q\to \gamma}}
\def\Gqp{\Gamma_{q\to \gamma}}
\def\Gqpp{\Gamma_{q'\to \gamma}}
\def\Ggp{\Gamma_{g\to \gamma}}
\def\Gqg{\Gamma_{q\to g}}
\def\Ggq{\Gamma_{g\to q}}
\def\Gij{\Gamma_{i\to j}}
\def\Gqq{\Gamma_{q\to q}}
\def\Ggg{\Gamma_{g\to g}}
\def\Gqpq{\Gamma_{q'\to q}}
\def\Gqqp{\Gamma_{q\to q'}}
\def\aqp{a_{q\to \gamma}}
\def\agp{a_{g\to \gamma}}
\def\calDq{{\cal D}_{q\to \gamma}}
\def\calDg{{\cal D}_{g\to \gamma}}
\def\aqed{\left(\frac{\alpha e_q^2}{2\pi}\right)}
\def\as{\left(\frac{\alpha_s C_F}{2\pi}\right)}
\def\4pi2{\left(\frac{4\pi\mu^2}{M^2}\right)^{2 \e}}
\def\coup{\frac{1}{\Gamma(1-\e)^2}\4pi2\as\aqed}

\begin{titlepage}
\vspace*{-1cm}
\begin{flushright}
DTP/97/26\\
July 1997\\
\end{flushright}                                
\vskip 1.cm
\begin{center}                                                             
{\Large\bf
A complete ${\cal O}(\alpha \alpha_s)$ calculation of the \\[2mm]
Photon +~1 Jet Rate 
in $e^{+}e^{-}$ annihilation}
\vskip 1.3cm
{\large A.~Gehrmann--De Ridder   and  E.~W.~N.~Glover}\\
\vspace{0.5cm}
{\it
Physics Department, University of Durham,\\ Durham DH1~3LE, England} \\
\vspace{0.5cm}
{\large \today}
\vspace{0.5cm}
\end{center}      

\begin{abstract}
We present a complete calculation 
of the photon +~1 jet rate in $e^+e^-$ annihilation 
up to ${\cal O}(\alpha \alpha_{s})$. Although formally 
of next-to-leading order in perturbation theory, this calculation 
contains several 
ingredients appropriate to a  next-to-next-to-leading order calculation 
of jet observables.
In
particular, we describe a generalization of the commonly used 
phase space slicing method to isolate the singularities present when 
more than one particle is unresolved. Within this approach,
we analytically evaluate the singularities associated with the following double unresolved regions;
triple collinear, soft/collinear  and double single collinear configurations as well as those from the collinear limit of virtual graphs. 
By comparing the results of our calculation with the existing data 
on the photon +~1 jet rate from the ALEPH Collaboration at CERN,
we make a next-to-leading order determination
of the process-independent non-perturbative quark-to-photon 
fragmentation function $D_{q \to \gamma}(z,\mu_{F})$ 
at ${\cal O}(\alpha \alpha_{s})$. As a first application of this measurement
allied with our improved perturbative calculation, we determine 
the dependence of the isolated photon +~1 jet cross section 
in a democratic clustering approach on the jet resolution parameter 
$\ycut$ at next-to-leading order. The next-to-leading 
order corrections to this observable are moderate
but improve the agreement
between theoretical prediction and experimental data. 
\end{abstract}                                                                
\vfill
\end{titlepage}

\section{Introduction} 
\setcounter{equation}{0}
\label{sec:intro}
Although the  
production dynamics of photons and of jets are both individually 
well understood theoretically, their interplay is not. Hadronic jets 
are initiated by the production of quarks and gluons which 
subsequently fragment into clusters of hadrons. In events 
where a photon is produced in addition to the jets, this photon
can have two possible origins: the {\it direct} radiation of a 
photon off a primary quark and the {\it fragmentation} of a 
hadronic jet into a photon carrying a sizeable fraction of the jet 
energy. While the former direct process can be calculated within the
framework of the Standard Model, the latter is described by 
the process-independent quark-to-photon 
and gluon-to-photon fragmentation functions \cite{phofrag}, which cannot  
be calculated using perturbative methods but have to be determined 
from experimental data. 

Directly emitted photons are usually well separated 
from all hadron jets produced in a particular event, while photons  
originating from fragmentation processes are primarily to be found 
inside hadronic jets. Consequently, by imposing some isolation 
criterion on the photon, one is in principle able to suppress (but 
not to eliminate) the fragmentation contribution to final state 
photon cross sections, and thus to define {\it isolated} photons. 
However, recent analyses of the production of isolated photons 
in electron-positron and proton-antiproton collisions 
have shown that 
the application of a geometrical isolation cone surrounding the 
photon does not lead to a reasonable agreement between theoretical 
prediction and experimental data. 
This discrepancy has important consequences.
Data on high $p_{T}$ photon production are commonly used 
to extract informations on the gluon distribution in the proton \cite{gluon}.
Furthermore, the understanding
of final state photon radiation in proton-proton 
collisions is even more crucial for 
searches for the two photon decay mode of the Higgs boson 
at the LHC \cite{LHC}.

An alternative approach to study final state photons produced in 
a hadronic environment is obtained by applying the so-called democratic 
clustering procedure \cite{andrew}. In this approach, the photon is 
treated like any other hadron and is clustered simultaneously 
with the other hadrons into jets. Consequently, one of the 
jets in the final state contains a photon and is labelled `photon jet'
if the fraction of electromagnetic energy within the jet is 
sufficiently large,
\begin{equation}
z=\frac{E_{EM}}{E_{EM} +E_{HAD}}>\zcut,
\label{eq:zdef}
\end{equation}
with $\zcut$ determined by the experimental conditions.
This photon is called  
{\it isolated} if it carries more than a certain fraction, typically 95\%,
of the jet energy and said to be non-isolated otherwise.
Note that this separation is made by studying the experimental $z$ 
distribution and is usually such that hadronisation effects,
which tend to reduce $z$, are minimized.
The 
cross section for the production of isolated photons defined in this 
approach thus receives sizeable contributions from both direct 
photon and fragmentation processes. 

Up to now, 
this democratic procedure has only been applied by the ALEPH
collaboration at CERN 
in an analysis of two jet events in electron-positron annihilation 
in which one of the jets contains a highly energetic photon
\cite{aleph}. 
In this analysis, 
ALEPH 
made a leading order determination of
the quark-to-photon fragmentation function
by comparing the 
photon +~1 jet rate calculated up to ${\cal O}(\alpha)$ 
\cite{andrew} with the data.
Then, by inserting their measurement into 
an ${\cal O}(\alpha\alpha_s)$ calculation of the `isolated' 
photon +~2 and 3 jet rates,
they found good agreement for a wide range of values of the jet 
resolution parameter  $\ycut$. 

In addition,
the ${\cal O}(\alpha)$ `isolated' photon +~1 jet rate agreed well with the 
ALEPH data.   This is remarkable, since with the use of an isolation criterion 
based on the application of a geometrical cone
previously considered \cite{iso}, the agreement between data and theory for this rate was poor and only improved 
at the expense of large negative radiative corrections  \cite{gs,ks,kt,cone}.
The main reason for the size and sign of these corrections was due to the
exclusion of soft gluons from the cone containing the photon \cite{softrad}.

Recently, we have presented a next-to-leading order determination of the quark to photon fragmentation function \cite{letter} based on
an ${\cal O}(\alpha\alpha_s)$
calculation of the photon +~1 jet rate and the ALEPH data.
Both theoretical and experimental analyses were performed 
using a democratic clustering approach;
in particular both analyses used the Durham jet recombination algorithm 
\cite{Durham}.
Not only is the agreement with the `isolated' photon +~1 jet rate improved,
but the higher order corrections 
obtained in this approach were found to be moderate, demonstrating the 
perturbative stability of this particular photon definition.
In this paper, we describe the calculation in more detail.
In particular, we extend the commonly used phase space slicing technique 
\cite{kramer,gg} 
developed to isolate the divergences when one of the final state particles
is unresolved to situations where two particles are unresolved.

The paper is organized as follows.
In section~\ref{sec:njet} we discuss the general structure of photon + $n$~jet
events in electron-positron collisions, with particular emphasis on where 
the singularities arise and how they are absorbed into the 
photon fragmentation functions.
In this fixed order approach, it is critical that the power counting of the various contributions is correct.  One consequence is that 
the commonly held view \cite{alphastr}
that the quark-to-photon fragmentation function is ${\cal O}(\alpha/\alpha_s)$ is
untenable.  The particular features of the photon +~1 jet rate are discussed in section~\ref{sec:1jet}.  We adopt the resolved parton philosophy of \cite{gg}
and introduce a small parameter $\ymin$ to separate the final 
state phase space into resolved and unresolved regions 
and to isolate the singularities lying in the unresolved regions.
Within this approach,
it is vital that the various singular regions precisely match onto each other.
The evaluation of the  associated  
different divergent contributions to the photon +~1 jet rate 
is presented in sections~\ref{sec:single}--\ref{sec:frag}.
Fully resolved and single unresolved contributions from the four parton process $\gamma^*\to q\bar q \gamma g$ are described in
section~\ref{sec:single} while the double unresolved contributions that arise when, for example, three particles are simultaneously collinear are
analytically evaluated in section~\ref{sec:double}.
Contributions from the one-loop process with collinear final state partons are
presented in section~\ref{sec:virt} and the fragmentation contribution is presented in section~\ref{sec:frag}.
All divergent contributions are combined in section~\ref{sec:total},
where we explicitly show that after factorization of collinear singularities, no divergences remain.    
Although the individual contributions are finite, they do depend on the 
parameter $\ymin$.  Clearly, the physical cross section cannot depend on
$\ymin$, and in section~\ref{sec:ymin} we numerically show that, 
provided  $\ymin$ is taken small enough, this is indeed the case.
We combine  our numerical results with the experimental data 
in section~\ref{sec:results} and determine 
the next-to-leading order quark-to-photon fragmentation function 
for two values of $\alpha_s(M_Z)$.  
Finally, section~\ref{sec:conc} contains our conclusions.

\section{The $n$ jet + photon cross section}
\setcounter{equation}{0}
\label{sec:njet}

Let us first consider the general structure of the
$e^+e^- \to n$ jet + photon cross section, fully 
differential in all quantities,
\begin{equation}
{\rm d}\sigma( n~{\rm jets} + ``\gamma  ")
=
{\rm d}\hat\sigma_\gamma+ \sum_a 
{\rm d}\hat\sigma_a \otimes \DBa.
\label{eq:gstruc1}
\end{equation}
There are two contributions, first the `prompt' photon production 
where the photon is produced directly in the hard interaction 
and second, the longer distance
fragmentation process where one of the partons fragments into a photon 
and transfers a fraction
of the parent momentum to the photon.
Each type of parton, $a$, contributes according to the process independent 
parton-to-photon 
fragmentation functions $\DBa$ and the sum runs over all partons.
Note that although the fragmentation functions are non-perturbative, 
we can nominally assign a power of coupling constants, based on counting the couplings necessary to radiate a photon.  
Since the photon couples directly to the quark, $\Dq$ is naively of ${\cal O}(\alpha)$ while the gluon can only couple to the photon via a quark 
and we might expect that $\Dg$ is of ${\cal O}(\alpha\alpha_s)$.
This simplistic argument is supported by models of the fragmentation
function \cite{grv} which suggest that gluon fragmentation is a much smaller effect than quark fragmentation.

The individual terms in eq.~(\ref{eq:gstruc1}) may be divergent and are
denoted by hatted quantities.
However, we can reorganise them as follows.
In the most general case we have,
\begin{eqnarray}
{\rm d}\hat\sigma_\gamma  &=& {\rm d}\sigma_\gamma 
+ \sum_q {\rm d}\sigma_q \otimes \Gqp
+ {\rm d}\sigma_g \otimes \Ggp, \nonumber \\
{\rm d}\hat\sigma_q  &=& \sum_{q'} {\rm d}\sigma_{q'}\otimes \Gqpq 
+ {\rm d}\sigma_g\otimes \Ggq, \nonumber \\
{\rm d}\hat\sigma_g  &=& \sum_{q} {\rm d}\sigma_{q}\otimes \Gqg + 
{\rm d}\sigma_g\otimes \Ggg, 
\label{eq:gstruc4}
\end{eqnarray}
where all of the divergences are concentrated in the factorization scale
dependent 
transition functions $\Gij$. The sum covers all active quark and antiquark 
flavours.
The process specific cross sections ${\rm d}\sigma_\gamma$,  
${\rm d}\sigma_q$, ${\rm d}\sigma_g$ are now finite.
Inserting these definitions back into eq.~(\ref{eq:gstruc1}), 
yields a physical cross section in terms of finite quantities, 
\begin{equation}
{\rm d}\sigma( n~{\rm jets} + ``\gamma  ")
=
{\rm d}\sigma_\gamma+ \sum_q 
{\rm d}\sigma_q \otimes \Dq + {\rm d}\sigma_g \otimes \Dg
\label{eq:gstruc2}
\end{equation}
where the physical (and factorization scale dependent)
fragmentation functions are given by,
\begin{eqnarray}
\Dq & = & \Gqp + \sum_{q'} \Gqqp \otimes \DBqpp + \Gqg \otimes \DBgp ,
\nonumber\\
\Dg & = & \Ggp + \sum_{q} \Ggq \otimes \DBqp + \Ggg \otimes \DBgp .
\end{eqnarray}

While the singular parts of the transition functions are process independent
and well known, it is sometimes convenient to carry some finite 
perturbative pieces as well \cite{andrew}, so that,
\begin{eqnarray}
{\rm d}\sigma_\gamma^R &=& {\rm d}\sigma_\gamma - \sum_q {\rm d}\sigma_q \otimes \aqp - {\rm d}\sigma_g \otimes \agp,\nonumber\\
\Gqp^R &=& \Gqp  + \aqp, \nonumber\\
\Ggp^R &=& \Ggp  + \agp, 
\end{eqnarray}
and to define the `effective' parton-to-photon fragmentation functions, 
\begin{eqnarray}
\calDq &=& \Dq + \aqp,\nonumber\\
\calDg &=& \Dg + \agp.
\end{eqnarray}
With these modifications that are particularly suited to the 
resolved parton philosophy \cite{gg}
we shall employ later on to isolate the singularities in ${\rm d}\hat\sigma_\gamma$ and 
${\rm d}\hat\sigma_a$, the cross section is given by,
\begin{equation}
{\rm d}\sigma( n~{\rm jets} + ``\gamma  ")
=
{\rm d}\sigma_\gamma^R+ \sum_q 
{\rm d}\sigma_q^R \otimes \calDq 
+{\rm d}\sigma_g^R \otimes \calDg,
\label{eq:gstruc3}
\end{equation}
where the superscript $R$ indicates that the partons are `resolved' 
\cite{gg}.
Within the fixed order approach,  
only the `prompt' photon process contributes at lowest order, and,
\begin{eqnarray}
{\rm d}\hat\sigma_\gamma^{(LO)} &=& {\rm d}\sigma_\gamma^{R(LO)}
= \Theta ~{\rm d}\hat\sigma_0(n~p+\gamma), \nonumber\\
{\rm d}\hat\sigma_q^{(LO)} &=& 0,\nonumber\\
{\rm d}\hat\sigma_g^{(LO)} &=& 0.
\end{eqnarray}
The $n$~parton + 
photon cross section ${\rm d}\hat\sigma_0(n~p+\gamma)$ is 
evaluated in the tree approximation and $\Theta$ represents the
experimental jet and photon definition cuts.  In this way the theoretical
cross section can be matched onto the specific experimental details.
At this order each parton is identified as a jet and the photon 
as a photon.

At next-to-leading order, both `prompt' photon production and 
the quark-to-photon fragmentation process contribute, 
\begin{eqnarray}
{\rm d}\hat\sigma_\gamma^{(NLO)}  &=& \Theta
 \left(    {\rm d}\hat\sigma_1(n~p+\gamma)
     +   \int {\rm d}\hat\sigma_0((n+1)~p+\gamma) \right), \nonumber\\
{\rm d}\hat\sigma_q^{(NLO)}&=&    \Theta ~{\rm d}\hat\sigma_0((n+1)~p),\nonumber\\
{\rm d}\hat\sigma_g^{(NLO)} &=& 0.
\end{eqnarray}
`Prompt' photon production may occur via the one loop virtual 
corrections to the 
$n$~parton + photon process $ {\rm d}\hat\sigma_1(n~p+\gamma)$
or by the tree level
emission of an additional parton ${\rm d}\hat\sigma_0((n+1)~p+\gamma)$.
The integral sign represents the integration over the additional phase space 
variables due to the presence of an extra parton in the final state.
The fragmentation  contribution 
arises from the lowest order $n+1$ parton process
with one quark fragmenting into a photon.
The remaining unfragmented partons are directly identified as jets.
Although the physical cross section is finite, 
individual
contributions are divergent.  
In ${\rm d}\hat\sigma_\gamma^{(NLO)}$,  there are infrared singularities 
arising from configurations where one of the partons is theoretically 
`unresolved'.
For example,
the virtual graphs contain 
singularities
due to soft gluons or collinear partons while  similar 
divergences occur in the bremsstrahlung process.
The correct treatment of infrared divergences is well known \cite{BN,KLN} 
and has been discussed widely in the literature.
Many general approaches have been developed to isolate the infrared 
divergences at next-to-leading order \cite{gg,cs,fks,nt}.
In \cite{andrew}, the approach of  \cite{gg} was used to isolate 
the divergences present in the bremsstrahlung contributions and to 
combine them 
with those arising from the virtual graphs.
While the leading infrared singularities due to soft gluon
emission should
cancel within ${\rm d}\hat\sigma_\gamma^{(NLO)}$, some collinear
mass singularities remain.
These singularities, which occur
as the quark and photon become collinear, 
are proportional to ${\rm d}\hat\sigma_q^{(NLO)}$,
are factorizable and can be absorbed by a redefinition of the
fragmentation function as described above.
Working within the $\overline{MS}$ scheme \cite{msbar},
the transition functions for a quark of charge $e_q$ are up to 
${\cal O}(\alpha)$ given by,
\begin{eqnarray}
\Gqqp & = & \delta_{qq'},\nonumber\\
\Gqp &=& \left(\frac{\alpha  e_q^2}{2\pi}\right) 
\frac{1}{\Gamma(1-\epsilon)}\left(\frac{4\pi\mu^2}{\mu_{F}^2}\right)^{\e}\,
\left[-\frac{1}{\e}\Pqpzero\right],
\end{eqnarray}
where $\Pqpzero$ is the $\e \to 0$ part
of the the lowest order splitting function in $(4-2\e)$-dimensions \cite{AP},
\begin{equation}
\Pqp(z) = \frac{1+(1-z)^2-\e z^2}{z}.
\label{eq:pqp}
\end{equation}
Here, the variable $z$ represents the fraction of the `photon' energy
that is genuinely electromagnetic in origin in the quark-photon cluster.

Each term in the `prompt' photon contribution is
finite and may be evaluated numerically,
\begin{equation}
{\rm d}\sigma_\gamma^{R(NLO)}  = \Theta 
 \left(    {\rm d}\sigma^R_1(n~p+\gamma)
     +   \int {\rm d}\sigma^R_0((n+1)~p+\gamma) \right),
\end{equation}
where the superscript $R$ indicates that the partons are `resolved' 
\cite{gg}.
For the numerical evaluation of the cross section 
within the resolved parton approach it is 
moreover useful to identify a universal  
constant piece and to absorb it into the `effective' fragmentation 
function~\cite{andrew}, 
\begin{equation}
\calDq(z) = D_{q \to \gamma}(z,\mu_F)
+ \left(\frac{\alpha  e_q^2}{2\pi}\right)\left(
\log\left(\frac{\smin s_{q\bar q}z (1-z)}{M^2 \mu_F^2}\right)\Pqpzero(z)
 + z \right),
\label{eq:calDq}
\end{equation}
where M is the mass of the final state and $s_{q \bar q}$
is the invariant mass of the radiating quark-antiquark system.
We see that the effective quark-to-photon fragmentation function, $\calDq$,
 depends
on the unphysical parton resolution parameter $\smin$ introduced 
to isolate the singularities.   Physical
cross sections cannot depend on $\smin$, and there must be a cancellation between the resolved prompt photon and 
fragmentation contributions.  In ref.~\cite{andrew} this was 
shown to be the case.
The genuine non-perturbative fragmentation function 
$D_{q \to \gamma}(z,\mu_F)$
cannot be calculated and needs to be extracted from data. 

At next-to-next-to-leading order, 
the `prompt' photon contribution now contains
contributions from
the two-loop $n$~parton + photon process, as well
as the one-loop $(n+1)$-parton + photon and 
tree level $(n+2)$-parton + photon processes.
Similarly, the quark-to-photon fragmentation term receives
contributions from the 
one-loop $(n+1)$-parton process and the 
tree level  $(n+2)$-parton process, while the gluon-to-photon fragmentation 
term, appearing for the first time at this order, only receives a 
contribution from the tree level $(n+1)$-parton process,
\begin{eqnarray}
{\rm d}\hat\sigma_\gamma^{(NNLO)}  &=& \Theta 
 \Biggl(    {\rm d}\hat\sigma_2(n~p+\gamma)
     +   \int {\rm d}\hat\sigma_1((n+1)~p+\gamma)  \nonumber \\
&& \qquad + \int\int {\rm d}\hat\sigma_0((n+2)~p+\gamma) \Biggr ),
\nonumber \\
{\rm d}\hat\sigma_q^{(NNLO)}&=&    \Theta 
\left (
{\rm d}\hat\sigma_1((n+1)~p) + \int {\rm d}\hat\sigma_0((n+2)~p)
\right ), \nonumber \\
{\rm d}\hat\sigma_g^{(NNLO)}&=&    \Theta 
\;{\rm d}\hat\sigma_0((n+1)~p) .
\end{eqnarray}
As at next-to-leading order, the singularities present in the `prompt'
photon bremsstrahlung and in the  
fragmentation processes can be isolated. 
However, unlike at next-to-leading order,
there are contributions to the `prompt' photon 
bremsstrahlung process where {\em two} of the 
final state partons can be simultaneously theoretically unresolved.
These contributions are expected to appear in the divergent part of
any jet cross section evaluated 
at next-to-next-to-leading order. 
In the most general case, the following four double unresolved 
configurations can occur: 
\begin{itemize}
\item The {\it triple collinear} configuration, where three partons 
become simultaneously collinear to each other, but none are soft.
\item The {\it soft/collinear} configuration, where two partons become 
collinear while a third parton becomes soft.
\item The {\it double single collinear} configuration, where 
two distinct, independent pairs of partons become simultaneously 
collinear. 
\item The {\it double soft} configuration, where two partons become 
simultaneously soft.
\end{itemize}
The photon +~1 jet cross section at ${\cal O}(\alpha \alpha_s)$, calculated 
in the remainder of this paper, receives only contributions 
from three of these 
double unresolved configurations (triple collinear quark-photon-gluon, 
soft gluon/collinear photon and double single collinear
 quark-photon/antiquark-gluon) and the corresponding double unresolved 
factors will be calculated analytically in section~\ref{sec:double}.
There are also singular contributions arising when a gluon is virtual while another 
parton is collinear or soft. In our calculation, we 
have only a {\em virtual collinear}
contribution, studied in section~\ref{sec:virt}, as the photon cannot be soft.
Finally, another new class of singularities occurs
when together with the remnants 
of the quark/photon fragmentation cluster an unresolved 
or virtual gluon is emitted.
We will explicitly evaluate this  {\em fragmentation}
contribution in section~\ref{sec:frag}.
Having isolated the singular contributions in this way allows us to define 
finite `resolved' parton cross sections for 
both `prompt' photon and fragmentation processes which 
in the most general case reads, 
\begin{eqnarray}
{\rm d}\sigma_\gamma^{R(NNLO)}  &=& \Theta
 \Biggl(    {\rm d}\sigma_2^R(n~p+\gamma)
     +   \int {\rm d}\sigma_1^R((n+1)~p+\gamma)  \nonumber \\
&& \qquad + \int\int {\rm d}\sigma_0^R((n+2)~p+\gamma) \Biggr ),\nonumber\\
{\rm d}\sigma_q^{R(NNLO)}&=&    \Theta 
\left (
{\rm d}\sigma_1^R((n+1)~p) + \int {\rm d}\sigma_0^R((n+2)~p)
\right ),\nonumber\\
{\rm d}\sigma_g^{R(NNLO)}&=&    \Theta 
\;{\rm d}\sigma_0^R((n+1)~p)\label{eq:rnnlog}. 
\end{eqnarray}
Each resolved cross section is finite and can be numerically 
evaluated for arbitrary experimental constraints.

The most singular divergences are due to soft gluon radiation and
cancel between the virtual and real processes.   However, 
double and single poles in $\e$ due to  
collinear singularities 
remain, whose form is determined by the known 
next-to-leading order transition functions.
Keeping only terms that are necessary to expand 
eq.~(\ref{eq:gstruc4}) up to ${\cal O}(\alpha\alpha_s)$ and bearing in mind that $\Dg$ is 
already  ${\cal O}(\alpha\alpha_s)$, then in the 
$\overline{MS}$--scheme,  
\begin{eqnarray}
\Gqp&=&\aqed   
\frac{1}{\Gamma(1-\epsilon)}\left(\frac{4\pi\mu^2}{\mu_{F}^2}\right)^{\e}\,
\left[-\frac{1}{\e}\Pqpzero\right]\nonumber
\\
& & +\aqed\as
\frac{1}{\Gamma^2(1-\epsilon)}\left(\frac{4\pi\mu^2}{\mu_{F}^2}\right)^{2\e}\,
\left[ \frac{1}{2\e^2}\Pqqzero\otimes \Pqpzero  
-\frac{1}{2 \e}\Pqpone\right],\nonumber \\
\Ggp&=&  \aqed\left(\frac{\alpha_s T_R}{2\pi}\right)
\frac{1}{\Gamma^2(1-\epsilon)}\left(\frac{4\pi\mu^2}{\mu_{F}^2}\right)^{2\e}\,
\left[ \frac{1}{2\e^2}\sum_q \Pgqzero\otimes \Pqpzero  
-\frac{1}{2 \e}\Pgpone\right],\nonumber\\
\Gqqp&=& \left( 1
+\as
\frac{1}{\Gamma(1-\e)}
\left(\frac{4\pi\mu^2}{\mu_{F}^2}\right)^{\e}
\left[-\frac{1}{\e}\Pqqzero\right]\right) \delta_{qq'},\nonumber\\
\Ggq & = & \left(\frac{\alpha_s T_R}{2\pi}\right)
\frac{1}{\Gamma(1-\e)}
\left(\frac{4\pi\mu^2}{\mu_{F}^2}\right)^{\e}
\left[-\frac{1}{\e}\Pgqzero\right],\nonumber\\
\Gqg & = & 0,\nonumber\\
\Ggg & = & 1,
\end{eqnarray}
where the QCD casimirs $C_F$ and $T_R$ are,
\begin{displaymath}
C_F = \frac{N^2-1}{2N},\qquad
T_R = \frac{1}{2}.
\end{displaymath}
The lowest order quark-to-quark and gluon-to-quark 
splitting functions are \cite{AP}, 
\begin{eqnarray}
\Pqqzero(z) &=& \frac{1+z^2}{(1-z)_+}+\frac{3}{2}\delta(1-z),
\label{eq:pqq}\\
\Pgqzero(z) &=& z^2 + (1-z)^2,
\label{eq:pgq} 
\end{eqnarray}
while the next-to-leading order quark-to-photon and gluon-to-photon
splitting functions are given by~\cite{curci,rijken}, 
\begin{eqnarray}
\Pqpone(z)&=&  
-\frac{1}{2} +\frac{9}{2}z + \left(-8 +\frac{1}{2}z \right) \ln z 
+2z\ln(1-z)+\left(1-\frac{1}{2}z\right)\ln^2 z  
\nonumber\\
& &  
+\left[\ln^2(1-z)+4 \ln z \ln(1-z) +8\Li_2(1-z) -\frac{4}{3}\pi^2\right]
\Pqpzero(z), \\
\Pgpone(z) &=&
-2 + 6z -\frac{82}{9} z^2 + \frac{46}{9z} + \left( 
5 +7z + \frac{8}{3} z^2 + \frac{8}{3z} \right) \ln z +  (1+z) \ln^2
z .
\end{eqnarray}
The  transition functions $\Gij$ 
determine the singularity structure of the bare fragmentation functions
up to ${\cal O}(\alpha \alpha_s)$,
\begin{eqnarray}
\DBqp(z)&=&
\sum_{q'} \Gqqp^{-1} \otimes \Dqp (z,\mu_F)
-\sum_{q'} \Gqqp^{-1} \otimes \Gqpp
+\Gqg^{-1} \otimes \Dg (z,\mu_F)
\nonumber\\
&=&\Dq(z,\mu_{F})\,
+\aqed  
\frac{1}{\Gamma(1-\e)}
\left(\frac{4\pi\mu^2}{\mu_{F}^2}\right)^{\e}
\frac{1}{\e}\Pqpzero(z) 
\nonumber\\
& & +\as  
\frac{1}{\Gamma(1-\e)}
\left(\frac{4\pi\mu^2}{\mu_{F}^2}\right)^{\e}
\frac{1}{\e}\Pqqzero\otimes \Dq(z,\mu_{F})
\nonumber\\
& & +
\aqed\as
\frac{1}{\Gamma^2(1-\e)}
\left(\frac{4\pi\mu^2}{\mu_{F}^2}\right)^{2\e}
\;\left[\frac{1}{2\e^2}\Pqqzero\otimes \Pqpzero
+\frac{1}{2\e}\Pqpone\right],
\nonumber\\
\label{eq:counter}\\
\DBgp(z)&=&
\Ggg^{-1} \otimes \Dg(z,\mu_F)  -\Ggg^{-1} \otimes \Ggp 
+ \sum_q \Ggq^{-1} \otimes \Dq (z,\mu_F)
\nonumber\\
& = & \Dg(z,\mu_{F})\, + \sum_q
  \left(\frac{\alpha_s T_R}{2\pi}\right)    
\frac{1}{\Gamma(1-\e)}
\left(\frac{4\pi\mu^2}{\mu_{F}^2}\right)^{\e}
\frac{1}{\e}\Pgqzero\otimes \Dq(z,\mu_{F})\nonumber\\
& & + \sum_q
\left(\frac{\alpha_s T_R}{2\pi}\right) \,  
\aqed
\frac{1}{\Gamma^2(1-\e)}
\left(\frac{4\pi\mu^2}{\mu_{F}^2}\right)^{2\e}
\;\left[\frac{1}{2\e^2}\Pgqzero\otimes \Pqpzero
+\frac{1}{2\e}\Pgpone\right],
\nonumber\\
\end{eqnarray}
which must absorb the remaining collinear singularities present in 
the $\gamma +n$~jet cross section at next-to-next-to-leading order.
Note that in expanding the perturbative counterterms, we have systematically assumed that the gluon-to-photon fragmentation function is 
${\cal O}(\alpha\alpha_s)$, and therefore does not contribute to
$\DBq$ at this order.

Concentrating on the evaluation of the photon +~1 jet rate at fixed order, 
${\cal O}(\alpha \alpha_{s})$,
the production of gluons is suppressed 
by a power of $\alpha_{s}$ compared to the quark production. 
Therefore, the contribution from the gluon fragmentation to
the photon +~1 jet rate is of 
${\cal O}(\alpha \alpha_{s}^2)$ and must be neglected 
in a consistent fixed order framework.
Consequently, the quark-to-photon fragmentation function 
{\it alone} must absorb the remaining singularities and 
demanding that this is so provides a 
powerful check that the singularities have been correctly isolated.
This factorization procedure will be explicitly described in section~\ref{sec:total}.
Moreover,  
we note that evaluating the real double unresolved contributions
using a strong ordered approach, i.e. requiring that 
one unresolved parton is much less unresolved than the other,
fails to produce
the necessary singularity structure \cite{aude}.

\section{The `photon' + 1 jet cross section} 
\setcounter{equation}{0}
\label{sec:1jet}

We now concentrate on the case where only a single jet is produced 
in addition to the `photon'.    
The photon + 1~jet rate in electron-positron annihilation 
is rather particular, since the leading order 
process,  1~parton + `direct' photon production, vanishes. The lowest 
non-vanishing order~\cite{andrew}, ${\cal O}(\alpha)$,
is therefore equivalent to the next-to-leading order outlined 
above, where `direct' and fragmentation processes contribute at equal   
level. This observable is therefore 
particularly suitable  for determining the 
non-perturbative component of the quark-to-photon fragmentation function.

\begin{figure}[t]
\vspace{8cm}
\begin{center}
~\includegraphics{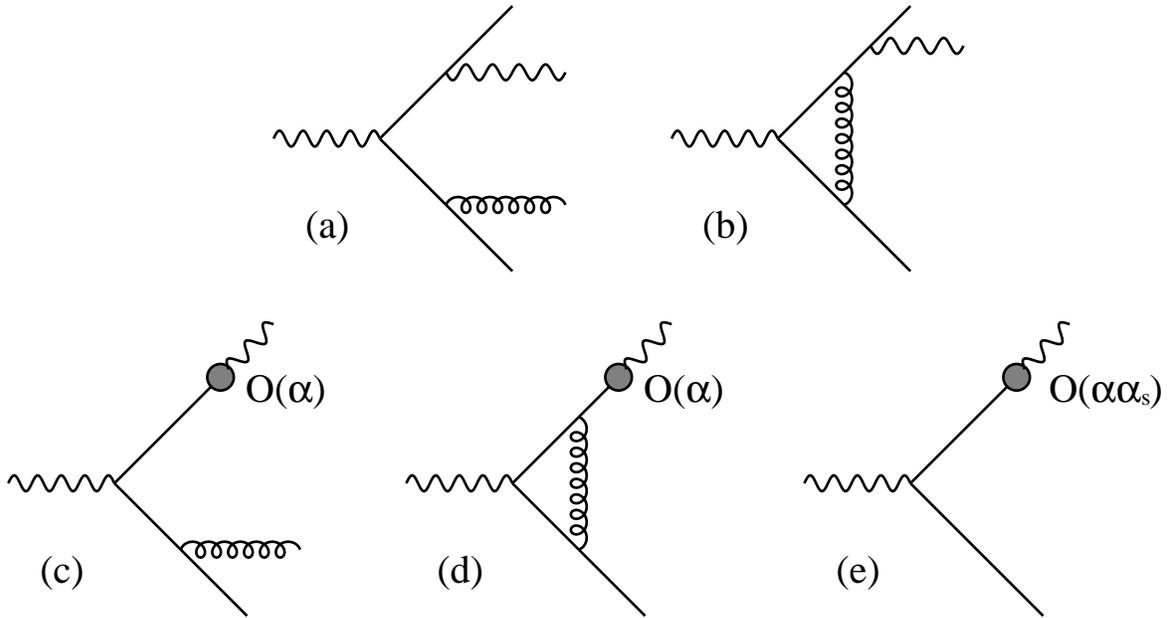}
\caption{Parton level subprocesses contributing to the photon +~1 jet 
rate at ${\cal O}(\alpha\alpha_s)$.} 
\label{fig:class}
\end{center}
\end{figure}

The parton level subprocesses contributing to the photon +~1 jet 
rate at ${\cal O}(\alpha\alpha_s)$ are shown in Fig.~\ref{fig:class}.
\begin{itemize}
\item[{(a)}] The tree level process $\gamma^* \to q \bar{q}g\gamma$,
where the final state particles are clustered together such that a
``photon jet'' and one additional jet are observed in the final state.
As the photon must be identified in the final state it cannot be soft.
\item[{(b)}] The one loop gluon correction to the $\gamma^* \to q
\bar{q}\gamma$ process, where the photon and one of the quarks are clustered
together or the photon is isolated while both quarks form a single jet.
\item[{(c)}] The process $\gamma^* \to q
\bar{q}g$, where one of the quarks fragments into a photon while
the remaining partons form only a single jet.
\item[{(d)}] The one loop gluon correction to the $\gamma^* \to q
\bar{q}$ process, where one of the quarks fragments into a photon.
\item[{(e)}] The tree level process $\gamma^* \to q\bar{q}$ with a
generic ${\cal O}(\alpha \alpha_{s})$ counterterm present in the {\it bare}
quark-to-photon fragmentation function.
Inclusion of this contribution absorbs all left over singularities of
the processes cited above.
\end{itemize}

Each of these processes gives a singular contribution to the final
photon +~1 jet rate.   Therefore, to combine them  together numerically
in a way that can match onto the precise experimental cuts, we must first
isolate the divergences analytically, so that the remaining finite
contributions may be numerically evaluated.  
We choose to employ the hybrid subtraction method
\cite{eec} which extends the phase space slicing method described in \cite{gg}.
In this approach, we introduce a parton resolution parameter $\ymin$ to decide 
whether or not two partons are theoretically resolved; if, for example,
 $y_{ij} < \ymin$,
then partons $i$ and $j$ are not resolved.  
In these unresolved regions, the matrix elements are singular, typically proportional to $1/y_{ij}$, and we analytically integrate approximate forms for the 
exact matrix elements over the restricted phase space.  
The essence of this approach can best be illustrated using the 
simple one-dimensional integral as suggested by 
Kunszt and Soper in \cite{soper},
\begin{equation}
{\cal I} = \lim_{\epsilon \to 0} \left \lbrace
\int^1_0 \frac{dx}{x} x^\epsilon F(x) -
\frac{1}{\epsilon} F(0)\right \rbrace.
\end{equation}
$F(x)$ is a complicated function, which renders the evaluation of 
${\cal I}$ analytically impossible.
${\cal I}$ could represent a $n$-jet cross section while  
$F(x)$ could stand for the
$(n+1)$-parton bremsstrahlung matrix elements and $x$ for a scaled invariant mass
$y_{ij}$.
As $x\to 0$, which corresponds to 
the case when one of the final state particles becomes soft or collinear, 
the integrand is regularized by the $x^\epsilon$ factor
as in dimensional regularization. The first term is however still
divergent as $\epsilon \to 0$.  This divergence is cancelled by the
second term - which is the equivalent of the $n$-parton one-loop contribution 
- so that the
integral is finite. 

In the {\it hybrid subtraction} method \cite{eec}, we choose to evaluate the integral by adding and subtracting the approximate matrix elements 
denoted by $F(0)$ in
the unresolved regions of phase space, so that,
$$
{\cal I}  \sim   \lim_{\epsilon \to 0} \left \lbrace
\int^1_0 \frac{dx}{x} x^\epsilon F(x)
-F(0) \int^{\ymin}_0 \frac{dx}{x} x^\epsilon
+F(0) \int^{\ymin}_0 \frac{dx}{x} x^\epsilon
- \frac{1}{\epsilon} F(0)\right \rbrace. 
$$
The approximate matrix elements are both simpler and process independent
and the integrations over the unresolved phase space can be carried out analytically,
$$
F(0) \int^{\ymin}_0 \frac{dx}{x} x^\epsilon = \frac{1}{\epsilon} 
F(0)\ymin^{\e},
$$
so that,
\begin{eqnarray}
{\cal I}&\sim &
\lim_{\epsilon \to 0} \left \lbrace
\int_{\ymin }^1 \frac{dx}{x} x^\epsilon F(x)
+\int_0^{\ymin}  \frac{dx}{x} x^\epsilon \left [ F(x)-F(0)\right ]\right.
\nonumber \\
& & \left.
+\frac{1}{\epsilon} 
F(0)\ymin^{\e}
- \frac{1}{\epsilon} F(0) \right \rbrace \nonumber \\
&\sim &
\left \lbrace
\int_{\ymin }^1 \frac{dx}{x} F(x)
+\int_0^{\ymin}  \frac{dx}{x}\left[F(x)-F(0) \right ]
+F(0)\ln(\ymin )\right \rbrace.
\label{eq:hybrid}
\end{eqnarray}
All three terms are now finite so that the $\e \to 0$ limit may be 
safely taken and, furthermore, are in a form suitable for numerical evaluation.

Of course, the parton resolution parameter $\ymin$ is unphysical and 
the integral ${\cal I}$ or equivalently any physical process should not
depend on $\ymin$.
The $\ymin$ dependence of the three terms in
eq.~(\ref{eq:hybrid}) should therefore cancel. 
In the evaluation of a jet cross section, this cancellation
is usually realised numerically by a Monte Carlo program.
This is not a straightforward point for the following reasons.
The matrix element approximations used in the analytic 
part of the calculation are reliable only when
$\ymin$ is small and are best when $\ymin$ is the smallest possible.
On the other hand, smaller values of $\ymin$ generate larger cancellations amongst the terms, possibly giving rise to 
numerical instability problems.
In practise $\ymin$ is chosen in such a way 
that the approximations performed in the analytic calculation are valid 
and that the numerical errors are minimized.

This approach, or the more basic phase space slicing method where the second term in eq.~(\ref{eq:hybrid}) is neglected, has been applied to a wide variety of processes;
$e^+e^- \to 2 ~{\rm jets}$, $e^+e^- \to 3 ~{\rm jets}$~\cite{gg},
$p\bar p \to W,Z +~1~{\rm jet}$, $p \bar p \to 2 ~{\rm jets}$
~\cite{ggk} and $e p \to e +~2 ~{\rm jets}$ ~\cite{mirkes}.
In all of these next-to-leading order calculations, there can be at most one unresolved particle; one soft gluon or two collinear partons.
In our calculation of the photon +~1 jet rate at 
${\cal O}(\alpha \alpha_{s})$, we will be concerned, 
for the first time, with situations where two partons are unresolved.

So far we have described the various parton level processes 
contributing to the photon +~1 jet rate at ${\cal O}(\alpha \alpha_{s})$ 
and merely outlined the general structure of the contributions associated 
with these processes.
Each of these processes contains different singular contributions 
that arise when one or more particles are theoretically unresolved.
We now consider each process in turn, and use the parton resolution parameter 
$\ymin$ to define for each of them, the resolved and unresolved contributions
from the resolved and unresolved regions of the phase space 
respectively.
As we will see in sections \ref{sec:single}-\ref{sec:frag}, 
all theoretically unresolved contributions have the common structure already encountered in \cite{gg}.
They can be written as the product of an unresolved factor
containing all the singularities and a resolved 
cross section. 

Note that in isolating the singularities, we are not concerned with the 
precise experimental jet definition. Provided the theoretical resolution parameter $\ymin$ is sufficiently small, any jet algorithm can be imposed numerically
on the resolved parton cross sections, ${\rm d}\sigma^R$.  

\begin{figure}[t]
\vspace{10cm}\begin{center}
~ \includegraphics{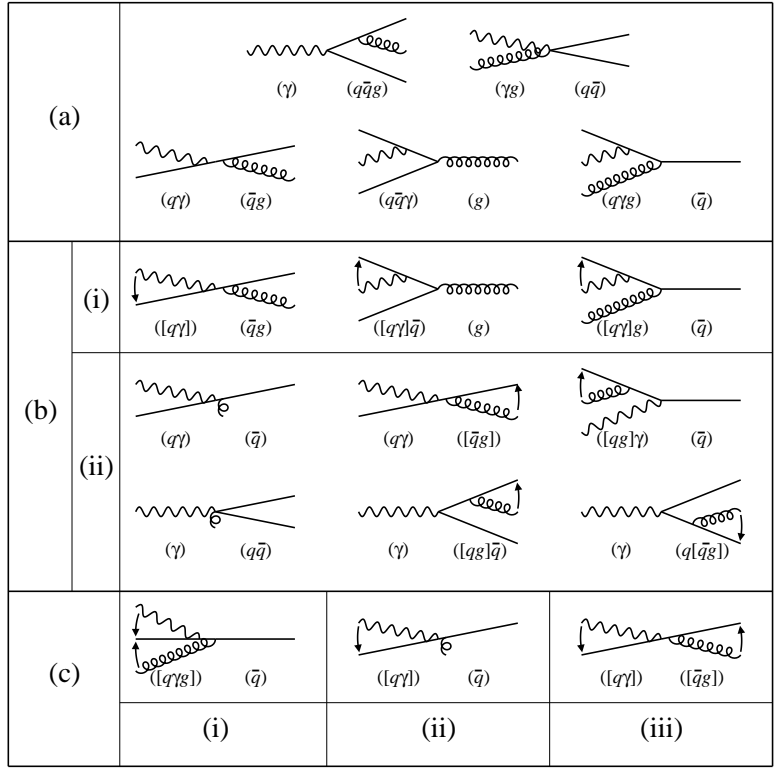}
\caption{Different final state `photon' + 1~jet topologies arising  
from the tree level  $\gamma^* \to q
\bar{q}\gamma g$ process.  The `photon' jet is moving to the left while the recoiling hadronic jet moves to the right.
Square brackets denote theoretically
unresolved particles, round brackets experimental clusters.}
\label{fig:clus4}
\end{center}
\end{figure}
\subsection{$\gamma^* \to q \bar{q}\gamma$ with real gluon bremsstrahlung}

The various configurations where the tree level process $\gamma^* \to q \bar{q}g\gamma$ contributes
to the photon +~1 jet rate are illustrated in Fig.~\ref{fig:clus4}.
Note that topologies where the role of quark and antiquark
are exchanged are also present, but are not shown.
Jets formed when the experimental jet algorithm clusters particles together
are denoted by $(ij)$, while theoretically unresolved clusters 
are denoted by $[ij]$.
The associated real contributions can be separated into three categories: either
theoretically resolved, single unresolved or double unresolved.
Care must be taken to divide the phase space so that the different regions 
match onto each other. We must ensure that the difference of exact and approximate matrix elements tends to zero as the singularity is approached and equally, there is no overcounting of the singularities. 
This can be done by constraining
the variables,
$$
y_{q\gamma},~~~~~y_{\bar q\gamma},~~~~~y_{qg},~~~~~y_{\bar qg},
$$
although in the {\em double unresolved} regions, 
we shall choose to constrain the
combinations,
$$
y_{q\gamma g},~~~~~~y_{\bar q\gamma g},
$$
for certain configurations. 
The matrix elements contain poles in each of these six invariants,
and are therefore singular as any of them tends to zero.
There are two invariants, $y_{g\gamma}$ and $y_{q\bar q}$, which are not associated with any singularities and are therefore completely unconstrained.

The individual configurations can be structured as follows:
\begin{itemize}
\item[{(a)}] {\bf Theoretically resolved contributions}
\newline
If all particles are resolved, a $\gamma$ +~1 jet event can only be
formed if some final state particles are clustered together
by the jet algorithm.
The possible configurations yielding a photon +~1 jet event are
displayed in Fig.~\ref{fig:clus4}.a.
In principle this region is defined by requiring that all invariants 
are greater than the theoretical parton resolution parameter, $\ymin$, i.e.
by requiring that,
\begin{equation}
y_{q\gamma}>\ymin,~~~~~y_{\bar q\gamma}>\ymin,
{}~~~~~y_{qg}>\ymin,~~~~~y_{\bar qg}>\ymin.
\label{eq:resolreg}
\end{equation}
However, it turns out that the boundaries of this region are more subtle 
than that and must be chosen to match onto 
the boundaries of the unresolved regions.
Consequently, the fully resolved region is defined as being 
the remaining phase space region of the four parton phase space 
when all unresolved regions are excluded.

\item[{(b)}]{\bf Single unresolved contributions}
\newline
As shown in Fig.~\ref{fig:clus4}.b, there are two classes of
single (or one-particle) theoretically unresolved
contributions, depending whether the gluon or the photon is unresolved.
The single unresolved regions associated to these contributions 
are defined as follows:
\begin{itemize}
\item[{(i)}] The {\it collinear quark-photon} region\newline
If the photon is unresolved, it is collinear to the quark while
the
gluon is {\it hard}, i.e.~the gluon is theoretically resolved
but combined with the photon-quark cluster or with the antiquark
by the experimental jet algorithm. Alternatively, the gluon forms a
jet on its own while the antiquark is clustered into the photon jet.
This can be defined by the constraints,
\begin{equation}
y_{q\gamma}<\ymin\, y_{q \bar q \gamma} ,
~~~~~y_{\bar q\gamma}>\ymin\, y_{q \bar q \gamma} ,
{}~~~~~y_{q\gamma g}>\ymin,~~~~~y_{\bar qg}>\ymin,
\label{eq:singlecolp}
\end{equation}
where $y_{q \bar q \gamma}$ is the scaled invariant mass of the radiating
quark-antiquark-photon antenna, $s_{q\bar q \gamma}/M^2$.
\item[{(ii)}] The {\it unresolved gluon} region\newline
If the gluon is theoretically unresolved, it can be soft or
collinear to the quark or to the antiquark, while the photon is
experimentally combined with the
quark to form the photon jet or is isolated while all other partons form
a single jet.
The three possible regions are defined by:
\begin{itemize}
\item[{(1)}] The {\it collinear quark-gluon} region 
\begin{equation}
y_{qg}<\ymin\, y_{q \bar q g} ,
~~~~~y_{\bar qg}>\ymin\, y_{q \bar q g} ,
{}~~~~~y_{q\gamma g}>\ymin,~~~~~y_{\bar q\gamma}>\ymin,
\label{eq:singlecolg}
\end{equation}
\item[{(2)}] The {\it collinear antiquark-gluon} region 
\begin{equation}
y_{\bar qg}<\ymin\, y_{q \bar q g} ,
~~~~~y_{qg}>\ymin\, y_{q \bar q g} ,
{}~~~~~y_{q\gamma g}>\ymin,~~~~~y_{\bar q\gamma}>\ymin,
\label{eq:singlecolgbar}
\end{equation}
\item[{(3)}] The {\it soft gluon} region 
\begin{equation}
y_{qg}<\ymin\, y_{q \bar q g} ,
~~~~~y_{\bar qg}<\ymin\, y_{q \bar q g} ,
{}~~~~~y_{q\gamma}>\ymin,~~~~~y_{\bar q\gamma}>\ymin.
\label{eq:singlesoft}
\end{equation}
\end{itemize}
As in the unresolved photon case, the mass of the antenna radiating the gluon, $y_{q \bar q g}$ plays a role in determining when the approximate matrix elements are used.
\end{itemize}

\item[{(c)}]{\bf Double unresolved contributions}
\newline
These contributions arise when both the photon and the gluon are
theoretically ``unseen'' in the final state.
As the photon has to be seen in the final state,
it can only be collinear with the quark and cannot be soft.
The gluon on the other hand can be collinear with the quark or with the
antiquark or it can be soft.
Corresponding to these different final state configurations we 
define three double unresolved phase space regions:
\begin{itemize}
\item[{(i)}]The {\it triple collinear} region\newline
The photon and the gluon are simultaneously collinear with the quark.
Here we need
to constrain the ``triple'' invariant
$y_{q\gamma g} \equiv  y_{q\gamma}+y_{qg}+y_{\gamma g}$
since it appears in the denominator of the four-particle matrix elements,
\begin{equation}
y_{q\gamma g} < \ymin \hspace{0.5cm} {\rm and}\hspace{0.5cm}
y_{\bar{q}g} >\ymin.
\label{eq:tricolinv}
\end{equation}
Note that this implies $y_{q\gamma} < \ymin$ and $y_{qg} < \ymin$.
We require $y_{\bar{q}g}>\ymin$ since in this region
the gluon is collinear but not soft.
\item[{(ii)}]The {\it soft/collinear} region\newline 
The photon is collinear with the quark while the gluon is soft,
\begin{equation}
y_{q\gamma}<\ymin  \hspace{0.5cm} {\rm and}\hspace{0.5cm}
y_{qg} <\ymin \hspace{0.5cm} {\rm and} \hspace{0.5cm}
y_{\bar{q}g}  <\ymin.
\label{eq:softcolinv}
\end{equation}
\item[{(iii)}]The {\it double single collinear} region\newline 
The photon is collinear with the quark while the gluon is collinear with
the antiquark,
\begin{equation}
y_{q\gamma} <\ymin   \hspace{0.5cm} {\rm and}\hspace{0.5cm}
y_{qg} >\ymin \hspace{0.5cm} {\rm and}\hspace{0.5cm}
y_{\bar{q}g} <\ymin.
\label{eq:doucolinv}
\end{equation}
\end{itemize}
For these three contributions, the final state configuration
corresponds already to a photon +1 jet event.
Hence, the final state particles will not be clustered further by the
jet algorithm.
These contributions are schematically
displayed in Fig.~\ref{fig:clus4}.c.
As before, these regions are matched by three analogous double unresolved
regions where the photon clusters with the antiquark.
\end{itemize}

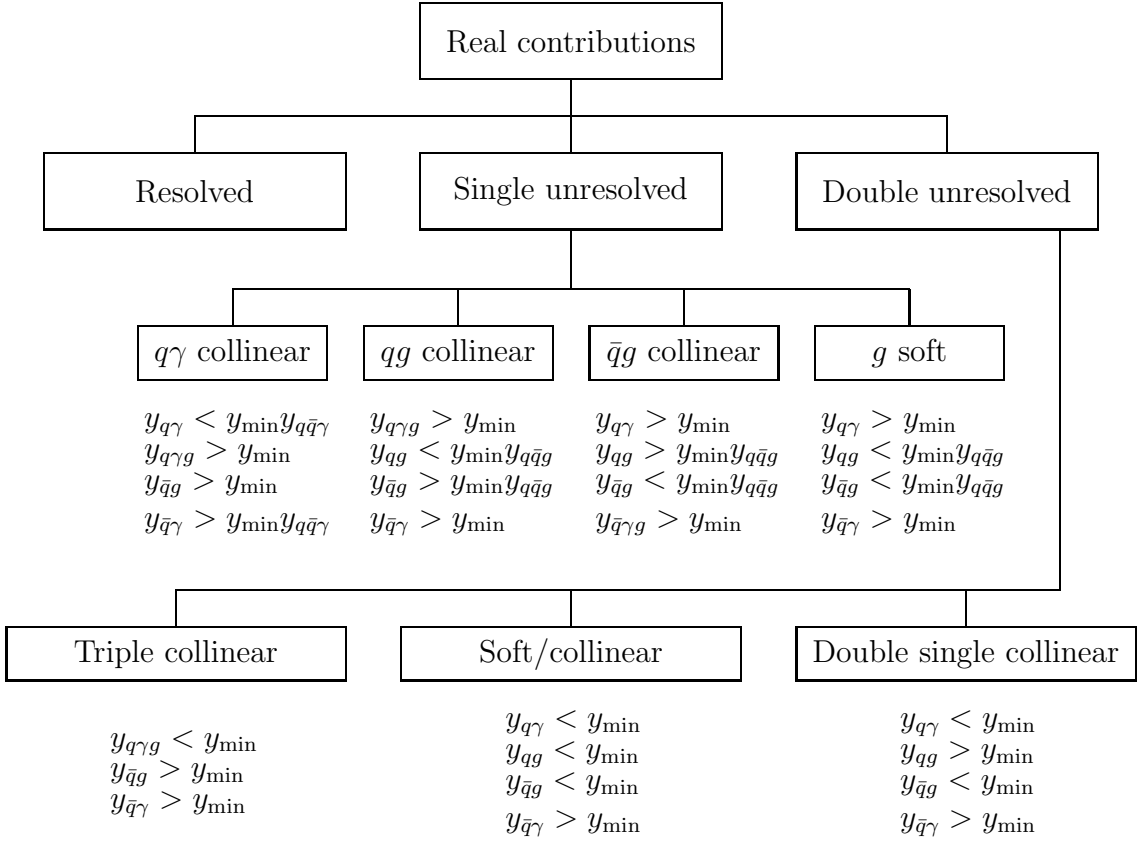
\begin{figure}[t]
\begin{center}
\unitlength1cm
\begin{picture}(16,10.5)
\put(6,9.5){\framebox(4,1){Real contributions}}
\put(1,7.5){\framebox(4,1){Resolved}}
\put(6,7.5){\framebox(4,1){Single unresolved}}
\put(11,7.5){\framebox(4,1){Double unresolved}}
\put(2.25,5.5){\framebox(2.5,0.7){$q\gamma$ collinear}}
\put(5.25,5.5){\framebox(2.5,0.7){$qg$ collinear}}
\put(8.25,5.5){\framebox(2.5,0.7){$\bar q g$ collinear}}
\put(11.25,5.5){\framebox(2.5,0.7){$g$ soft}}
\put(2.15,4.2){$\begin{array}{l} 
y_{q\gamma} < \ymin y_{q\bar q \gamma}
\vspace{-1mm} \\ 
y_{q\gamma g}> \ymin
\vspace{-1mm} \\ 
y_{\bar q g} > \ymin \\ 
y_{\bar q \gamma} > \ymin y_{q\bar q \gamma}\end{array}$}
\put(5.15,4.2){$\begin{array}{l} 
y_{q\gamma g}> \ymin\vspace{-1mm} \\ y_{qg} < \ymin y_{q\bar q g}
\vspace{-1mm} 
\\ y_{\bar q g} > \ymin y_{q\bar q g}
\\ y_{\bar q \gamma} > \ymin \end{array}$}
\put(8.15,4.2){$\begin{array}{l} 
y_{q\gamma}>\ymin\vspace{-1mm} \\ y_{qg} > \ymin y_{q\bar q g}
\vspace{-1mm} \\ y_{\bar q g} <  \ymin y_{q\bar q g}
\\ y_{\bar q \gamma g}> \ymin \end{array}$}
\put(11.15,4.2){$\begin{array}{l} 
y_{q\gamma} > \ymin\vspace{-1mm} \\ y_{qg} < \ymin y_{q\bar q g}
\vspace{-1mm} 
\\ y_{\bar q g}< \ymin y_{q\bar q g} \\ y_{\bar q \gamma}> \ymin \end{array}$}
\put(0.5,1.5){\framebox(4.5,0.7){Triple collinear}}
\put(5.75,1.5){\framebox(4.5,0.7){Soft/collinear}}
\put(11,1.5){\framebox(4.5,0.7){Double single collinear}}
\put(1.7,0.2){$\begin{array}{l} 
y_{q\gamma g}< \ymin\vspace{-1mm} \\ y_{\bar q g} > \ymin\vspace{-1mm} 
\\ y_{\bar q \gamma}>\ymin \end{array}$}
\put(6.95,0.2){$\begin{array}{l} 
y_{q\gamma}<\ymin\vspace{-1mm} \\ y_{qg} < \ymin\vspace{-1mm} 
\\ y_{\bar q g}<\ymin \\ y_{\bar q \gamma}>\ymin \end{array}$}
\put(12.2,0.2){$\begin{array}{l} 
y_{q\gamma}<\ymin\vspace{-1mm} \\ y_{qg} > \ymin\vspace{-1mm} 
\\ y_{\bar q g}<\ymin \\ y_{\bar q \gamma}>\ymin \end{array}$}
\put(8,9.5){\line(0,-1){1}}
\put(3,9){\line(1,0){10}}
\put(3,9){\line(0,-1){0.5}}
\put(13,9){\line(0,-1){0.5}}
\put(8,7.5){\line(0,-1){0.8}}
\put(3.5,6.7){\line(1,0){9}}
\put(3.5,6.7){\line(0,-1){0.5}}
\put(6.5,6.7){\line(0,-1){0.5}}
\put(9.5,6.7){\line(0,-1){0.5}}
\put(12.5,6.7){\line(0,-1){0.5}}
\put(14.5,7.5){\line(0,-1){4.8}}
\put(2.75,2.7){\line(1,0){11.75}}
\put(2.75,2.7){\line(0,-1){0.5}}
\put(8,2.7){\line(0,-1){0.5}}
\put(13.25,2.7){\line(0,-1){0.5}}
\end{picture}
\vspace{0.6cm}

\caption{Phase space decomposition
 of the real $\gamma^{\star}\to q\bar q \gamma g$
contributions. Note that the single and double unresolved
regions where the photon clusters with the antiquark are not shown.
For these regions, the necessary cuts are obtained by exchanging
$q$ and $\bar q$.
Altogether, there are five single unresolved and six double unresolved
regions.}
\label{fig:realps}
\end{center}
\end{figure}

The decomposition
of the four-particle phase space is summarized
in Fig.~\ref{fig:realps}.
In this table, we have specified which invariants are less than
$\ymin $ (or a cut proportional to $\ymin$) 
for each singular region of phase space.
We have also noted which invariants are greater than
$\ymin $ 
to eliminate overlaps between regions determined
by the same combinations of invariants less than $\ymin $.
Invariants that are not specified are completely unconstrained.

\subsection{$\gamma^* \to q \bar{q}\gamma$ with a virtual gluon}

The final state topology is similar to that for the lowest order
$\gamma^* \to q \bar{q}\gamma$ contribution discussed in 
\cite{andrew} and contributes
to the photon +~1 jet rate, if two of the final state partons 
are clustered together.
We distinguish the theoretically unresolved collinear
quark-photon contribution from the contributions where the 
theoretically resolved photon is
clustered with the quark by the experimental jet algorithm
to form the photon jet or 
isolated while quark and antiquark combine to form the jet.
The three particle phase space therefore divides as follows:
\begin{itemize}
\item[{(a)}]The {\it resolved photon} region \newline
\begin{equation}
y_{q\gamma}>\ymin, ~~~~~y_{\bar q \gamma}> \ymin.
\label{eq:resolvig}
\end{equation}
\item[{(b)}]The {\it quark-photon collinear} region \newline
\begin{equation}
y_{q\gamma}<\ymin, ~~~~~y_{\bar q \gamma}> \ymin,
\label{eq:unresolvig}
\end{equation}
plus a similar region for the collinear antiquark-photon configuration.
\end{itemize}

\subsection{$\gamma^* \to q \bar{q}(g)$ with fragmentation}
\begin{figure}[t]
\vspace{7cm}\begin{center}
~ \includegraphics{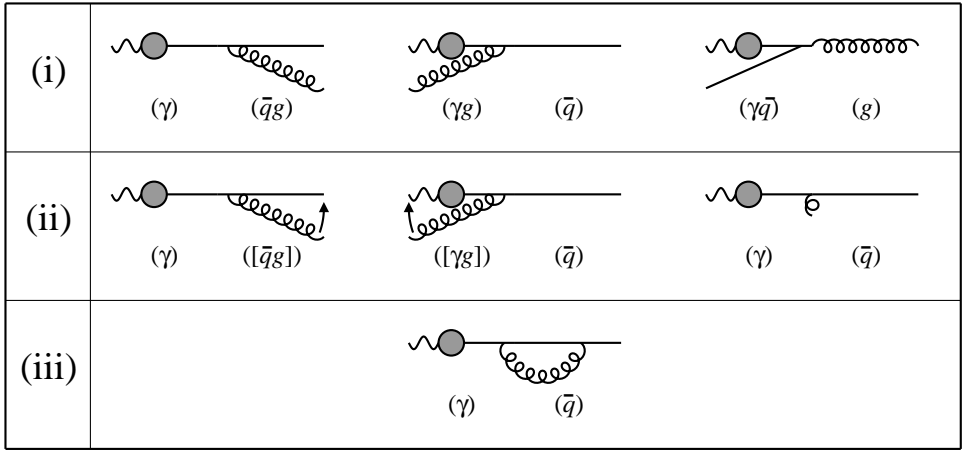}
\caption{Different final state `photon' + 1~jet topologies arising from the 
  $\gamma^* \to q
\bar{q} (g)$ process with subsequent fragmentation of the quark into a
photon. 
The  `photon'  jet is moving to the left while the recoiling hadronic jet moves to the right.
Square brackets denote theoretically
unresolved particles, round brackets represent experimental clusters.}
\label{fig:clusf}
\end{center}
\end{figure}

The ${\cal O}(\alpha_s)$ processes with associated fragmentation
shown in Fig.~\ref{fig:class}.c and \ref{fig:class}.d
contribute to the photon +~1 jet cross section if,
in addition to the photon-jet, only a single jet is formed.
As illustrated in Fig.~\ref{fig:clusf}, the gluon may be resolved, unresolved
or virtual, while the photon is produced via the fragmentation process. 
There are five distinct contributions:
\begin{itemize}
\item[{(i)}] The {\it  resolved gluon} region\newline
The real gluon is theoretically
resolved, but may be clustered by the experimental jet algorithm
with either the photon fragmentation cluster or with the
antiquark or it may form a jet on its own, while the antiquark is
combined with the photon/quark cluster.
\begin{equation}
y_{qg}>\ymin, ~~~~~y_{\bar q g}> \ymin.
\label{eq:resolregd}
\end{equation}
\item[{(ii)}] The {\it unresolved gluon region} \newline
If the gluon is theoretically unresolved, it can be soft or
collinear to the fragmenting quark or to the antiquark.
The three possible regions are defined by:
\begin{itemize}
\item[{(1)}]The {\it collinear quark-gluon} region\newline
\begin{equation}
y_{qg}<\ymin, ~~~~~y_{\bar q g}> \ymin.
\label{eq:fsinglecolg}
\end{equation}
\item[{(2)}]The {\it collinear antiquark-gluon} region\newline
\begin{equation}
y_{qg}>\ymin, ~~~~~y_{\bar q g}< \ymin.
\label{eq:fsinglecolgbar}
\end{equation}
\item[{(3)}]The {\it  soft gluon} region \newline
\begin{equation}
y_{qg}<\ymin, ~~~~~y_{\bar q g}< \ymin.
\label{eq:fsinglesoftg}
\end{equation}
\end{itemize}
\item[{(iii)}] The gluon is virtual.
\end{itemize}
Since this process is already of ${\cal O}(\alpha_s)$, only the ${\cal O}(\alpha)$
counterterm in the bare fragmentation function contributes.

\subsection{$\gamma^* \to q \bar{q}$ with fragmentation}
In addition to the processes described above involving a real or virtual
gluon, one has to consider a contribution to the photon +~1 jet rate
from the generic ${\cal O}(\alpha\alpha_{s})$ counterterm present in
the  bare fragmentation function of eq.~(\ref{eq:counter}).

\section{Resolved and Single unresolved contributions}
\setcounter{equation}{0}
\label{sec:single}

In the next two sections, we will discuss the contributions 
to the $\gamma $ +~1 jet rate at ${\cal O}(\alpha \alpha_{s})$ relevant to 
the process $\gamma^* \to q \bar {q}\gamma$ with real gluon bremsstrahlung.
As shown in Fig.~\ref{fig:clus4}, there are many different topologies 
contributing which can be divided according to the number of theoretically unresolved particles.
In this section, we consider the cases where at most one particle is unresolved, while the double unresolved contributions are studied in section~\ref{sec:double}.

The $\gamma^* \to q\bar q \gamma g$ process contributes to the $\gamma $ +~1 jet differential cross section if the final state configuration is such that only the photon jet and a single associated jet are observed.
The generic contribution of this four particle final state process
to the rate can be written as,
\begin{equation}
{\rm d}\sigma_{q\bar q \gamma g} 
= 
\as\aqed \mu^{4\e}
4 (2\pi)^4 \frac{1}{2M^2} 
\int |{\cal M}_{q\bar q \gamma g}|^2 {\rm d}P_4^{(d)}(M;p_{q},p_{\bar q},p_{\gamma},p_{g}).
\end{equation}
Here $|{\cal M}_{q\bar q \gamma g}|^2$ represents the squared matrix elements in $d$-dimensions, while the Lorentz invariant phase space for the decay of an off-shell 
photon with $p_{\gamma^*}^2 = M^2$  can be written,
\begin{eqnarray}
\int{\rm d}P_{4}^{(d)}(M;p_q,p_{\gamma},p_g,p_{\bar q}) & = & 
 \frac{M^{4-6\e}}{2^{7-4\e}(2\pi)^{8-6\e}}
\int {\rm d}\Omega_{d-1}\;{\rm d}\Omega_{d-2}\;{\rm d}\Omega_{d-3}
{\rm d}y_{q\bar q}\;{\rm d}y_{q\gamma}{\rm d}y_{qg}{\rm d}y_{\bar q\gamma}{\rm d}y_{\bar qg}
{\rm d}y_{\gamma g}
\nonumber\\
&\times & 
{(-\Delta_{4})}^{-1/2-\e} \;\delta(y_{q\bar q}+ y_{q\gamma}+y_{qg}
+y_{\bar q\gamma}+y_{\bar qg}+y_{\gamma g}-1),
\label{eq:4ps}
\end{eqnarray}
with,
\begin{equation}
\Delta_{4} = 
 {y_{q\bar q}}^{2}{y_{\gamma g}}^{2}\; +\;
{y_{q\gamma}}^{2}{y_{\bar qg}}^{2}\;+\;{y_{qg}}^{2}{y_{\bar q\gamma}}^{2}
-2\bigg(y_{q\bar q}y_{\bar q\gamma}y_{\gamma g}y_{qg}\;+\;y_{q\gamma}y_{\bar q\gamma}y_{\bar qg}y_{qg}\;
+\;y_{q\bar q}y_{\bar qg}y_{\gamma g}y_{q\gamma} \bigg) .
\end{equation}
The matrix elements are well known and contain terms up to ${\cal O}(\e^3)$.
These additional contributions, which vanish in 4-dimensions, are important in determining the contributions from the unresolved regions which may contain singularities up to ${\cal O}(1/\e^3)$.

\subsection{The resolved contributions}
\label{subsec:resol}

In the resolved phase space region which is the region of the four-particle 
phase space left over when all unresolved regions specified in 
Fig.~\ref{fig:realps} are excluded,
the matrix element squared is finite.
Thus we may take the $\e \to 0$ limit for both matrix elements and phase space,
so that
\begin{equation}
{\rm d}\sigma^{(R)}_{q\bar q \gamma g} 
= 
\as\aqed
4 (2\pi)^4 \frac{1}{2M^2} \int |{\cal M}_{q\bar q \gamma g}|^2 {\rm d}P_4^{(R)}(M;p_{q},p_{\bar q},p_{\gamma},p_{g}),
\end{equation}
where the phase space ${\rm d}P_4^{(R)}$ is restricted to this resolved region.
For these resolved contribution, the integration is 
carried out numerically and the photon and jet definitions 
applied directly to the final state particles.
As we are employing  the hybrid subtraction method to calculate 
the photon +1 jet cross section, 
there are additional contributions which are to be numerically calculated.
Those come from evaluating the difference between the exact 
and approximate squared matrix elements in the various unresolved regions of 
phase space.

\subsection{The single unresolved contributions}
\label{subsec:single}
There are two classes of single unresolved real contributions 
depending on whether the photon or the gluon is unresolved.
If the photon is unresolved, it is collinear to the 
quark while 
if the gluon is unresolved it can be collinear to the quark, 
collinear to the antiquark or it can be soft.
The possible final state configurations of single unresolved contributions yielding a $\gamma$ +1 jet event are displayed in Fig.~\ref{fig:clus4}.(b).
In each unresolved region of the four-particle phase space
we expect to be able to write the differential cross section 
as the product of a
{\it one particle unresolved } factor and a three-particle cross section
\cite{gg}.

\subsubsection{The unresolved gluon contribution}
\label{subsec:unresolg}

The phase space region where the quark and the gluon are collinear 
is defined by eq.~(\ref{eq:singlecolg}), 
\begin{displaymath}
y_{qg}<\ymin\, y_{q \bar q g} ,
~~~~~y_{\bar qg}>\ymin\, y_{q \bar q g} ,
{}~~~~~y_{q\gamma g}>\ymin,~~~~~y_{\bar q\gamma}>\ymin.
\end{displaymath}
Here, the quark and the gluon cluster to form a new parton
$Q$ such that,
\begin{displaymath}
p_{q}+p_{g}=p_{Q},
\end{displaymath}
and  carry respectively a fraction $y$ and $1-y$ of the parent
parton momentum $p_{Q}$.
The fractional momentum $y$ is defined with respect to the momenta carried 
by the colour connected particles: the quark, the antiquark and the gluon.
In particular $y$ is defined as,
\begin{equation}
y\equiv\frac{y_{\bar q g}}{y_{q\bar q g}},
\end{equation}
and, since $y_{\bar q g}>\ymin y_{q\bar q g}$,   
the lower boundary of the $y$ integral is $\ymin$.
In this limit the four particle invariants are related 
to the invariants of the three remaining particles,
\begin{equation} 
y_{q\bar q}=(1-y)\,y_{Q\bar q},\hspace{1cm} y_{\bar q g}=y\,y_{Q\bar q},\hspace{1cm}y_{q\bar q g}=y_{Q\bar q}
\end{equation}
while the invariants containing $p_{\gamma}$ become,
$$
y_{q \gamma}=(1-y)\,y_{Q\gamma},\hspace{1cm} y_{\gamma g}=y\,y_{Q\gamma}.
$$

The matrix elements and phase space exhibit an overall factorization,
\begin{equation}
|{\cal M}_{q\bar q \gamma g}|^2 \to P_{qg \to Q}(y, s_{qg})|{\cal M}_{Q\bar q \gamma}|^2,
\label{eq:singlecolM4}
\end{equation}
with,
$|{\cal M}_{Q\bar q \gamma}|^2$ the three-particle matrix element squared for the 
scattering of a quark-antiquark pair with a photon,
and,
\begin{equation}
P_{qg \to Q}(y, s_{qg})=\frac{2}{s_{qg}}\Pqg(y).
\label{eq:colfac}
\end{equation}
Similarly, the four particle phase space becomes,
\begin{equation}
{\rm d}P_{4}^{(d)}(M; p_{q}, p_{\bar q}, p_{\gamma},p_{g}) 
\to {\rm d}P_{3}^{(d)}(M;p_{Q},p_{\bar q},p_{\gamma})\,
{\rm d}P^{(d)}_{col}(p_{q},p_{g},y)
\label{eq:singlecolP4}
\end{equation}
where ${\rm d}P_{3}^{(d)}(M;p_Q,p_{\bar q},p_{\gamma})$ is the three-particle Lorentz invariant phase
space in $d$-dimensions.
The  collinear phase space factor reads \cite{gg},
\begin{equation}
{\rm d}P^{(d)}_{col}(p_{q},p_{g},y) =
\frac{(4\pi)^{\e}}{16\pi^2 \Gamma(1-\e)}{\rm d}s_{qg}\,
{\rm d}y \,s_{qg}^{-\e}y^{-\e}(1-y)^{-\e}.
\end{equation}

To evaluate the quark-gluon collinear factor, we need to integrate 
the collinear matrix element squared over the unresolved variables $y_{qg}$ and $y$.
Reinserting the overall coupling factor, we have,
\begin{eqnarray}
\tilde{C}_{F}(q) &=&
\int g_{s}^2 C_F \mu^{2\e}
P_{qg \to Q}(y,s_{qg}){\rm d}P^{(d)}_{col}(p_{q},p_{g},y)\nonumber \\
&=&
\as 
\left(\frac{4\pi\mu^2}{M^2}\right)^{\e} \frac{2}{\Gamma(1-\e)}
\int_{0}^{\ymin y_{q\bar q g}}{\rm d}y_{qg}\, y_{qg}^{-\e-1}
 \int_{\ymin}^{1} {\rm d}y y^{-\e}(1-y)^{-\e}\Pqg(y)
\nonumber\\
&=& -\frac{1}{\e}\as 
\left(\frac{4\pi\mu^2}{M^2}\right)^{\e}
\frac{1}{\Gamma(1-\e)}
\ymin^{-\e}y_{Q\bar q}^{-\e}
\left[\frac{2}{\epsilon}
\ymin^{-\e}\;-\;\frac{(1-\e)(4-\e)}
{2\e(1-2\e)}\;\frac{\Gamma^2(1-\e)}{\Gamma(1-2\e)}\;
\right].\nonumber \\
\label{eq:cfq}
\end{eqnarray}
We see that compared to the single quark-gluon collinear factor 
obtained considering the single quark-gluon collinear limit 
of a three parton cross section
found in \cite{gg}, the collinear factor given by  
eq.~(\ref{eq:cfq}) is multiplied by $y_{Q\bar q}^{-\e}$.
This slight modification is caused by the change in the boundary of the 
$y_{qg}$ integration.
The invariant $y_{qg}$ is here bounded to be less than 
$\ymin y_{q \bar{q}g}$ instead of $\ymin$ in the three parton case.

Putting all the factors together, we find that in the collinear 
$quark-gluon$ limit the
four particle differential cross section ${\rm d}\sigma_{4}$ becomes,  
\begin{equation}
{\rm d}\sigma_{4} 
\to
\tilde{C}_{F}(q)\times 
\aqed 2 (2\pi)^2
\frac{1}{2M^2} \int  |{\cal M}_{Q\bar q \gamma}|^2
{\rm d}P_{3}^{(d)}(M;p_{Q},p_{\bar q}, p_{\gamma}) = 
\tilde{C}_{F}(q)\times {\rm d}\sigma_{Q \bar q \gamma},
\label{eq:unresolvedcolq}
\end{equation}
where ${\rm d}\sigma_{Q \bar q \gamma}$ is the three-particle 
differential cross 
section for the scattering of a quark-antiquark 
with an additional hard photon.
All of the divergences are isolated in $\tilde{C}_{F}(q)$ and therefore 
${\rm d}\sigma_{Q \bar q \gamma}$ can be evaluated in 4-dimensions.

In the region where the antiquark and the gluon are collinear the roles of
quark and antiquark are exchanged.
Now, the antiquark and gluon form a parent parton $\bar Q$ and
the resulting contribution in this region of the four particle phase space
therefore yields,
\begin{equation}
{\rm d}\sigma_{q\bar q \gamma g} \to  \tilde{C}_{F}(\bar q)\;{\rm d}\sigma_{q\bar {Q} \gamma},
\label{eq:unresolvedcolqbar}
\end{equation}
where, since $y_{Q\bar q} \equiv y_{q\bar Q}$,
\begin{equation}
\tilde{C}_{F}(\bar q) = \tilde{C}_{F}(q).
\label{eq:cfqbar}
\end{equation}

In order to match onto the single collinear quark-gluon regions, 
the soft gluon region is defined in eq.~(\ref{eq:singlesoft}),
\begin{displaymath}
y_{qg}<\ymin\, y_{q \bar q g} ,
~~~~~y_{\bar qg}<\ymin\, y_{q \bar q g} ,
{}~~~~~y_{q\gamma}>\ymin,~~~~~y_{\bar q\gamma}>\ymin.
\end{displaymath}
As in the collinear regions, the 
matrix elements and phase space both factorise
in the soft gluon limit,
\begin{equation}
|{\cal M}_{q\bar q \gamma g}|^2 \to |{\cal M}_{q\bar q \gamma}|^2\,f_{q\bar q }(g),
\end{equation}
with the {\it eikonal factor}, 
\begin{displaymath}
f_{q\bar q}(g)=\frac{4s_{q\bar q}}{s_{qg}s_{\bar qg}},
\end{displaymath}
and,
\begin{equation}
{\rm d}P_{4}^{(d)}(M;p_{q},p_{\bar q},p_{\gamma}, p_{g})
\to {\rm d}P_{3}^{(d)}(M;p_{q},p_{\bar q},p_{\gamma})\,
{\rm d}P^{(d)}_{soft}(p_{q},p_{\bar q},p_{g}),
\end{equation}
where the soft phase space factor reads \cite{gg},
\begin{displaymath}
{\rm d}P^{(d)}_{soft}(p_{q},p_{\bar q},p_{g})=
\frac{(4\pi)^{\e}}{16 \pi^2 \Gamma(1-\e)}
\frac{{\rm d}s_{qg}{\rm d}s_{\bar qg}}{s_{q\bar q}}
\left[\frac{s_{qg}s_{\bar qg}}{s_{q\bar q}}\right]^{-\e}.
\end{displaymath}
As before, all of the dependence on the unresolved variables is collected 
into the soft approximations to the matrix elements and the 
phase space.
We find,
\begin{eqnarray}
\tilde{S}_F &=& \int g_{s}^2  C_F \mu^{2\e}
f_{q\bar q}(g){\rm d}P^{(d)}_{soft}(p_{q},p_{\bar q},p_{g})\nonumber \\
&=&\as\left(\frac{4 \pi \mu^2}{M^2}\right)^{\e}
\frac{1}{\Gamma(1-\e)} 
\,\frac{2}{y_{q\bar q}}
\int_{0}^{\ymin y_{q\bar q}}{\rm d}y_{qg}
\int_{0}^{\ymin y_{q\bar q}}{\rm d}y_{\bar qg}\, \; 
\left[\frac{y_{qg}y_{\bar qg}}{y_{q\bar q}}\right]^{-\e -1}
\nonumber\\
&=&  \as \left(\frac{4 \pi \mu^2}{M^2}\right)^{\e}
\frac{1}{\Gamma(1-\e)}
\,\frac{2}{\e^2}\;\,\ymin^{-2\e}y_{q\bar q}^{-\e}.
\label{eq:sf}
\end{eqnarray}
Again, apart from the changed boundaries, this is the same soft factor as given in eq.~(3.33) of \cite{gg}.
As usual, the contribution to the cross section from the single soft
singular region factorizes,
\begin{equation}
{\rm d}\sigma_{q\bar q \gamma g} \to  \tilde{S}_{F}\;{\rm d}\sigma_{q\bar q \gamma}.
\label{eq:unresolvedsoft}
\end{equation}

The sum of the single unresolved gluon contributions is then given by
combining eqs.~(\ref{eq:unresolvedcolq}), (\ref{eq:unresolvedcolqbar})
and (\ref{eq:unresolvedsoft}),
\begin{displaymath}
{\rm d} \sigma_{q\bar q \gamma g} \to \left(\tilde{C}_{F}(q) +\tilde{C}_{F}(\bar q) +\tilde{S}_{F}\right) 
{\rm d} \sigma_{q \bar q \gamma}
\equiv {R_{q \bar q \gamma}}
{\rm d} \sigma_{q \bar q \gamma},
\end{displaymath}
where the real unresolved factor $R_{q \bar q \gamma}$ depends on the 
invariant mass of the quark-antiquark pair and is given 
by,
\begin{eqnarray}
{R_{q\bar{q} \gamma}}&=&\as
\frac{1}{\Gamma(1-\e)}
\left(\frac{4\pi\mu^2}{M^2}\right)^{\e}
\nonumber\\
& & \times 
\left(\frac{2\;y_{q\bar q}^{-\e}}{\e^2}+\frac{3}{\e}  -2\ln^2(\ymin) 
-3\ln(y_{q\bar q}\;\ymin)  +7 -\frac{2\pi^2}{3}\right). 
\label{eq:Rfac}
\end{eqnarray}
This is the same real unresolved factor as defined in eq.~(3.79) 
of \cite{gg} with
$\ymin \to \ymin y_{q\bar q}$ reflecting the changed boundaries of the 
unresolved gluon region of the four parton phase space compared with the
unresolved gluon region of the three parton phase space.
The singularities present in these single unresolved gluon contributions 
will cancel with those from the resolved photon one-loop 
$\gamma^* \to q\bar q \gamma$ process discussed in section~\ref{sec:resolvirt}.  

\subsubsection{The unresolved photon contribution}
\label{subsec:unresolp}
In the region where the quark and the photon are collinear we have,
cf. eq.~(\ref{eq:singlecolp}), 
\begin{equation}
y_{q\gamma}<\ymin\, y_{q \bar q \gamma},
~~~~~y_{\bar q\gamma}>\ymin\, y_{q \bar q \gamma},
{}~~~~~y_{q\gamma g}>\ymin,~~~~~y_{\bar qg}>\ymin.
\end{equation}
The quark  and the photon
cluster to form a new parent parton
$Q$ such that
each carries respectively a fraction $z$ and $1-z$ of the parent
parton momentum $p_{Q}$.
The four-particle matrix elements and phase space 
factorize in exactly the same way as in the quark-gluon collinear limit
with the roles of photon and gluon being interchanged and $y$ 
replaced by $z$.
Unlike the quark-gluon case however, 
the photon is observed in the final state and  
hence only $y_{q\gamma}$ is an unresolved variable.
The momentum fraction $z$ carried by the photon 
inside the quark-photon cluster is defined with respect to the momenta 
carried by the electromagnetically connected particles,  
\begin{equation}
z=\frac{y_{\bar q \gamma}}{y_{q \bar q \gamma }}. 
\end{equation}

In this limit, the four particle differential cross section again
factorizes,
\begin{equation}
{\rm d}\sigma_{4} \to 
\tilde{C}_{F\gamma}(q){\rm d}z\times {\rm d}\sigma_{Q \bar q g} 
\end{equation}
where ${\rm d}\sigma_{Q \bar q g}$ is the  
differential cross section for the production of a quark-antiquark pair and a gluon. 
The singular factor $\tilde{C}_{F\gamma}(q)$ is,
\begin{equation}
\tilde{C}_{F\gamma}(q)= -\aqed 
\left(\frac{4\pi\mu^2}{M^2}\right)^{\e} \frac{1}{\Gamma(1-\e)}
\frac{2}{\e}\;\ymin^{-\e}\;y_{Q\bar q}^{-\e}\;z^{-\e}\;(1-z)^{-\e}\;\Pqp(z).
\label{eq:cfgamma}
\end{equation}
There is a similar contribution where the photon is collinear with the antiquark.

\subsubsection{Overlapping of single collinear regions}
\label{subsec:overlap}
It is worth noting that 
in both, $q-g$ and $q-\gamma$ collinear regions we have required  
$y_{q\gamma g}>\ymin$ in order to guarantee that these regions match onto 
the double unresolved triple collinear region defined 
by $y_{q\gamma g}<\ymin$. 
However this requirement has an important consequence.
We do not avoid the situation where both invariants 
$y_{q\gamma}$ and $y_{q g}$ are simultaneously less than $\ymin$ and
the constraints,
\begin{equation}
y_{q\gamma}<\ymin \hspace{1cm} y_{q g}<\ymin \hspace{1cm} 
{\rm but} \hspace{1cm} y_{q \gamma g}>\ymin,
\label{equation:overlap}
\end{equation}    
define an {\it overlapping} region of 
the two single collinear $q-\gamma$ and $q-g$ regions.
In this overlapping region, the matrix elements are correctly approximated 
by the sum of the two single collinear approximations
and the error in the analytic contribution resulting 
from evaluating the approximated 
matrix elements over this restricted phase space region 
is of ${\cal O}(\ymin)$ and therefore negligible.

However, as we mentioned before, 
the photon +~1 jet rate is to be evaluated numerically
applying the hybrid subtraction method.
It is then crucial to ensure that the matrix elements are 
correctly approximated in this overlapping region since here we
evaluate   
the difference between this approximation 
given by the sum of both collinear approximations 
and the full matrix elements.    
It is precisely to take into account these numerical contributions
(which generate terms proportional to $\ln(\ymin)$) 
that we are constrained 
to apply the hybrid subtraction scheme rather then 
the more commonly used phase space slicing method \cite{kramer,gg} 
where the two single collinear regions 
would have to be clearly distinct and not overlapping.  
However, in this region of phase space, either collinear limit 
on its own is a
very poor approximation to the full matrix elements.
As a consequence, using the phase space slicing approach,
we would fail to obtain the necessary 
$\ln(\ymin)$  cancellation in the physical photon +~1 jet cross
section.

\section{Double unresolved contributions}
\setcounter{equation}{0}
\label{sec:double}

In the previous section we have discussed the resolved and single unresolved real contributions relevant to the tree level
process $ \gamma^* \to q \bar q \gamma g$.
Each of these two classes of real contributions corresponds to final state
configurations where {\it more than two } particles are theoretically ``seen" 
and  a $ \gamma $ +~1 jet event can only
arise if some final state particles are clustered together by the jet
algorithm.
Therefore, to allow the adjustment of our results to any jet algorithm 
used in the experimental analysis, 
the finite contributions to the differential cross section will
be evaluated numerically.
The {\it two-particle unresolved} contributions, on the
other hand, already  correspond to $ \gamma$ +~1 jet events,
and there is no further clustering by the jet algorithm needed.
Because of this, the integrations can be performed analytically.

\subsection{The triple collinear factor}
\label{subsec:triple}

As we saw in section~3, the triple collinear contributions arise when the
gluon and the photon are collinear to the quark.
The triple collinear configuration is illustrated in Fig.~\ref{fig:clus4}.c.i.
In order to evaluate these contributions we need to determine
the appropriate approximations for the matrix element squared and
phase space in the
{\it triple collinear limit} and perform the phase space integrals
over the unresolved variables.

The triple collinear  region of phase space
is defined by eq.~(\ref{eq:tricolinv}),
\begin{displaymath}
y_{q\gamma g} < \ymin \hspace{0.5cm} {\rm and}\hspace{0.5cm}
y_{\bar{q}g} >\ymin.
\end{displaymath}
In this limit, $y_{qg\gamma}$ is small and
the photon, gluon and quark cluster to form
a new parent parton Q such that,
\begin{equation}
p_{q} +p_{\gamma} +p_{g}=p_{Q}.
\end{equation}
The photon, the gluon and the quark carry respectively a fraction
$z$, $y$ and $(1-y-z)$ of the parent parton momentum $p_{Q}$,
\begin{equation}
p_{\gamma} = z~p_Q,\hspace{1cm}p_{g} = y~p_Q,\hspace{1cm}
p_{q }= (1-z-y)~p_Q,
\label{eq:tripledef}
\end{equation}
so that the invariants are given by the following,
\begin{eqnarray}
y_{q\bar q}&=&(1-y-z)\,y_{Q\bar q}\equiv (1-y-z) ,\nonumber \\
y_{\bar q \gamma}&=&z\,y_{Q\bar q}\equiv z ,\\
y_{\bar{q} g}&=&y\,y_{Q\bar q}\equiv y .\nonumber 
\label{eq:sijdef}
\end{eqnarray}

The algebraic structure of these double unresolved
contributions is unique to the triple
collinear limit of the matrix element squared, and when analytically
integrated over the singular regions of phase space will form 
the {\it triple collinear factor}.
These contributions are expected to arise in  
analytic  calculations of   exclusive  quantities at the second
order in perturbation theory.
Such calculations have, to the best of our knowledge, not been performed
before in the literature.

We are interested in the triple collinear limit of the matrix
element squared for the scattering of a quark-antiquark pair with a
photon and a gluon.
In this  limit the $d$-dimensional four-particle matrix element squared 
factorises,
\begin{displaymath}
|{\cal M}_{q\bar q g \gamma}|^2 \to 
P_{qg\gamma\to Q}(z,y,s_{q\gamma},s_{qg},s_{qg\gamma})
|{\cal M}_{Q\bar q}|^2,
\end{displaymath}
where
$|{\cal M}_{Q\bar q}|^2$ is the two-particle matrix element squared and
$P_{qg\gamma\to Q}(z,y,s_{q\gamma},s_{qg},s_{qg\gamma})$ defines the {\it triple collinear
splitting function}.
It is obtained by keeping only the terms
containing  a pair of the unresolved invariants, $s_{q\gamma}$, $s_{qg}$ and $s_{qg\gamma}$, in the denominator 
of the ``full" four particle squared matrix elements $|{\cal M}_{q\bar q g \gamma}|^2$,
\begin{eqnarray}
\lefteqn{P_{qg\gamma\to Q}(z,y,s_{q\gamma},s_{qg},s_{qg\gamma}) = }\nonumber \\
&+&   \frac{4}{s_{q\gamma} s_{qg}}\frac{(1-z-y)(1+(1-z-y)^2
-\epsilon (z^2+zy+y^2)-\epsilon^2 z y)}{zy}\nonumber\\
&+& \frac{4}{s_{q\gamma}s_{qg\gamma}}\frac{(1-z-y)(1-z+\epsilon^2 z y)+(1-y)^3
-\epsilon (1-y)(z^2+zy+y^2)+\epsilon^2 z y}{zy} \nonumber\\
&+& \frac{4}{s_{qg}s_{qg\gamma}}\frac{(1-z-y)(1-y+\epsilon^2 z y)+(1-z)^3
-\epsilon (1-z)(z^2+zy+y^2)+\epsilon^2 z y}{zy} \nonumber\\
&-&  \frac{4(1-\epsilon)}{s_{qg\gamma}^2}\left((1-\epsilon)\frac{s_{q\gamma}}{s_{qg}}+
(1-\epsilon)\frac{s_{qg}}{s_{q\gamma}}-2\epsilon\right).
\label{eq:P134}
\end{eqnarray}
This triple collinear splitting function is
the generalization of the single collinear
factor with three collinear particles instead of two and is as universal
as
the single soft and single collinear factors encountered earlier in section 4.

In the triple collinear limit, we can reorganise the $d$-dimensional 
four particle phase space given in eq.~(\ref{eq:4ps})
into one part appropriate for the remnant $Q-\bar q$ pair multiplied by the integral over all of the unresolved variables,
\begin{displaymath}
{\rm d}P_{4}^{(d)}(M;p_{q},p_{\gamma},p_{g},p_{\bar q})\;\to\;{\rm
d}P_{2}^{(d)}(M;p_{Q},p_{\bar q})
\times {\rm d}P_{tricol}^{(d)}(p_{Q},p_{q},p_{\gamma},p_{g}),
\label{eq:tricolph}
\end{displaymath} 
where ${\rm d}P_{2}^{(d)}(M;p_{Q},p_{\bar q})$ is the usual two body phase space and,
\begin{eqnarray}
{\rm d}P_{tricol}^{(d)}(p_{Q},p_{q},p_{\gamma},p_{g})  &=&
\frac{1}{\Gamma(1-2\epsilon)}
\frac{1}{4\pi}\frac{M^4}{4(2\pi)^4} 
\left(\frac{4\pi}{M^2}\right)^{2\e}\left[-\Delta_{4}\right]^{-\frac{1}{2}-\e}
\nonumber\\
& \times& {\rm d}y_{q\gamma g}\;{\rm d}y_{q\gamma}\;{\rm d}y_{qg}\;{\rm d}y_{\gamma g}\;{\rm d}z\;
{\rm d}y\;\delta(y_{q\gamma} + y_{qg}+y_{\gamma g}-y_{q\gamma g}).
\label{eq:triplephase}
\end{eqnarray}
Here the angular integrations for
the rotation of the $qg\gamma$-system around the $\bar{q}$ axis 
and for the parity of the $qg\gamma$-system 
have been performed.
In terms of the unresolved variables, the Gram determinant 
is,
\begin{displaymath}
\Delta_{4} \to  
\big((1-y-z)\;y_{\gamma g}\;-\;y\;y_{q\gamma}\;-\;z\;y_{qg}\big)^2 -4zy\;y_{q\gamma}\;y_{qg}.
\end{displaymath}

Reinserting all coupling factors into the matrix elements, we find that 
in the triple collinear limit, the cross section factorises,
\begin{eqnarray}
{\rm d}\sigma_{4}&\equiv&
\as\aqed
4 (2\pi)^4
 \mu^{4\epsilon}\;
 \int |{\cal M}_{q\bar q g \gamma}|^2\;
{\rm d}P_{4}^{(d)}(M; p_{q}, p_{\bar q}, p_{\gamma}, p_{g}) \nonumber \\
&\to &
TC_{F\gamma}{\rm d}z\times  \int   |{\cal M}_{Q\bar q}|^2
{\rm d}P_{2}^{(d)}(M;p_{Q},p_{\bar q})  \nonumber \\
&\equiv& TC_{F\gamma}{\rm d}z \times  \sigma_0.
\end{eqnarray}
As usual, $\sigma_{0}$ is the  two-particle cross section while
the dimensionless factor $TC_{F\gamma}{\rm d}z$ 
containing all the singularities 
is formally given by,
\begin{eqnarray}
TC_{F\gamma}{\rm d}z&\equiv&
\as\aqed\4pi2
\frac{1}{\Gamma(1-2\epsilon)}
\frac{1}{4\pi}  
\nonumber \\
& \times&
\int \left[-\Delta_{4}\right]^{-\frac{1}{2}-\e}
 P_{qg\gamma\to Q}(z,y,y_{q\gamma},y_{qg},y_{qg\gamma})
\nonumber \\
& &\times
 {\rm d}y_{q\gamma g}\;{\rm d}y_{q\gamma}\;{\rm d}y_{qg}\;{\rm d}y_{\gamma g}\;{\rm d}z\;
{\rm d}y\;\delta(y_{q\gamma} + y_{qg}+y_{\gamma g}-y_{q\gamma g}).
\end{eqnarray}
To evaluate $TC_{F\gamma}{\rm d}z$, we must integrate out the unresolved variables over the phase space region where the triple collinear approximation is appropriate.
This is,
\begin{displaymath}
0 < y_{q\gamma g} < \ymin,\qquad\qquad\ymin < y < 1-z,
\end{displaymath}
while the other variables are constrained by the Gram determinant.
Using the delta function to eliminate $y_{\gamma g}$, we find that the bounds on $y_{q\gamma}$ are given by solving the quadratic equation $\Delta_4 = 0$, which generates the additional constraint $y_{qg} < (1-z)y_{q\gamma g}$.
It is always most convenient to integrate the invariant mass that does not appear in $P_{qg\gamma\to Q}$ first.  This removes the Gram determinant and 
generates factors that regulate the other singularities.
The first term in $P_{qg\gamma\to Q}$ does not allow this approach and is rather more tricky, yielding Hypergeometric functions with complicated arguments after the first integration \cite{aude}.  However, the integrals can be carried through and we find that after making a series expansion in $\e$, the integrated triple collinear factor is given by,
\begin{eqnarray}
TC_{F\gamma}& = &
\coup
\nonumber\\ 
& & \hspace{-1.3cm}
\times \Bigg\{ \frac{1}{\e^2} \Bigg[ \ln (z)\left (1-{\frac {z}{2}}
\right )+1-{\frac {z}{4}}-{\frac {3\,\Pqpzero(z)}{2}}+2\,\ln (1-z)\Pqpzero(z)
-2\,\ln (\ymin)\Pqpzero(z) \Bigg] \nonumber\\
& & \hspace{-0.7cm} + \frac{1}{\e}
\Bigg[\ln (z)\ln (1-z)\left (-2+z-2\,\Pqpzero(z)\right )+\ln (z)\left (-1-{\frac {5
\,z}{4}}+{\frac {3\,\Pqpzero(z)}{2}}\right ) \nonumber \\
& & \hspace{0.2cm} +\ln^2 (z)\left
(-\frac{3}{2}+{\frac {3\,z}{4}}\right )-3\,\ln^2 (1-z)\Pqpzero(z)-\frac{1}{4}
+{\frac {11\,z}{4}}-{\frac {7\,\Pqpzero(z)}{2}}\nonumber \\
& & \hspace{0.2cm}
+\ln (1-z)\left (-2-{\frac {3
\,z}{2}}+3\,\Pqpzero(z)\right )+\Li_2(1-z)\left (-2+z-\Pqpzero(z)\right )+{
\frac {{\pi}^{2}\Pqpzero(z)}{2}}
\nonumber\\
& & \hspace{0.2cm} + \ln(\ymin) \Bigg (-2+{\frac {5\,z}{2}}+3\,\Pqpzero(z)+\ln
(z)\left (-2+z+2\, 
\Pqpzero(z)\right ) \nonumber \\
& & \hspace{0.2cm}
-2\,\ln (1-z)\Pqpzero(z)\Bigg ) + \ln^2(\ymin)\, 5 \Pqpzero(z) \Bigg] \nonumber \\ 
& & \hspace{-0.7cm} -1+{\pi}^{2}\left (-\frac{1}{3}
-{\frac {5\,z}{12}}+{\frac {\Pqpzero(z)}{
2}}\right )+\ln (z)\left ({\frac {13}{4}}-{\frac {17\,z}{4}}+{\frac {7
\,\Pqpzero(z)}{2}}\right )\nonumber \\
& & \hspace{-0.7cm}
+\ln (z){\pi}^{2}\left (-\frac{1}{3}+{\frac {z}{6}}-{\frac
{\Pqpzero(z)}{3}}\right )+\ln^2 (z)\left (\frac{1}{2}+{\frac {17\,z}
{8}}-{\frac {3\,\Pqpzero(z)}{4}}\right ) \nonumber\\
& & \hspace{-0.7cm}
+\ln (z)\ln (1-z)\left (2+{\frac {9\,
z}{2}}-3\,\Pqpzero(z)\right )+\ln (z)\Li_2(1-z)\left (4-2\,z\right )
\nonumber \\
& & \hspace{-0.7cm}
+\ln^2 (1-z) \left (2+{\frac {5\,z}{2}}-3\,\Pqpzero(z)\right )
+\ln (1-z)\Li_2(1-z)\left (4-2\,z+2\,\Pqpzero(z)\right )\nonumber \\
& & \hspace{-0.7cm}
+{\frac {25\,z}{4
}}-7\,\Pqpzero(z)+\ln^3 (z) \left (\frac{7}{6}-{\frac {7\,z}{12}}
\right )+\ln^2 (z)\ln (1-z)\left (3-{\frac {3\,z}{2}}
+\Pqpzero(z)\right ) \nonumber \\
& &  \hspace{-0.7cm}
+\ln (z) \ln^2 (1-z)\left (2-z+3\,\Pqpzero(z)
\right )+\ln (1-z)\left (\frac{1}{2}-{\frac {11\,z}{2}}+7\,\Pqpzero(z)\right
)\nonumber \\
& &  \hspace{-0.7cm} 
+\Li_3(1-z)\left (-4+2\,z-2\,\Pqpzero(z)\right )+S_{12} (1-z)\left (2-z-3\,\Pqpzero(
z)\right )\nonumber \\
& & \hspace{-0.7cm} 
+{\frac {7\,
\ln^3 (1-z) \Pqpzero(z)}{3}}+4z\,\Li_2(1-z)+9\,\Pqpzero(z) \zeta(3) -{\frac {4\,\ln
(1-z){\pi}^{2}\Pqpzero(z)}{3}}  \nonumber \\
& &  \hspace{-0.7cm} 
+\ln(\ymin) \Bigg (\Li_2(1-z)\left (4-2\,z+2\,\Pqpzero(z)\right )
+\ln (z)\ln (1-
z)\left (4-2\,z+2\,\Pqpzero(z)\right )\nonumber \\
& & \hspace{0.3cm}
+\ln^2 (z)\left (3-{
\frac {3\,z}{2}}-\Pqpzero(z)\right )-{\frac {2\,{\pi}^{2}\Pqpzero(z)}{3}}+\ln (1-z)
\left (4+z-6\,\Pqpzero(z)\right ) \nonumber \\
& & \hspace{0.3cm}
+\ln (z)\left (2+{\frac {z}{2}}-3\,\Pqpzero(z)
\right )+5\,\left (\ln (1-z)\right )^{2}\Pqpzero(z)+\frac{1}{2}
-{\frac {11\,z}{2}}+7
\,\Pqpzero(z)\Bigg ) \nonumber \\
& & \hspace{-0.7cm} 
+ \ln^2(\ymin) \left (\ln (z)\left (2-z-5\,\Pqpzero(z)\right )+2-{
\frac {11\,z}{2}}-3\,\Pqpzero(z)-\ln (1-z)\Pqpzero(z)\right ) \nonumber \\
& &  \hspace{-0.7cm} 
-\frac {19\, \ln^3 (\ymin)\Pqpzero(z)}{3}.
\label{eq:tripleex}
\end{eqnarray}

\subsection{The soft/collinear factor}
\label{subsec:softcol}

The soft/collinear configuration arises when the
photon is collinear to the quark and the gluon is soft
as illustrated in Fig.~\ref{fig:clus4}.c.ii.
As before, to isolate the singularities we need to determine
the appropriate approximations for the matrix element squared and
phase space in the
{\it soft/collinear limit} and perform the phase space integrals
over the unresolved variables.

The soft/collinear region of phase space
is defined by eq.~(\ref{eq:softcolinv}),
\begin{displaymath}
y_{q\gamma}<\ymin  \hspace{0.5cm} {\rm and}\hspace{0.5cm}
y_{qg} <\ymin \hspace{0.5cm} {\rm and} \hspace{0.5cm}
y_{\bar{q}g}  <\ymin.
\end{displaymath}
In this limit, the photon and the quark cluster to form
a new parent parton $Q$ such that,
\begin{equation}
p_{q} +p_{\gamma} =p_{Q},
\end{equation}
while the energy of the gluon tends to $0$ ($p_{g}\to 0$).
The photon and the quark carry respectively a fraction
$z$ and $(1-z)$ of the parent parton momentum $p_{Q}$,
\begin{equation}
p_{\gamma} = z~p_Q,\hspace{0.5cm}p_{q}= (1-z)~p_Q,
\label{eq:softdef}
\end{equation}
and the invariants are given by the following,
\begin{equation}
y_{q\bar q}=(1-z)\,y_{Q\bar q}\equiv (1-z) \qquad {\rm and} \qquad
y_{\bar q\gamma}=z\,y_{Q\bar q}\equiv z.
\label{eq:sijsoftdef}
\end{equation}

Once again, the matrix elements factorise 
in the soft/collinear limit defined above, 
\begin{displaymath}
|{\cal M}_{q\bar q g \gamma}|^2 \to 
 P_{qg\gamma\to Q}^{soft/col}(z,
,s_{q\gamma},s_{qg},s_{\bar q g},s_{qg\gamma})~
|{\cal M}_{Q\bar q}|^2,
\end{displaymath}
As usual,
$|{\cal M}_{Q\bar q}|^2$ is the two-particle matrix element squared 
while
$P_{qg\gamma\to Q}^{soft/col}(z,s_{q\gamma},s_{qg},s_{\bar q g},s_{qg\gamma})$ defines
the soft/collinear approximation to the squared matrix elements.
This is obtained by setting $y= 0$
in the triple collinear splitting function, $P_{qg\gamma\to Q}$,
given in eq.~(\ref{eq:P134})
and is therefore,
\begin{equation}
 P_{qg\gamma\to Q}^{soft/col}(z,s_{q\gamma},s_{qg},s_{\bar q g},s_{qg\gamma}) \equiv
\frac{4}{s_{q\gamma}s_{qg}s_{\bar qg}}   \left(  (1-z)  +
\frac{s_{qg}+(1-z)s_{q\gamma}}{s_{qg\gamma} } \right)\Pqp(z) .
\label{eq:Msoft}
\end{equation}

In the soft/collinear limit, we can again rewrite the phase space as 
the phase space for the $Q\bar q$ pair multiplied by an integral over the unresolved variables,
\begin{displaymath}
{\rm d}P_{4}^{(d)}(M;p_{q},p_{\gamma},p_{g},p_{\bar q})\;\to\;{\rm
d}P_{2}^{(d)}(M;p_{Q},p_{\bar q})
\;{\rm d}P_{soft/col}^{(d)}(p_{Q},p_{q},p_{\gamma},p_{g}),
\end{displaymath} 
where,
\begin{eqnarray}
{\rm d}P_{soft/col}^{(d)}(p_{Q},p_{q},p_{\gamma},p_{g})  &=&
\frac{1}{\Gamma(1-2\epsilon)}
\frac{1}{4\pi}\frac{M^6}{4(2\pi)^4} 
\left(\frac{4\pi}{M^2}\right)^{2\e}\left[-\Delta_{4}\right]^{-\frac{1}{2}-\e}
\nonumber\\
& \times& {\rm d}y_{q\gamma g}\;{\rm d}y_{q\gamma}\;{\rm d}y_{qg}\;{\rm d}y_{\gamma g}\;{\rm d}y_{\bar q g}\;{\rm d}z\;
\delta(y_{q\gamma} + y_{qg}+y_{\gamma g}-y_{q\gamma g}).
\label{eq:softcolphase}
\end{eqnarray}
This factor is similar to the triple collinear phase space factor given by
eq.~(\ref{eq:tricolph}) with $y$ replaced by $y_{24}$.

As $y_{q\gamma g}$ is unconstrained in this region of phase space,
we choose to rewrite $\Delta_{4}$ as a quadratic in $y_{q\gamma g}$,
and, when performing the phase space integrals, will first
integrate over $y_{q\gamma g}$.
With the definitions of the invariants $y_{q\bar q}$ and $y_{\bar q \gamma}$
in the soft/collinear region of phase space,
\begin{displaymath}
-\Delta_{4}= 
 (1-z)^2(y_{q\gamma gb}-y_{q\gamma g})(y_{q\gamma g}-y_{q\gamma ga}),
\end{displaymath}
with $y_{q\gamma ga,b}$ given by,
\begin{displaymath}
y_{q\gamma ga,b}\;=\;\frac{1}{1-z}\Bigg(y_{q\gamma}(1-z)\,+\,y_{qg}\,+\,y_{q\gamma}y_{qg}
\pm 2 \sqrt{y_{q\gamma}y_{qg}y_{\bar qg}z}\Bigg).
\end{displaymath}

Once again, the cross section factorises,
\begin{eqnarray}
{\rm d}\sigma_{4}&\equiv&
\as\aqed
4 (2\pi)^4
\mu^{4\e}\;
 \int |{\cal M}_{q\bar q g \gamma}|^2\;
{\rm d}P_{4}^{(d)}(M; p_{q}, p_{\bar q}, p_{\gamma}, p_{g}) \nonumber \\
&\to &
SC_{F\gamma}{\rm d}z\times  \int   |{\cal M}_{Q\bar q}|^2
{\rm d}P_{2}^{(d)}(M;p_{Q},p_{\bar q})  \nonumber \\
&\equiv& SC_{F\gamma}{\rm d}z \times  \sigma_0.
\end{eqnarray}
The singular dimensionless factor $SC_{F\gamma}{\rm d}z$ 
is given by,
\begin{eqnarray}
SC_{F\gamma}{\rm d}z&\equiv&
\as\aqed 4 (2\pi)^4 \left(\mu^2\right)^{2\e}
\int  {\rm d}P_{soft/col}^{(d)}   P_{qg\gamma\to Q}^{soft/col} \nonumber \\
&=& 
\as\aqed\4pi2
\frac{1}{\Gamma(1-2\epsilon)}
\frac{1}{4\pi}  
\nonumber \\
& \times&
 \int \left[-\Delta_{4}\right]^{-\frac{1}{2}-\e}
 P_{qg\gamma\to Q}^{soft/col}(z,y_{q\gamma},y_{qg},y_{\bar qg},y_{qg\gamma})
\nonumber \\
& &\times
 {\rm d}y_{q\gamma g}\;{\rm d}y_{q\gamma}\;{\rm d}y_{qg}\;{\rm d}y_{\gamma g}\;y_{\bar q g}\;{\rm d}z\;
\delta(y_{q\gamma} + y_{qg}+y_{\gamma g}-y_{q\gamma g}).
\end{eqnarray}
To evaluate the soft/collinear differential factor $SC_{F\gamma}{\rm d}z$
we need to integrate
$P_{qg\gamma\to Q}^{soft/col}$, given by eq.~(\ref{eq:Msoft})
over the soft/collinear phase space given by eq.~(\ref{eq:softcolphase}).

The  Gram determinant fixes the allowed range of $y_{q\gamma g}$, while the other three unresolved variables are constrained to be less than $\ymin$.
As expected, the $y_{q\gamma g}$ integral generates factors 
regulating the other phase space integrals and we find,
\begin{equation}
SC_{F\gamma}=
\coup
\left(-\frac{2}{\e^3}\;\ymin^{-3\e} \;z^{-\e} \Pqp(z)\right).
\label{eq:socolresult}
\end{equation}

\subsection{The double collinear factor}
\label{subsec:doublecol}

The double single collinear region of phase space
is defined by the following constraints on the invariants,
\begin{equation}
y_{q\gamma} < \ymin, \qquad
y_{\bar{q}g} <\ymin,
\end{equation}
with the additional requirement,
\begin{equation}
y_{q g}  >\ymin,
\end{equation}
because the gluon is collinear to the antiquark but is {\em not} soft.
This configuration occurs
when the photon and the quark
form a collinear pair
simultaneously with the gluon and the antiquark being collinear and
is illustrated in Fig.~\ref{fig:clus4}.c.iii.

In this limit, the photon and the quark cluster to form
a new parent parton $Q$ such that,
\begin{equation}
p_{q} +p_{\gamma} =p_{Q},
\end{equation}
while the gluon and the antiquark cluster to form a new parent
parton, $\bar{Q}$ with,
\begin{equation}
p_{\bar q} +p_{g} =p_{\bar Q},
\end{equation}
The photon and the quark carry respectively a fraction
$z$ and $(1-z)$ of the parent parton momentum $p_{Q}$,
\begin{equation}
p_{\gamma} = z~p_Q,\qquad p_{q}= (1-z)~p_Q.
\label{eq:doubledef1}
\end{equation}
whereas the gluon and the antiquark each carry a fraction $y$ and $1-y$
of the parent momentum $p_{\bar{Q}}$ such that,
\begin{equation}
p_{g} = y~p_{\bar Q},\qquad p_{\bar q}= (1-y)~p_{\bar Q}.
\label{eq:doubledef2}
\end{equation}
The invariants can be redefined as follows,
\begin{eqnarray}
y_{q\bar q}&=&(1-y)(1-z)\,y_{Q\bar Q}\equiv (1-y)(1-z)\nonumber \\
y_{qg}&=&y(1-z)\,y_{Q\bar{Q}}\equiv y(1-z)\nonumber\\
y_{\bar \gamma}&=&z(1-y)\,y_{Q\bar Q}\equiv z(1-y)\nonumber\\
y_{\gamma g}&=&y z\,y_{Q\bar Q}\equiv y z. 
\label{eq:sijdef2}
\end{eqnarray}

Using the redefinitions  of the invariants given by eq.~(\ref{eq:sijdef2}),
the four particle matrix element squared factorizes in the double single collinear
limit as follows,
\begin{equation}
|{\cal M}_{q\bar q \gamma g}|^2 \to P_{q\gamma\to Q;\bar q g\to\bar{Q}}(z,y,s_{q\gamma},s_{\bar q g})
|{\cal M}_{Q \bar{Q}}|^2
\label{eq:doublem2}
\end{equation}
with,
\begin{equation}
P_{q\gamma\to Q;\bar q g\to\bar{Q}}(z,y,s_{q\gamma},s_{\bar q g}) =
P_{q\gamma\to Q}(z,s_{q\gamma})
P_{\bar q g\to \bar Q}(y,s_{\bar q g}).
\end{equation}
In other words,
the double single collinear factor
is the product of two
single collinear factors.

In this limit
the four-particle
phase space again factorises,
\begin{equation}
{\rm d}P_4^{(d)}(M;p_{q},p_{\bar q},p_{\gamma},p_{g}) \to {\rm d}P_2^{(d)}(M;p_Q,p_{\bar Q})
\times  {\rm d}P_{double}^{(d)}(p_{q},p_{\bar q},p_{\gamma},p_{g}).
\end{equation}
As before, the unresolved phase space factor contains integrals over five unresolved variables.
Note that unlike in the soft/collinear region, $y_{\gamma g}$ is precisely
defined in eq.~(\ref{eq:sijdef2}). Consequently
the triple invariant $y_{q\gamma g}$ is also fixed,
$y_{q\gamma g} = y_{q\gamma}+y_{qg}+y_{\gamma g} = y$.
This appears to reduce the number of independent variables by
one. 
However, a closer look enables us to assert that there is no inconsistency in this
procedure. In fact,
the boundaries of the $y_{q\gamma g}$ integration generated by the Gram determinant
turn out to be $y_{q\gamma ga,b}=y \pm {\cal O}(\ymin)$.
Hence by replacing $y_{q\gamma g}$ by $y$  in order to obtain
the double single collinear matrix elements we make an error of ${\cal O}(\ymin)$,
which we do throughout the calculation and it is therefore a consistent
approximation to make.
Integrating out $y_{q\gamma g}$ and the unresolved angular
variables,
we find that,
\begin{eqnarray}
{\rm d}P_{double}^{(d)}(p_{q},p_{\bar q},p_{\gamma},p_{g}) 
&=&
\frac{1}{\Gamma^2(1-\e)}
\left(\frac{4\pi}{M^2}\right)^{2\e}
\frac{M^4}{16(2\pi)^4}
\bigg({\rm d}y_{q\gamma}\;{\rm d}z
\Big[z(1-z)y_{q\gamma}\Big]^{-\epsilon}\bigg)\nonumber \\
&&\times 
\bigg({\rm d}y_{\bar qg}\;{\rm d}y
\Big[y(1-y)y_{\bar qg}\Big]^{-\epsilon}\bigg)\,,
\label{eq:doublephase}
\end{eqnarray}
which is exactly the product of two single collinear phase space
factors as one could have expected.

As with the previous double unresolved contributions, the integration of the
resolved
two particle matrix elements over the two particle phase space yields a factor
of $\sigma_0$.
This is multiplied by the integral of the approximation
of the unresolved matrix elements over the unresolved phase space.
Explicitly, we have,
\begin{equation}
{\rm d}\sigma_{q\bar q \gamma g} \to DC_{F\gamma}{\rm d}z \times \sigma_0,
\end{equation}
where,
\begin{equation}
DC_{F\gamma}{\rm d}z = 
\as\aqed
4 (2\pi)^4
\left(\mu^2\right)^{2\epsilon}\;
\int{\rm d}P_{double}^{(d)} P_{q\gamma\to Q;\bar q g\to\bar{Q}}(z,y,s_{q\gamma},s_{\bar q g}).
\end{equation}
The unresolved region is specified by eq.~(\ref{eq:doucolinv}) and
the constraint $y_{qg}>\ymin$ fixes the lower boundary of
the $y$ integral to be $\ymin/(1-z)$ since in the
double single collinear region $y_{qg}=y(1-z)$.
The integrals are straightforward, and we find,
\begin{eqnarray}
DC_{F\gamma}= 
\coup 
z^{-\e}(1-z)^{-\e}\Pqp(z) 
\nonumber\\
\times
\frac{\ymin^{-2\epsilon}}{\epsilon^3}
\left(\;2
\ymin^{-\epsilon}(1-z)^{\e}\;-\;\frac{(1-\epsilon)(4-\epsilon)}
{2(1-2\epsilon)}\;\frac{\Gamma^2(1-\epsilon)}{\Gamma(1-2\epsilon)} 
\right).
\label{eq:doubleresult}
\end{eqnarray}

\subsection{Strong Ordering}
\label{subsec:strong}
As a check of our calculation of the real two particle
unresolved contributions to the
differential cross section, we have rederived them in
the strongly ordered
limits \cite{aude}. Instead of considering particle 1 and particle 2 to be
collinear {\it  at the same time} to particle 3, we consider the
two different contributions;
{\bf either} particle 1 is collinear to particle 3 {\it followed}
by particle 2 collinear to the cluster of particles 1 and 3, 
(denoted by (13)), so that
({\it $y_{13}\ll y_{23} $}),
{\bf  or} particle 2 is collinear
to particle 3 {\it followed } by particle 1 being collinear to particle (23)
where we have ({\it $y_{23}\ll y_{13} $}).
In general, within the strongly ordered approximation,
each of the unresolved real contributions ({\it triple collinear},
{\it soft/collinear} and {\it double single collinear}), 
gets ``replaced'' by the sum of two
{\it strongly ordered} contributions.
In addition to changing the approximations to the matrix elements,
the phase space is also reorganized to be the product of two single unresolved factors.
However, although the strongly ordered
approximation correctly reproduces the {\it leading divergences}
--those proportional to
${\cal O}(1/\e^3)$ or ${\cal O}(1/\e^2)$ which
are associated with the leading and 
next-to-leading logarithms -- 
it does not generate the correct {\it subleading divergences}
proportional to
${\cal O} (1/\e)$.
The finite terms of ${\cal O} (1)$ are also incorrectly reproduced.
We understand this as follows:
The poles in $1/\e^2,~1/\e^3$ arise from the evaluation
of successive phase space integrals at the lower boundaries where
the strongly ordered approximation 
is very close to the ``full'' approximation of the matrix elements.
On the other hand, terms proportional 
to $1/\e$  arise
when evaluating only one of the phase space integrals at its lower boundary
while the other phase space integrals contain significant contributions close to their upper boundaries.
At these upper boundaries, the  two
invariants defining the strongly ordered limit are no longer strongly ordered
and the approximation is invalid.

\section{Virtual contributions}
\setcounter{equation}{0}
\label{sec:virt}
In the previous two sections we have decomposed
the four-particle phase space 
and extracted the divergences present  
in the ${\cal O}(\alpha\alpha_s)$ four-parton 
process $\gamma^* \to q \bar q \gamma g$ 
where one or two particles are theoretically unresolved. 
In other words, only two or three particles are {\em theoretically}
identified in the final state.
If three particles are theoretically well separated,
the experimental cuts will combine these particles further to select 
photon +~1 jet events.
In this section we will take into account the exchange of
a virtual gluon
in the $\gamma^{*}\to q\bar{q} \gamma$ process,
which when interfered with the tree level process also gives rise
to contributions of ${\cal O}(\alpha\alpha_s)$.

As discussed in section~\ref{sec:1jet}, the calculation naturally
divides into two parts,
depending on whether or not the three particles are resolved.
Both resolved and unresolved contributions
are divergent and need to be combined with the appropriate
real contributions described earlier in sections 4 and 5.
For the {\it resolved} virtual contribution, the quark, the antiquark and the photon
are clearly distinguishable and the divergences will cancel
when combined with the real contributions where the gluon is either
collinear with one of the quarks or is soft 
and the photon is theoretically resolved (c.f.~section \ref{sec:single}).
On the other hand, in the {\it unresolved} virtual contributions, 
the quark and
photon are collinear and form a single pseudo particle, $Q$, the
parent quark.
The leading singularity occurs 
from a soft gluon being internally exchanged simultaneously 
with the collinear emission of the photon from a quark and  
is proportional to $\Pqp(z)/\e^3$ .
These most singular poles will cancel with those present in
the soft/collinear contributions calculated in the previous section.

\subsection{The resolved contribution}
\label{sec:resolvirt}
The ``squared'' matrix elements arising from the
$\gamma^{*}\rightarrow q\bar{q}\gamma$ process at one loop
is part of the ${\cal O}(\alpha_{s}^2)$ corrections
to the three-jet rate in $e^+e^-$ annihilation, which was
originally derived by Ellis, Ross and Terrano in \cite{ert}.
As we are interested in the virtual contributions with an outgoing
photon instead of an outgoing gluon, we need to replace the colour
factors  in eq.~(2.20) of~\cite{ert} with,
\begin{displaymath}
C_{A}\rightarrow 0, \hspace{1cm}N_{F}\rightarrow 0, \hspace{1cm}
C^{2}_{F}\rightarrow C_{F},
\end{displaymath}
as well as the replacement,
\begin{displaymath}
\alpha_s^2 \to \alpha_s\alpha e_q^2,
\end{displaymath}
when the quark has charge $e_q$.
The contribution to the cross section can be written as,
\begin{equation}
{\rm d}\sigma^V_{q\bar q \gamma} =
\as\aqed 4 (2\pi)^4 \mu^{4\e} \frac{1}{2M^2} \int 
|{\cal M}_{q\bar q \gamma}|^2_V \;
{\rm d}P_3^{(d)}(M;p_q,p_{\bar q},p_{\gamma}),
\end{equation}
where,
\begin{equation}
g_s^2 C_F \mu^{2\e} \; |{\cal M}_{q\bar q \gamma}|^2_V =
V_{q \bar q \gamma} \times  
|{\cal M}_{q\bar{q}\gamma}|^2   
+F(y_{q\bar q},y_{q\gamma},y_{\bar q\gamma}).
\label{eq:ertm2}
\end{equation}
Here,
\begin{equation}
V_{q\bar q\gamma} =
\as  \left(\frac{4\pi\mu^2}{M^2}\right)^{\e}
\frac{\Gamma^2(1-\e)\Gamma(1+\e)}{\Gamma(1-2\e)}
\left(-\frac{2y_{q\bar q}^{-\e}}{\e^2}\,-\frac{3}{\e}\,+\pi^2\,-8
\right),
\label{eq:Vfac}
\end{equation}
and,
\begin{eqnarray}
F(y_{q\bar q},y_{q\gamma},y_{\bar q\gamma})&=&
\as 
\Bigg \{\frac{y_{q\bar q}}{y_{q\bar q}+y_{q\gamma}}\,+
\frac{y_{q\bar q}}{y_{q\bar q}+y_{\bar q\gamma}}\,+\frac{y_{q\bar q}+y_{\bar q\gamma}}{y_{q\gamma}}\,
+\frac{y_{q\bar q}+y_{q\gamma}}{y_{\bar q\gamma}}\,
\nonumber\\
& & \quad +\ln y_{q\gamma} \left[
\frac{4y^2_{q\bar q}+2y_{q\bar q}y_{q\gamma}+4y_{q\bar q}y_{\bar q\gamma}+y_{q\gamma}y_{\bar q\gamma}}
{(y_{q\bar q}+y_{\bar q\gamma})^2}\right]
\nonumber\\
& & \quad +\ln y_{\bar q\gamma} \left[
\frac{4y^2_{q\bar q}+2y_{q\bar q}y_{\bar q\gamma}+4y_{q\bar q}y_{q\gamma}+y_{q\gamma}y_{\bar q\gamma}}
{(y_{q\bar q}+y_{q\gamma})^2}\right]
\nonumber\\
& & \quad -2 \Bigg[\frac{y^2_{q\bar q}+(y_{q\bar q}+y_{q\gamma})^2}{y_{q\gamma}y_{\bar q\gamma}}
\,R(y_{q\bar q},y_{\bar q\gamma})\;
+\frac{y^2_{q\bar q}+(y_{q\bar q}+y_{\bar q\gamma})^2}{y_{q\gamma}y_{\bar q\gamma}}
\,R(y_{q\bar q},y_{q\gamma})\;
\nonumber\\
& & \quad +
\frac{y^2_{q\gamma}+y^2_{\bar q\gamma}}{y_{q\gamma}y_{\bar q\gamma}(y_{q\gamma}+y_{\bar q\gamma})}
\;-2\ln y_{q\bar q}\left(\frac{y^2_{q\bar q}}{(y_{q\gamma}+y_{\bar q\gamma})^2}
+\frac{2y_{q\bar q}}{y_{q\gamma}+y_{\bar q\gamma}}\right)\Bigg]\Bigg\}.
\label{eq:Fert}
\end{eqnarray}
The function $R$ is defined as,
\begin{equation}
R(x,y)\,= \ln x\ln y -\ln x\ln (1-x)-\ln y\ln (1-y)
+\frac{1}{6}\pi^2 -\Li_{2}(x)-\Li_{2}(y) .
\label{eq:Rert}
\end{equation}

To ensure that the photon is resolved from the quark and antiquark,
we define the {\it resolved} three parton phase space to be,
$$
y_{q \gamma} > \ymin, \qquad y_{\bar{q}\gamma} > \ymin.
$$
In this region 
the resolved virtual cross section can be written as,
\begin{equation}
{\rm d}\sigma^{V}_{q \bar q \gamma}=V_{q \bar q \gamma }
{\rm d}\sigma^{R}_{q \bar q \gamma} + {\rm d}\sigma^{F}_{q \bar q \gamma},
\end{equation}
with,
\begin{equation}
{\rm d}\sigma^{F}_{q \bar q \gamma} = \aqed 2(2\pi)^2 \frac{1}{2M^2}
\int F(y_{q\bar q},y_{q\gamma},y_{\bar q\gamma})\; {\rm d}P_3(M;p_q,p_{\bar q},p_{\gamma}).
\end{equation}
The singularities present in $V_{q \bar q \gamma}$ precisely cancel with those from the 
$\gamma^* \to q \bar q \gamma g$ process when the gluon is unresolved
given in eq.~(\ref{eq:Rfac}).
The contribution from the finite function ${\rm d}\sigma^{F}_{q \bar q \gamma}$  will be evaluated numerically 
using the expression of Ellis, Ross and Terrano   
in eq.~(\ref{eq:Fert}) and the experimental jet algorithm 
to select photon +~1 jet final state events. 
The lowest order resolved cross section
${\rm d}\sigma^R_{q \bar q \gamma}$ will also be evaluated numerically. 
The more interesting problem lies in the {\it unresolved} region
as we shall see in the next subsection.

\subsection{The unresolved contribution}
\label{sec:unresolvirt}

In the {\it unresolved} region of phase space, the quark becomes
collinear with the photon so that, defining a new parent quark,
$Q$, with momentum $p_Q$ we have,
$$
 p_q + p_\gamma= p_Q.
$$
As usual, we introduce the variable $z$,
\begin{equation}
p_q= (1-z)p_Q, \hspace{1cm} p_{\gamma}=z p_Q.
 \end{equation}
The photon carries then a fraction $z$ of the composite quark momentum.
In this single collinear limit the three particle phase space
factorizes into a single collinear phase space factor, as we saw in
section~\ref{sec:single},
\begin{equation}
{\rm d}P_{3}^{(d)}(M; p_{q}, p_{\bar q}, p_{\gamma}) 
\to {\rm d}P_{2}^{(d)}(M;p_{Q},p_{\bar q})\,
{\rm d}P^{(d)}_{col}(p_{q},p_{\gamma},z).  
\end{equation}

At this stage we would like to
take the corresponding limit of the virtual matrix elements.
However, we note that the form given in eq.~(\ref{eq:Fert}) is
unsuitable for taking the collinear limit,
since as $y_{q\gamma}\to 0$, terms of the form,
$$
\frac{\log(y_{q\gamma})}{y_{q\gamma}},
$$
are present.
Such terms are problematic and are generated by taking the
$y_{q\gamma}\to 0$ limit {\bf after}
an expansion of the virtual matrix elements as a series in $\e$.
The correct procedure is to take the collinear limit first
and then expand the matrix elements as power series in $\e$.
Doing this, we find that the matrix elements factorize and are proportional to the lowest order $\gamma^* \to Q\bar q$ matrix elements,
\begin{equation}
|{\cal M}_{q\bar q \gamma}|^2_V \to  V_{col}(z,s_{q\gamma}) |{\cal M}_{Q\bar q}|^2,
\end{equation}
where,
\begin{eqnarray}
V_{col}(z,s_{q\gamma}) &=&
\left( \frac{4\pi }{M^2} \right)^{\e} \Re (-1)^{-\e}
\,\frac{\Gamma(1+\e)\Gamma^2(1-\e)}{\Gamma(1-2\e)}
\frac{1}{s_{q\gamma}}\frac{1}{(2\pi)^2}
\nonumber \\
&\times& 
\left(
\Pqp(z) \left(-\frac{2}{\e^2}
+\frac{2}{\e^2}\,y_{q\gamma}^{-\e}\,-\,
\frac{2}{\e^2}\,y_{q\gamma}^{-\e}\,(1-z)^{-\e}
\,F_{21}\left(-\e, -\e; 1-\e; z\right)-\frac{1}{\e}\frac{(3+2\e)}{(1-2\e)} \right)\right.
\nonumber\\
& & \left.  \hspace{1.5cm}
+ \frac{(\e z -1)}{(1-2\e)}\,y_{q\gamma}^{-\e} \right).
\end{eqnarray}
The Hypergeometric function $F_{21}(a, b; c; z)$
can be expanded as a series in $\e$,
\begin{displaymath}
F_{21}(-\e, -\e; 1-\e; z)\,=\,1 +\e^2\,\Li_{2}(z)
+\e^3\left(\Li_{3}(z)-{\rm S}_{12}(z)\right) + {\cal O}(\e^4).
\end{displaymath}

The collinear behaviour of one-loop amplitudes has been
studied by Bern, Dixon, Dunbar and Kosower \cite{bddk}
using helicity amplitudes.
We have checked our result for the
collinear limit of the squared matrix element by comparing with their results.

Integrating the unresolved squared matrix element 
over the corresponding unresolved phase space region yields
a dimensionless {\em virtual collinear} factor $VC_{F\gamma}{\rm d}z$
multiplying the
lowest order two particle cross section,
\begin{equation}
{\rm d}\sigma^V_{q\bar q\gamma} \to VC_{F\gamma}{\rm d}z \times \sigma_0,
\end{equation}
where,
\begin{equation}
VC_{F\gamma} \equiv
\as\aqed\left(\frac{4\pi\mu^2}{M^2}\right)^{\e} \mu^{2\e} (2\pi)^2
\frac{1}{\Gamma(1-\e)} 
\nonumber\\
\;z^{-\e}(1-z)^{-\e}  \int_{0}^{\ymin} {\rm d}y_{q\gamma}
y_{q\gamma}^{-\e} V_{col}(z,y_{q\gamma}).
\label{eq:sigV}
\end{equation}
Performing the $y_{q\gamma}$ integration,
we find,
\begin{eqnarray}
VC_{F\gamma}& = & 
\as\aqed\4pi2
\frac{\Gamma(1+\e) \Gamma(1-\e)}{\Gamma(1-2\e)}\Re (-1)^{-\e} \;\ymin^{-2\e}\;\;z^{-\e}(1-z)^{-\e}\;
\nonumber \\
&\times &\Bigg(
\Pqp(z)\bigg[\frac{2}{\e^3}\;\ymin^{\e}\;-\frac{1}{\e^3}
+\frac{1}{\e^2}\;\ymin^{\e}\;\left(\frac{3+2\e}{1-2\e}\right)
+\frac{1}{\e^3}\;(1-z)^{-\e}\;F_{21}(-\e,-\e;1-\e;z)
\bigg]\nonumber \\
& &\hspace{1cm}+\frac{(1-\e z)}{2\e(1-2\e)} \Bigg ).
\end{eqnarray}
As expected, the most divergent term in this expression
is proportional to $\Pqp(z)$ and precisely cancels the
leading singularity present in the two-particle unresolved contributions
from the four parton process discussed in the previous section, namely
the leading singularity in the soft/collinear contribution $SC_{F\gamma}$
(c.f.~section~\ref{subsec:softcol}).
The subleading poles do not cancel and are ultimately
factorized into the ${\cal O}(\alpha \alpha_{s})$ counterterm of the quark-to-photon fragmentation function, 
as will be discussed in section~\ref{sec:total}.

\section{Fragmentation contributions}
\setcounter{equation}{0}
\label{sec:frag}
In addition to the real and virtual contributions derived in the three
preceding sections, we need to consider contributions associated with 
the fragmentation processes
shown in Figs.~\ref{fig:class}.c and \ref{fig:class}.d.
These arise 
when  a quark-antiquark pair associated with a real or
virtual gluon are produced,
 followed by the fragmentation of a quark into a photon.
The contribution of these processes to the differential photon +~1 jet 
cross section
is determined by the convolution of the tree level
$\gamma^*\rightarrow~q\bar{q}g$ or the one-loop
$\gamma^* \to q\bar q$ cross
section with the {\it bare} quark-to-photon fragmentation function,
$\DBq(x)$, and has the symbolic form,
\begin{equation}
{\rm d}\sigma_{q\bar{q}(g)}^{D}\;=\;{\rm d}\sigma_{q\bar{q}(g)}\;
D^{B}_{q\rightarrow \gamma}(x)\;{\rm d}x.
\label{eq:structureD}
\end{equation}
Here, ${\rm d}\sigma_{q\bar{q}(g)}^{D}$, and 
${\rm d}\sigma_{q\bar{q}(g)}$
are the fully differential cross sections and
$x$ is the ratio between the photon and the {\it parent quark}
momenta.

The {\it bare} quark-to-photon fragmentation function
$\DBq(x)$ is the sum of a
{\it non perturbative} part, $\Dq(x,\mu_{F})$
which depends on the factorization scale $\mu_{F}$ and can only be
determined by experiment, and a {\it
perturbative} counterterm. Since the underlying
$\gamma^* \to q\bar q (g)$ process is already of
${\cal O}(\alpha_s)$, only the ${\cal O}(\alpha)$ counterterm needs to
be considered and we have cf. eq.~(\ref{eq:counter}),
\begin{equation}
\DBq(x)\,=\,
\Dq(x,\mu_{F}) +\frac{1}{\e}\aqed
\left(\frac{4\pi\mu^2}{\mu_{F}^2}\right)^{\e}\frac{1}{\Gamma(1-\e)}
\Pqpzero(x).
\label{eq:Ddef}
\end{equation}
As usual, this separation introduces a dependence on the fragmentation scale
$\mu_F$ in both the counterterm and the physical fragmentation function
$\Dq(x,\mu_{F})$.

As discussed in section~\ref{sec:1jet}, the fragmentation contributions
separate into three categories,
depending whether the gluon is resolved, unresolved or virtual.
If the gluon is identified in the final state, we will
find that
the singularities present in this resolved fragmentation contribution
are exactly cancelled  by the real {\it collinear
photon/resolved gluon} contribution from the
$\gamma^* \rightarrow~q\bar{q}g\gamma $ process.
On the other hand, if the gluon is unresolved, it may be collinear with
the quark or with the antiquark or it may be soft.
In the absence of the quark-to-photon fragmentation function, 
the infrared singularities
from the $\gamma^* \to q \bar q g$ process exactly cancel against those
from the one-loop $\gamma^* \to q\bar q$ process.
However, because of the fragmentation function, this is no longer the case.
When the gluon is collinear to the quark which subsequently
fragments into a photon,
the parent quark momentum is shared between the quark and the gluon 
and the fractional momenta carried by the photon and the gluon 
are related to each other.
This introduces a convolution between fragmentation function 
and parton level cross section.
As we shall see  
in section~\ref{subsec:unresolD},
a large part of the divergences present in this contribution
cancels against the singularities present in the double 
unresolved contributions discussed in section~\ref{sec:double}.

\subsection{Resolved contributions}
\label{subsec:resolD}

To ensure that the gluon is resolved from the quark and antiquark,
we define the {\it resolved} three parton phase space to be,
$$
y_{q g} > \ymin, \qquad y_{\bar{q}g} > \ymin.
$$
Events satisfying this constraint
contribute to the $\gamma +1 $~jet differential cross section in the
following two cases:
\begin{itemize}
\item[{(i)}]The gluon is clustered together with the quark which
fragments into a photon.
\item[{(ii)}]The gluon is clustered to the antiquark or it is
isolated while the antiquark is clustered with the photon jet.
\end{itemize}
In both cases the cross section has the form given by
eq.~(\ref{eq:structureD}) with $x$,  the fractional momentum carried by
the photon inside the {\it quark-photon} collinear cluster.
Note that $x$ is a {\it theoretical}
parameter which is only related to the momenta of quark and photon.
It does not necessarily coincide with the fractional momenta
carried by the photon inside the photon jet, $z$, which is reconstructed
by the jet algorithm.
In particular $x=z$ only holds if the photon jet only contains the
quark and photon, while the
antiquark and gluon are combined to form the second jet.
If on the other hand, the antiquark or the gluon are clustered
by the jet algorithm into the photon jet, one will generally find
$z<x$.
Ultimately, it is the {\it experimental} $z$, which is 
required to be greater than the experimental cut $\zcut$.

We note that the singularity structure from the
$q\bar q g \gamma$ final state in the limit where the
quark and photon are collinear (discussed in section~\ref{subsec:unresolp})
is proportional to
$\Pqp(x)$ and depends only on the theoretical $x$ value.
In fact, when the gluon is resolved,
the cancellation of the singularities
between the $q\bar q g$ final state with fragmentation counterterm
and those generated in the $q\bar q g \gamma$ final state
when the quark and photon are collinear
is unaffected by
the possible discrepancy between $x$ and $z$.
This explicit cancellation will be demonstrated in section~\ref{sec:total}.

\subsection{Unresolved contributions}
\label{subsec:unresolD}
In the previous subsection, the precise value of $z$ was determined by
the jet algorithm and is not necessarily the
same as $x$.
Similarly, when the gluon is unresolved,
$z$ and $x$ do not necessarily coincide.
If the gluon is virtual, soft or collinear to the antiquark,  we can
identify the ratio between the photon and the quark momenta
$x$ by $z$, since only fragmenting quark and photon form the ``photon'' jet.
and the cross section has the form given by
eq.~(\ref{eq:structureD}).
In each case the partonic cross section
${\rm d}\sigma_{q\bar{q}g}$
factorizes into a single unresolved factor multiplying
the tree level
cross section $\sigma_{0}$.
As these factors have already been derived before \cite{gg},
we will merely quote their results.
If a gluon is exchanged internally, the contribution to the $\gamma^*
\to \gamma$ +~1 jet rate reads,
\begin{equation}
{\rm d}\sigma_{V}^{D}=\sigma_{0}\;V_{q\bar{q}}\DBq(z){\rm d}z,
\end{equation}
with,
\begin{equation}
V_{q\bar{q}} = \as\left(\frac{4\pi\mu^2}{M^2}\right)^{\e}
\frac{\Gamma(1+\e)\Gamma^2(1-\e)}{\Gamma(1-2\e)}
\left(-\frac{2}{\e^2}\,-\,\frac{3}{\e}\,-8 +\pi^2 -16\e +  \frac{3}{2}\pi^2\e +
{\cal O}(\e^2)\right).
\label{eq:Vconv}
\end{equation}
When the gluon is real but soft, the invariants $y_{qg}$  and $y_{\bar{q}g}$
are both less than the theoretical cut $\ymin$ we find,
\begin{equation}
{\rm d}\sigma_{S}^{D}=\sigma_{0}\;S_{F}\DBq(z){\rm d}z,
\end{equation}
with the soft gluon factor being,
\begin{equation}
S_{F} = \as\left(\frac{4\pi\mu^2}{M^2}\right)^{\e}
\frac{1}{\Gamma(1-\e)}
\left(-\frac{2}{\e^2}\;\,\ymin^{-2\e}\right).
\label{eq:softconv}
\end{equation}
Similarly, when the gluon is collinear to the antiquark, $y_{\bar{q}g}<\ymin$
but $y_{qg}>\ymin$, we have,
\begin{equation}
{\rm d}\sigma_{C(\bar q)}^{D}=\sigma_{0}\;C_{F}(\bar q)\DBq(z){\rm d}z,
\end{equation}
with,
\begin{equation}
C_{F}(\bar{q})=\as\left(\frac{4\pi\mu^2}{M^2}\right)^{\e}
\frac{1}{\Gamma(1-\e)}
\;\ymin^{-\e}
\left(-\frac{2}{\epsilon^2}
\;\,\ymin^{-\epsilon}\;+\;\frac{(1-\epsilon)(4-\epsilon)}
{2\epsilon^2(1-2\epsilon)}\;\frac{\Gamma^2(1-\epsilon)}{\Gamma(1-2\epsilon)}\;
\right).
\label{eq:qbarconv}
\end{equation}

On the other hand, if the gluon is collinear to the quark,
so that the gluon carries a fraction $y$ of the quark/gluon cluster
momentum,
$z$ is no longer equal to $x$.
In fact, $z$ is given by
the product of the
momentum fraction carried by the quark, $1-y$ and the ratio of the
photon and quark momenta $x$, so that,
$$
z = x (1-y).
$$
We therefore introduce the constraint, $\int_{0}^{1}{\rm d}z
\;\delta \left(x(1-y)-z\right)$ and integrate over $x$ so that,
in the collinear region, 
$$y_{qg} < \ymin \qquad {\rm and}\qquad y_{\bar q g} \equiv y > \ymin,$$ 
we have,
\begin{eqnarray}
{\rm d}\sigma_{C(q)}^{D}
&=& -\frac{1}{\e}\sigma_{0}\;\as
\left(\frac{4\pi\mu^2}{M^2}\right)^{\e}
\frac{1}{\Gamma(1-\e)}\;\ymin^{-\e}
\nonumber\\
& &
\times \int_{\ymin}^{1-z}\frac{{\rm d}y}{1-y}\,[y(1-y)]^{-\e}\Pqg(y)
\DBq\left(\frac{z}{1-y}\right){\rm d}z.
\label{eq:quarkconv}
\end{eqnarray}
The trivial integral over $y_{qg}$ has been carried out while the factor of $(1-y)$ comes from the $x$ integration.
The $y$ integral now involves the fragmentation function
and requires some work to evaluate. 
The resulting expression will involve a convolution of the
splitting function with the fragmentation function.
However, the convolution integral present in eq.~(\ref{eq:quarkconv}) 
appears to have
an explicit $\ymin$
dependence coming from the lower boundary of the $y$ integral.
We know that since $\ymin$ is an artificial parameter
which cannot influence the physical cross section for any choice
of fragmentation function, the $\ymin$ dependence must
only act
multiplicatively on  the fragmentation function.

We therefore add and subtract
the contribution where a gluon
is collinear to a quark multiplied by $\DBq(z)$. 
The convolution integral can be rewritten in the following way,
\begin{eqnarray}
{\rm d}\sigma_{C(q)}^{D}&=&{\rm d}\sigma_{C(q)}^{D}
+\frac{1}{\e}\sigma_{0} \as
\left(\frac{4\pi\mu^2}{M^2}\right)^{\e}
\frac{1}{\Gamma(1-\e)}\;\ymin^{-\e}\;\DBq(z) {\rm d}z
\nonumber\\
& \times & \left (
\int_{\ymin}^{1}{\rm d}y\;[y(1-y)]^{-\e}\Pqg(y)
-\left[\;\frac{2}{\epsilon}
\ymin^{-\epsilon}\;-\;\frac{(1-\epsilon)(4-\epsilon)}
{2\epsilon(1-2\epsilon)}\;\frac{\Gamma^2(1-\epsilon)}{\Gamma(1-2\epsilon)}\;
\right] \right )\nonumber \\
&=&{\rm d}\sigma^{D}_{C(q)'}\,+\,
\sigma_{0}\;C_{F}(q)\,\DBq(z){\rm d}z,
\label{eq:qcon}
\end{eqnarray}
so that $C_F(q) \equiv C_F(\bar q)$ and 
with ${\rm d}\sigma^{D}_{C(q)'}$ given by,
\begin{eqnarray*}
{\rm d}\sigma^{D}_{C(q)'}
&=& -\frac{1}{\e}\sigma_{0}
\as\left(\frac{4\pi\mu^2}{M^2}\right)^{\e}
\frac{1}{\Gamma(1-\e)}\;\ymin^{-\e}{\rm d}z
\nonumber\\
& & \times \Bigg(
\int_{\ymin}^{1-z}{\rm d}y \,[y(1-y)]^{-\e}\Pqg(y)
\left[\frac{\DBq\left(\frac{z}{1-y}\right)}{1-y}-\DBq(z)\right]
\nonumber\\
& & \quad -\int_{1-z}^{1}{\rm d}y\,[y(1-y)]^{-\e}\Pqg(y)\DBq(z)\Bigg).
\end{eqnarray*}
In the first integral in  the expression for ${\rm d}\sigma^{D}_{C(q)'}$,
the integrand vanishes when
$y\rightarrow 0$, and we can safely extend the range of integration to 0.
By doing so, the convolution contribution itself becomes $\ymin$
independent as required.

Recalling the definition of the Altarelli-Parisi splitting function \cite{AP},
$$
\Pqg(y) = \frac{1+(1-y)^2-\e y^2}{y},
$$
making the change of variables $ y = 1-t$, and using the ``+''-prescription
to evaluate the singular parts of the integrals,
we can recast the expression for ${\rm d}\sigma^{D}_{C(q)'}$ 
 in a more suggestive form,
\begin{eqnarray}
{\rm d}\sigma^{D}_{C(q)'}&=&
\sigma_{0}{\rm d}z
\as\left(\frac{4\pi\mu^2}{M^2}\right)^{\e}
\frac{1}{\Gamma(1-\e)}\;\ymin^{-\e}
\nonumber\\
& \times & 
\int_{z}^{1}\frac{{\rm d}t}{t}\,D\left(\frac{z}{t}\right)\;
\left \{  -\frac{1}{\e} \Pqqzero(t) \right.
\nonumber\\
& &  \left. +
\left[\left(\frac{\ln(1-t)}{1-t}\right)_{+}\,(1+t^2)\;+
\frac{\ln(t)}{1-t}(1+t^2) +(1-t) 
- \left(3-\frac{\pi^2}{3}\right)\delta(1-t)
\right] \right.
\nonumber\\
& &  \left.-\e
\left[\frac{1}{2}\left(\frac{\ln^2(1-t)}{1-t}\right)_{+}\,(1+t^2)\;+
\;\frac{1}{2}\frac{\ln^2(t)}{1-t}\,(1+t^2)\;
+\;\frac{\ln(t)\ln(1-t)}{1-t}\,(1+t^2)
\right.\right.
\nonumber\\
& & \left. \left. \quad
\:+\;(1-t)\left[\ln(t)+\ln(1-t)\right] 
+ \left(7-\frac{\pi^2}{4}-4\zeta(3)\right)\delta(1-t)
\right] \right \}.
\end{eqnarray}
As expected, the coefficient of the leading pole term
is the universal lowest order Altarelli-Parisi \cite{AP}
quark-to-quark splitting function in four dimensions, see eq.~(\ref{eq:pqq}).

The contributions directly proportional to $\DBq$ given 
in eqs.~(\ref{eq:Vconv}), (\ref{eq:softconv}), (\ref{eq:qbarconv}) 
and by the second term of eq.~(\ref{eq:qcon}) can be combined together and yield a finite result,
\begin{eqnarray*}
{\rm d}\sigma^{D}_{K}&=&
\sigma_{0}\, \left(C_{F}(q) + C_F(\bar q)\,+\,S_{F}+V_{q\bar{q}}\right)
\,\DBq(z){\rm d}z
\nonumber\\
&=& \sigma_{0}\, {\cal K}_{q\bar{q}}\,\DBq(z){\rm d}z,
\end{eqnarray*}
where ${\cal K}_{q\bar{q}}$ is the finite two quark ${\cal K}$-factor given in eq.~(4.22) of Ref.~\cite{gg}.
Since the ${\cal O}(\alpha)$ bare fragmentation function counterterm 
is proportional to $1/\e$, we need to expand ${\cal K}_{q\bar{q}}$
up to ${\cal O}(\e)$, 
\begin{eqnarray}
{\cal K}_{q\bar{q}}&=& \as 
\left(\frac{4\pi\mu^2}{M^2}\right)^{\e}
\frac{1}{\Gamma(1-\e)} \nonumber\\
& & \times \left[ \left(-2\ln^2(\ymin)\,-3\ln(\ymin)
+\frac{\pi^2}{3} -1\right)
+ \e \bigg(2\ln^3(\ymin)\,+\frac{3}{2}\ln^2(\ymin) \right.
\nonumber\\
& &  \left. \quad
+\left(\frac{2\pi^2}{3}-7 \bigg)\,\ln(\ymin)\,
-2\,+\pi^2\,-4\zeta(3)\right)\right].
\label{eq:sigKD}
\end{eqnarray}

Finally the sum of all unresolved contributions involving $\DBq$ 
can be written in terms of the dimensionless
{\it fragmentation collinear} factor $FC_{F\gamma}$,
\begin{eqnarray}
{\rm d}\sigma_{q\bar{q} (g)}^{D}&=&
{\rm d}\sigma^{D}_{K} + {\rm d}\sigma^{D}_{C(q)'}\nonumber \\
&\equiv& \,FC_{F\gamma}{\rm d}z \times \sigma_{0}.
\end{eqnarray}
Expanding up to ${\cal O}(\e)$, we find,
\begin{equation}
FC_{F\gamma}=
\frac{1}{\Gamma(1-\e)}
\as
\left(\frac{4\pi\mu^2}{M^2}\right)^{\e}
\int_{z}^{1}\frac{{\rm d}t}{t}\,
\DBq\left(\frac{z}{t}\right){\rm d}z
\left[-\frac{1}{\e}\Pqqzero(t) +c_{q}^{(1)} +\e c_{q}^{(2)} \right], 
\label{eq:sumD1}
\end{equation}
where,
\begin{eqnarray}
c_{q}^{(1)} &=&
 \left(2\ln^2(\ymin)\,+\frac{3}{2}\ln(\ymin)
- \left(\frac{2\pi^2}{3}+\frac{9}{2}\right) \right)\delta(1-t) 
\nonumber\\
& &   -\frac{(1+t^2)}{(1-t)_{+}}\ln(\ymin)
    -\left(\frac{\ln(1-t)}{1-t}\right)_{+} (1+t^2)
    -\frac{\ln(t)}{1-t}(1+t^2) -(1-t),\\
c_{q}^{(2)} &=& 
\left(2\ln^3(\ymin)\,+\frac{3}{2}\ln^2(\ymin)
\,+\left(\frac{2\pi^2}{3}-7\right )\ln(\ymin) +5\,+\frac{3}{4}\pi^2\,-8\zeta(3)\right)\,\delta(1-t)
\nonumber\\
& &   \qquad  +
 \frac{1}{2}\,\left(\frac{\ln^2(1-t)}{1-t}\right)_{+}\,(1+t^2)\;+
\;\frac{1}{2}\frac{\ln^2(t)}{1-t}(1+t^2)
+\;\frac{\ln(t)\ln(1-t)}{1-t}\,(1+t^2)
\nonumber\\
& &     \qquad 
\;+\;(1-t)\ln(t)+(1-t)\ln(1-t).   
\label{eq:cq}
\end{eqnarray}

The final step is to insert the decomposition of the bare
fragmentation function given in eq.~(\ref{eq:Ddef}) 
into the sum of all fragmentation contributions given in eq.~(\ref{eq:sumD1}).
We divide the fragmentation collinear factor into two pieces,
one depending on the non-perturbative fragmentation function and
one involving the perturbative counterterm,
\begin{equation}
FC_{F\gamma}= FC^{np}_{F\gamma}+FC^{p}_{F\gamma},
\end{equation}
with,
\begin{eqnarray}
FC^{np}_{F\gamma}&=&
\as 
\left(\frac{4\pi\mu^2}{M^2}\right)^{\e}
\frac{1}{\Gamma(1-\e)}  
\left[-\frac{1}{\e}\Pqqzero + c_{q}^{(1)}\right]\otimes \Dq(z,\mu_{F}),
\nonumber \\
FC^{p}_{F\gamma}&=&
\as \aqed
\left(\frac{4\pi\mu^2}{M^2}\right)^{\e}
\left(\frac{4\pi\mu^2}{\mu_F^2}\right)^{\e}
\frac{1}{\Gamma^2(1-\e)}
\left [-\frac{1}{\e}\Pqqzero
+c_{q}^{(1)} +\e c_{q}^{(2)}\right]
\otimes \frac{1}{\e}\Pqpzero.\nonumber \\
\end{eqnarray}
We see that this  unresolved fragmentation contribution   
contains $1/\e^2$ poles as leading singularities 
which must be combined with the virtual and the double unresolved
singularities presented in sections~\ref{sec:double} and \ref{sec:virt}.

\section{Sum of all contributions}
\setcounter{equation}{0}
\label{sec:total}
So far we have isolated all  the divergent contributions 
to the $\gamma$ +~1 jet rate
arising in the unresolved regions of the phase space
analytically.
However, to obtain a physical (finite) 
next-to-leading order prediction for the photon +~1 jet rate,
we must combine these divergent contributions together and 
absorb the remaining singularities 
into the perturbative counterterm of the 
{\it bare} quark-to-photon fragmentation function. Following a fixed order
approach to evaluate the photon +~1 jet differential cross section, 
we evaluate this rate order by order and consequently determine 
the perturbative counterterm of the bare fragmentation function 
order by order.
The ${\cal O}(\alpha)$ counterterm has been determined while performing
the calculation up to order $\alpha$ \cite{andrew}, we shall now utilise
the ${\cal O}(\alpha \alpha_{s})$ perturbative counterterm given in eq.~(\ref{eq:counter}).

Recalling the generic structure for the photon +~1 jet rate 
given in eq.~(\ref{eq:gstruc1}) we have,
\begin{equation}
\frac{1}{\sigma_0}
\frac{{\rm d}\sigma( 1~{\rm jet} + ``\gamma")}{{\rm d}z}
=
\frac{1}{\sigma_0}
\frac{{\rm d}\hat\sigma_\gamma}{{\rm d}z}+ 
\frac{2}{\sigma_0} 
\frac{{\rm d}\hat\sigma_q }{{\rm d}z}\otimes \DBq.
\end{equation}
Here, the factor of 2 reflects the fact that by charge conjugation, the 
quark and antiquark fragmentation contributions are equal.
We recall that at ${\cal O}(\alpha)$, the differential cross section is given by,
\begin{equation}
\frac{1}{\sigma_0}
\frac{{\rm d}\sigma^{LO}(1~{\rm jet} + ``\gamma")}{{\rm d}z}
= 2\calDq + \frac{1}{\sigma_0}
\frac{{\rm d}\sigma^{R}_{q \bar q \gamma}}{{\rm d}z}
\end{equation}
where $\calDq$ is the effective fragmentation function of eq.~(\ref{eq:calDq}) \cite{andrew}. 
Then at ${\cal O}(\alpha\alpha_s)$, ${\rm d}\hat\sigma_\gamma$ receives contributions from the tree level four parton process and the one-loop three parton process.
In sections~\ref{sec:single} and \ref{sec:double}, the singularities in the four parton process were isolated so that,
\begin{equation}
\frac{1}{\sigma_0}\frac{{\rm d}\hat\sigma_{q\bar q \gamma g}}{{\rm d}z}
=
 \frac{1}{\sigma_0}\frac{{\rm d}\sigma_{q\bar q \gamma g}^{R}}{{\rm d}z}
+ R_{q\bar q \gamma}\frac{1}{\sigma_0} 
\frac{{\rm d}\sigma^{R}_{q \bar q \gamma}}{{\rm d}z} 
+2\left (TC_{F\gamma} + SC_{F\gamma} + DC_{F\gamma}\right)
+ 2 \tilde{C}_{F\gamma}\frac{{\rm d}\sigma^{R}_{q\bar q g}}{\sigma_0}.
\end{equation}
The divergences are concentrated in factors multiplying finite resolved parton differential cross sections.
Similarly, the singularities from the one-loop process were reorganized in section~\ref{sec:virt},
\begin{equation}
\frac{1}{\sigma_0}\frac{ {\rm d}\hat\sigma^V_{q\bar q \gamma}}{{\rm d}z}
=
V_{q\bar q \gamma}\frac{1}{\sigma_0} 
\frac{{\rm d}\sigma^{R}_{q \bar q \gamma}}{{\rm d}z} 
+\frac{1}{\sigma_0}\frac{{\rm d} 
\sigma^{F}_{q \bar q \gamma}}{{\rm d}z}
+2 VC_{F\gamma}. 
\end{equation}
The fragmentation contribution from the real $\gamma^*\to q\bar q g$ and
$\gamma^* \to q\bar q$ processes discussed in section~\ref{sec:frag} can be written,
\begin{equation}
\frac{1}{\sigma_0}\frac{{\rm d}\sigma_{q\bar q (g)}}{{\rm d}z}\otimes 
 \DBq  =
\frac{2}{\sigma_0}{\rm d}\sigma^{R}_{q\bar q g}\DBq(z) 
+ 2 FC_{F\gamma}. 
\end{equation}
Finally, there is a contribution from the ${\cal O}(\alpha\alpha_s)$ perturbative counterterm in $\DBq$ multiplying the lowest order 
$\gamma^*\to q\bar q$ cross section, cf. eq.~(\ref{eq:counter}),
\begin{equation}
\frac{1}{\sigma_0}\frac{{\rm d}\sigma_{q\bar q}}{{\rm d}z}\otimes \DBq  =
2 \Dq^{\alpha\alpha_s}(z). 
\end{equation}

The overall result for the photon +~1 jet cross section is obtained by summing these four contributions.
Regrouping the ${\cal O}(\alpha\alpha_s)$ terms according to the 
resolved parton cross section they are proportional to,
we obtain,
\begin{eqnarray}
\frac{1}{\sigma_0}
\frac{{\rm d}\sigma^{NLO}(1~{\rm jet} + ``\gamma")}{{\rm d}z}
&=&
2\calDq^{\alpha\alpha_s}
\nonumber \\
&&   +  
 {\cal K}_{q\bar q \gamma}  
\frac{1}{\sigma_0}
\frac{{\rm d}\sigma^{R}_{q \bar q \gamma}}{{\rm d}z}
+\frac{1}{\sigma_0}
\frac{{\rm d}\sigma^{F}_{q \bar q \gamma}}{{\rm d}z}
\nonumber \\
&&   +  
2  \calDq  
\frac{{\rm d}\sigma^{R}_{q\bar q g}}{\sigma_0}
\nonumber \\
& & + \frac{1}{\sigma_0}
\frac{{\rm d}\sigma_{q\bar q \gamma g}^{R}}{{\rm d}z}.
\label{eq:sumjet3}
\end{eqnarray}
Recalling the definition of $\tilde{C}_{F\gamma}$ together with the
${\cal O}(\alpha)$ perturbative counterterm given in eq.~(\ref{eq:calDq})
yields the effective quark-to-photon fragmentation function up 
to ${\cal O}(\alpha)$,
\begin{eqnarray}
\calDq &=& 
\tilde{C}_{F\gamma} +\DBq  \nonumber \\
&=& \Dq(z,\mu_F) +
\aqed \left
(\Pqpzero \ln \left (\frac {\smin y_{q\bar q} 
\,z(1-z)}{\mu_{F}^2}\right )\, +z \right),
\end{eqnarray}
while,
\begin{eqnarray}
{\cal K}_{q\bar q \gamma} &=& V_{q\bar q \gamma} + R_{q\bar q \gamma} \nonumber \\
&=&\as \left(-2 \log^2(\ymin) -3 \log(\ymin y_{q\bar q}) 
+\frac{\pi^2}{3} -1\right).
\end{eqnarray}
These, and the other contributions 
given in the last three lines of 
eq.~(\ref{eq:sumjet3}) are finite and
will be evaluated numerically.
The sum of the double unresolved factors, together with 
the fragmentation counterterm is also finite and provides the
${\cal O}(\alpha\alpha_s)$ contribution to the effective fragmentation function,
\begin{eqnarray}
\lefteqn{\calDq^{\alpha\alpha_s} =TC_{F\gamma}+SC_{F\gamma}+DC_{F\gamma}+VC_{F\gamma}+FC_{F\gamma}
+\Dq^{\alpha\alpha_s}=\as\aqed} 
\nonumber \\
& & \times \Bigg\{ 
-6+{\frac {z}{4}}
+{\pi}^{2}\left (-\frac{1}{3}+{\frac {z}{12}}+{\frac {\Pqpzero(z)}{2}}
\right )+\ln (z)\left ({\frac {31}{4}}-{\frac {27\,z}{4}}-\Pqpzero(z)\right )
\nonumber \\
& & \quad
+\ln (z){\pi}^{2}\left (-\frac{1}{3}
+{\frac {z}{6}}+{\frac {\Pqpzero(z)}{3}}\right )+
\ln^2 (z)\left (-2+{\frac {13\,z}{8}}\right )
\nonumber \\
& & \quad
+\ln (z)\ln (1-z)\left (-3+{\frac {7\,z}{2}}-{\frac {3\,\Pqpzero(z)}{2}}\right )+\ln 
(z){\rm Li}_2(1-z)\left (4-2\,z\right )
\nonumber \\
& & \quad
+\ln^2 (1-z)
\left (1+{\frac {5\,z}{4}}-{\frac {3\,\Pqpzero(z)}{2}}\right )+\ln (1-z)
{\rm Li}_2(1-z)\left (2-z+5\,\Pqpzero(z)\right )
\nonumber \\
& & \quad
+\ln^3 (z)\left (\frac{5}{6}
-{\frac {5\,z}{12}}\right )+\ln^2 (z)
\ln (1-z)\left (2-z+\Pqpzero(z)\right )-{\frac {\ln (1
-z){\pi}^{2}\Pqpzero(z)}{2}}
\nonumber \\
& & \quad
+\ln (z)\ln^2 (1-z)
\left (1-{\frac {z}{2}}+3\,\Pqpzero(z)\right )+\ln (1-z)\left (-{
\frac {z}{2}}-\Pqpzero(z)\right )
\nonumber \\
& & \quad
+{\rm Li}_2(1-z)\left (-3+{\frac {7\,z}{2}}-{
\frac {3\,\Pqpzero(z)}{2}}\right )+{\rm Li}_3(1-z)\left (-2+z-3\,\Pqpzero(z)\right )
\nonumber \\
& & \quad
+{\rm S}_{12}(1-z)\left (4-2\,z-6\,\Pqpzero(z)\right ) 
+{\frac {5\,\ln^3 (1-z)\Pqpzero(z)}{6}}+9\,\Pqpzero(z)\zeta (3)
\nonumber \\
& & \quad
+\ln \left(\frac{\mu_F^2}{M^2}\right)
\Bigg[ -2\,\ln^2 (1-z) \Pqpzero(z)+\ln 
(1-z)\left (-2-{\frac {3\,z}{2}}+{\frac {3\,\Pqpzero(z)}{2}}\right ) \nonumber \\
& & \qquad
+{\rm Li}_2(1-z)\left (-2+z-6\,\Pqpzero(z)\right )+\ln (z)\left (3-{\frac {3\,z}{2}}
\right )+\frac{1}{2}-z+\Pqpzero(z) \nonumber \\
& & \qquad
+\ln^2 (z)\left (-2+z\right )+\ln 
(z)\ln (1-z)\left (-2+z-4\,\Pqpzero(z)\right )+{\frac {2\,{\pi}^{2}\Pqpzero(z)}{3}}
\Bigg] \nonumber\\
& & \quad
+\ln^2 \left(\frac{\mu_F^2}{M^2}\right)
\Bigg[
\ln (1-z)\Pqpzero(z)+1-{\frac {z}{4}}+\ln (z)\left (1-{
\frac {z}{2}}\right )\Bigg] \nonumber \\
& & \quad
+\ln (\ymin) \Bigg[
-\frac{1}{2}-2\,z-\Pqpzero(z)+2\,\ln^2 (1-z)\Pqpzero(z)-{\frac {{\pi}^{2}
\Pqpzero(z)}{3}}+\ln^2 (z)\left (2-z\right ) \nonumber \\
& & \qquad
+\ln (z)\ln (1-z)
\left (2-z+4\,\Pqpzero(z)\right )+\ln (1-z)\left (2+{\frac {3\,z}{2}}-{\frac 
{9\,\Pqpzero(z)}{2}}\right )
\nonumber \\
& & \qquad
+{\rm Li}_2(1-z)\left (2-z+6\,\Pqpzero(z)\right )+\ln (z)
\left (-3+{\frac {3\,z}{2}}-3\,\Pqpzero(z)\right )
\Bigg]\nonumber \\
& & \quad
+\ln^2 (\ymin) \Bigg[
\ln (z)\left (1-{\frac {z}{2}}-2\,\Pqpzero(z)
\right )+1-{\frac {9\,z}{4}}-3\,\Pqpzero(z)-\ln (1-z)\Pqpzero(z)\Bigg]
\nonumber \\
& & \quad
-2\,\ln^3 (\ymin)\Pqpzero(z) \nonumber \\
& & \quad
+\ln \left(\frac{\mu_F^2}{M^2}\right) \ln (\ymin)
\Bigg[-2+{
\frac {z}{2}}+3\,\Pqpzero(z)+\ln (z)\left (-2+z\right )-2\,\ln (1-z)\Pqpzero(z)
\Bigg] \nonumber\\
& & \quad
+2\,\ln \left(\frac{\mu_F^2}{M^2}\right)\ln^2(\ymin)\Pqpzero(z) \Bigg\}
\nonumber \\
& & \quad +\as 
\left (-\Pqqzero\ln \left(\frac{\mu_{F}^2}{M^2}\right)+c_{q}^{(1)}\right)\otimes
\Dq(z,\mu_{F}),
\label{eq:total}
\end{eqnarray}
where $c_{q}^{(1)}$  is given by eq.~(\ref{eq:cq}).

Note that although each contribution given 
in eq.~(\ref{eq:sumjet3})
is now finite the individual contributions depend on both the factorization scale $\mu_F$ and the parton resolution parameter $\ymin$.
When combined numerically 
this $\ymin$ dependence must cancel as we will explicitly 
show in the next section.

\subsection{The evolution equation for $ D_{q \to \gamma}(z,\mu_{F})$}
\label{subsec:evol}
In order to obtain a finite differential cross section we have 
factorized the collinear singularities in the perturbative counterterm 
of the bare quark-to-photon fragmentation function at some factorization 
scale $\mu_{F}$.
The bare quark-to-photon fragmentation function 
should however not depend on the scale at which 
the factorization procedure takes place.
Requiring that it is independent of 
the factorization scale $\mu_{F}$ yields the evolution equation for 
the renormalized non-perturbative fragmentation function
$D_{q \to \gamma}(z,\mu_{F})$.
We can see this by differentiating $\DBq$ with respect to $\ln(\mu_F^2)$,
\begin{equation}
\frac{{\rm d}\DBq(z)}{{\rm d} \ln(\mu_{F}^2)}=0,
\end{equation}
and using eq.~(\ref{eq:counter}),
\begin{eqnarray}
\frac{\partial\Dq(z,\mu_{F})}{\partial \ln(\mu_{F}^2)}&=&
    \aqed \Pqpzero
+\as\aqed \Pqpone
\nonumber\\
& & 
-\as\aqed \frac{(4\pi)^{2\e}}{\Gamma^2(1-\e)}
\Pqqzero\otimes \Pqpzero
\left[-\frac{1}{\e}+2 \ln\left(\frac{\mu_{F}^2}{\mu^2}\right)\right]
\nonumber\\
& & 
+\as \frac{(4\pi)^{\e}}{\Gamma(1-\e)}\Pqqzero \otimes \frac{\partial \Dq (z,\mu_{F})}
{\partial \ln(\mu_{F}^2)}\left[-\frac{1}{\e}+
\ln\left(\frac{\mu_{F}^2}{\mu^2}\right)\right]
\nonumber\\
& & 
+\as\Pqqzero \otimes \Dq(z,\mu_{F}).
\label{eq:evolnlo1}
\end{eqnarray}
For terms in the third line of this equation, 
which are proportional to $\alpha_{s}$,
the variation of the non-perturbative fragmentation function 
with respect to $\mu_{F}$,  
$\partial D_{q \to \gamma}(z,\mu_{F})/
\partial \ln(\mu_{F}^2)$ 
is given by the lowest order evolution equation for 
$D_{q \to \gamma}(z,\mu_{F})$.
To be more precise, at ${\cal O}(\alpha)$ and keeping terms of ${\cal O}(\e)$, we have,
\begin{equation}
\frac{\partial \Dq(z,\mu_{F})}
{\partial \ln(\mu_{F}^2)}= \aqed \frac{(4\pi)^{\e}}{\Gamma(1-\e)} 
\left[ 1-\e\ln\left(\frac{\mu_{F}^2}{\mu^2}\right)\right]  \Pqpzero.
\end{equation}
Inserting this into eq.~(\ref{eq:evolnlo1}),
we find the expected, $\mu$ independent, result,
\begin{equation}
\frac{\partial \Dq(z,\mu_{F})}
{\partial \ln(\mu_{F}^2)}=
\aqed \Pqpzero  
+\as\aqed \Pqpone
+\as\Pqqzero\otimes \Dq(z,\mu_{F}).
\label{eq:evolnlo}
\end{equation}

\subsection{An exact solution of the  evolution equation}
\label{subsec:evolsol}

The evolution equation (\ref{eq:evolnlo}) is insufficient to uniquely 
determine 
the non-perturbative quark-to-photon fragmentation function 
$\Dq(z,\mu_{F})$.
However it is possible to give an exact 
(up to ${\cal O}(\alpha \alpha_{s})$) solution 
of this next-to-leading order evolution equation.
This solution is a first step leading to the ultimate determination of 
$D_{q \to \gamma}(z,\mu_{F})$. In the same way, the exact (up to ${\cal O}(\alpha)$) 
solution of the leading 
order evolution equation \cite{andrew} led to 
a determination of the quark-to-photon fragmentation function by a comparison 
between the LO calculation of the photon +~1 jet rate and the data \cite{aleph}.

We construct the exact  ${\cal O}(\alpha \alpha_{s})$ solution by imposing that it takes the following 
general form,
\begin{equation}
\Dq(z,\mu_{F})=\left[1+\as A\right]\otimes 
\Dq^{(LO)}(z,\mu_{F}) +\as\aqed B,
\label{eq:gDnlo}
\end{equation}
where $A$, $B$ are unknown functions of $z$, $\mu_{F}$
and the constant of integration $\mu_{0}$.
Here, 
$\Dq^{(LO)}(z,\mu_{F})$ is the exact solution of the lowest order evolution 
equation obtained by ignoring the terms proportional to $\alpha_s$ in 
eq.~(\ref{eq:evolnlo}),
\begin{equation}
\Dq^{(LO)}(z,\mu_{F})=\aqed\Pqpzero 
\ln\left(\frac{\mu^2_{F}}{\mu_{0}^2}\right) +\Dq^{np}(z,\mu_{0}).
\label{eq:Dlo}
\end{equation}
where the non-perturbative input 
fragmentation function $\Dq^{np}(z,\mu_{0})$ is to be determined by data.
Inserting eq.~(\ref{eq:Dlo})
in the general form for the exact solution 
of the next-to-leading evolution equation   
and neglecting all terms which have more 
than one power of $\alpha_{s}$, we obtain,
\begin{eqnarray}
A &=&\Pqqzero
\ln\left(\frac{\mu^2_{F}}{\mu^2_{0}}\right), \nonumber \\
B &=&\Pqpone  
\ln\left(\frac{\mu^2_{F}}{\mu^2_{0}}\right )
-\frac{1}{2}\Pqqzero\otimes \Pqpzero
\ln ^2\left(\frac{\mu^2_{F}}{\mu_{0}^2}\right), 
\label{eq:ab}
\end{eqnarray}
so that the exact solution of the next-to-leading order evolution equation reads,
\begin{eqnarray}
\Dq^{(NLO)}(z,\mu_{F})&=&
\Dq^{np}(z,\mu_{0}) + \aqed\Pqpzero 
\ln\left(\frac{\mu^2_{F}}{\mu_{0}^2}\right) 
+\as\aqed \Pqpone 
\ln \left(\frac{\mu^2_{F}}{\mu_{0}^2}\right)
\nonumber\\
& & +
\as
\ln \left(\frac{\mu^2_{F}}{\mu_{0}^2}\right) \Pqqzero\otimes 
\left[\aqed\frac{1}{2}\Pqpzero\ln \left(\frac{\mu^2_{F}}{\mu_{0}^2}\right)
+\Dq^{np}(z,\mu_{0})\right].
\nonumber\\
\label{eq:Dnlo}
\end{eqnarray}

This solution has some interesting properties.
First, 
it is exact at the order of the calculation (${\cal O}(\alpha \alpha_s)$), 
and yields no terms of higher orders. 
The photon +~1 jet rate with this solution implemented is therefore 
independent of the choice of the factorization scale $\mu_{F}$,
 as it should be in a 
calculation at fixed order in perturbation theory. 
Second, we have not resummed terms proportional to 
$\ln \mu_F^2$.  We choose to do this for a variety of reasons.
Such resummations are only 
unambigous if the resummed logarithm is the only large logarithm   
in the kinematical region under consideration. 
If logarithms of different arguments can become simultaneously large, 
resummation of one of these logarithms at a given order implies that 
all other potentially large logarithms are shifted into a higher order 
of the perturbative expansion, i.e.~have to be neglected. 
In our calculation, we encounter three different potentially large
logarithms, $\ln \mu_F^2$, $\ln (1-z)$ and 
$\ln y_{{\rm cut}}$, and resummation of one of these would 
immediately imply that effects from the other logarithms at the given  
order of the calculation are neglected, a procedure which appears to 
be clearly inconsistent. 
Furthermore, since the overall photon +~1 jet rate does not depend on
$\mu_F$,  it appears   
to be more than doubtful that such logarithms should be resummed.
As a final point, while $\mu_F$ is an artificial parameter, it does represent 
the boundary between the perturbative and non-perturbative contributions.
It corresponds approximately to the transverse 
momentum of the photon with respect to the jet axis, 
a physical scale, typically of the order of a few GeV,
 and {\em not} a large scale such as the $e^+e^-$ centre-of-mass energy.

\section{Dependence on $\ymin$}
\setcounter{equation}{0}
\label{sec:ymin}
We have now collected all necessary ingredients to evaluate 
the ${\cal O}(\alpha \alpha_{s})$ $\gamma$ +~1 jet 
differential cross section numerically.
There are four separate contributions specified by eq.~(\ref{eq:sumjet3})
and determined according to the number of partons in the final state and 
the presence or absence of the quark-to-photon fragmentation function.
For each contribution, the appropriate matrix elements  
are integrated over the corresponding phase space using 
Monte Carlo techniques, i.e.~with {\tt VEGAS}  \cite{vegas}.
In particular, for each event the invariants $y_{ij}$ are defined allowing 
the reconstruction of the four-momenta $p_{i}^{\mu}$ 
of the particles in the events.
The Durham jet clustering algorithm, 
with a particular jet resolution parameter $\ycut$,
is then applied to these momenta
to select $\gamma$ +~1 jet events.
One of the clusters is identified as a photon if the fraction of 
electromagnetic energy $z$ inside the jet is greater 
than $\zcut$, a value fixed experimentally.
Moreover, the fragmentation function mentioned above is the effective 
fragmentation function which at ${\cal O}(\alpha)$ is given by 
eq.~(\ref{eq:counter}) and at ${\cal O}(\alpha \alpha_{s})$ is the sum of the 
$\mu_{F}$ dependent quark-to-photon fragmentation function and a finite contribution coming from the unresolved photon contributions given 
by eq.~(\ref{eq:total}).
The four contributions are:
\begin{itemize}
\item[{(A)}] {\bf 2 partons + photon} \newline
This contribution corresponds to the process $\gamma^*
\to q \bar q \gamma$ with a hard photon in the final state. 
It is present at ${\cal O}(\alpha)$ and ${\cal O}(\alpha\alpha_s)$  
(due to the presence of a theoretically 
unresolved real or virtual gluon). 
\item[{(B)}] {\bf 2 partons + ``fragmentation''} \newline
The   
$\gamma^* \to q \bar q \otimes D_{q\to \gamma}$ process is present 
at ${\cal O}(\alpha)$ and ${\cal O}(\alpha\alpha_s)$.
In particular, it contains the finite terms corresponding to the
$q-\gamma$ collinear region at ${\cal O}(\alpha)$ and the finite parts associated  
with all double unresolved regions at ${\cal O}(\alpha\alpha_s)$.
\item[{(C)}] {\bf 3 partons + photon} \newline
This contribution is only present at ${\cal O}(\alpha \alpha_s)$ and 
describes the $\gamma^* \to q \bar q \gamma g$ process where 
both photon and gluon are theoretically resolved.
\item[{(D)}] {\bf 3 partons + ``fragmentation''} \newline
The  
$\gamma^* \to q \bar q g \otimes D_{q\to \gamma}$ process
with a hard gluon in the final state, 
is only present at ${\cal O}(\alpha \alpha_s)$. 
\end{itemize}

\begin{figure}[t]
\vspace{10cm}
\begin{center} 
~ \includegraphics{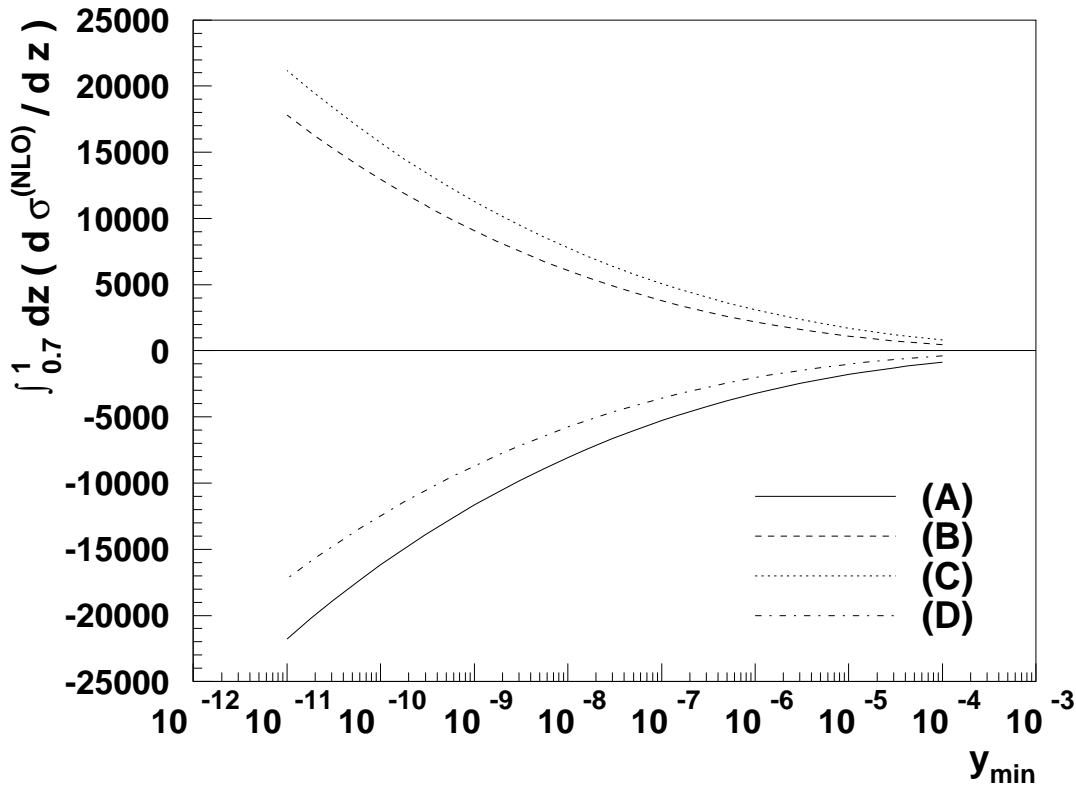}
\caption{Contributions of the individual terms ({\tt (A),(B),(C),(D)}) 
to the total cross section as function of $\ymin$ 
for $\ycut=0.1$ and $\zcut=0.7$.
For clarity, only the next-to-leading order contributions are shown. 
Furthermore we take $\alpha e_{q}^2=2\pi$ and $\alpha_s C_F =2\pi$.}
\label{fig:ymindep1}
\end{center}
\end{figure}

Each of the contributions depends on the theoretical parton resolution parameter $\ymin$. 
In Fig.~\ref{fig:ymindep1} we show the various contributions to the 
integrated cross section $0.7 < z < 1$, for a single quark of unit charge.
The jet resolution parameter is $\ycut=0.1$ and only the ${\cal O}(\alpha\alpha_s)$ contributions are shown.
As can be seen in Fig.~\ref{fig:ymindep1}, the magnitude of the various 
contributions to the differential cross 
section increases dramatically as $\ymin$ becomes smaller.
This rapid rise is due to the presence of logarithms of $\ymin$ due to 
expanding the residues of the poles in $\e$.
Since the leading poles are ${\cal O}(1/\e^3)$, powers of logarithms up to
$\ln^3(\ymin)$ are present.

\begin{figure}[t]
\vspace{10cm}
\begin{center}
~ \includegraphics{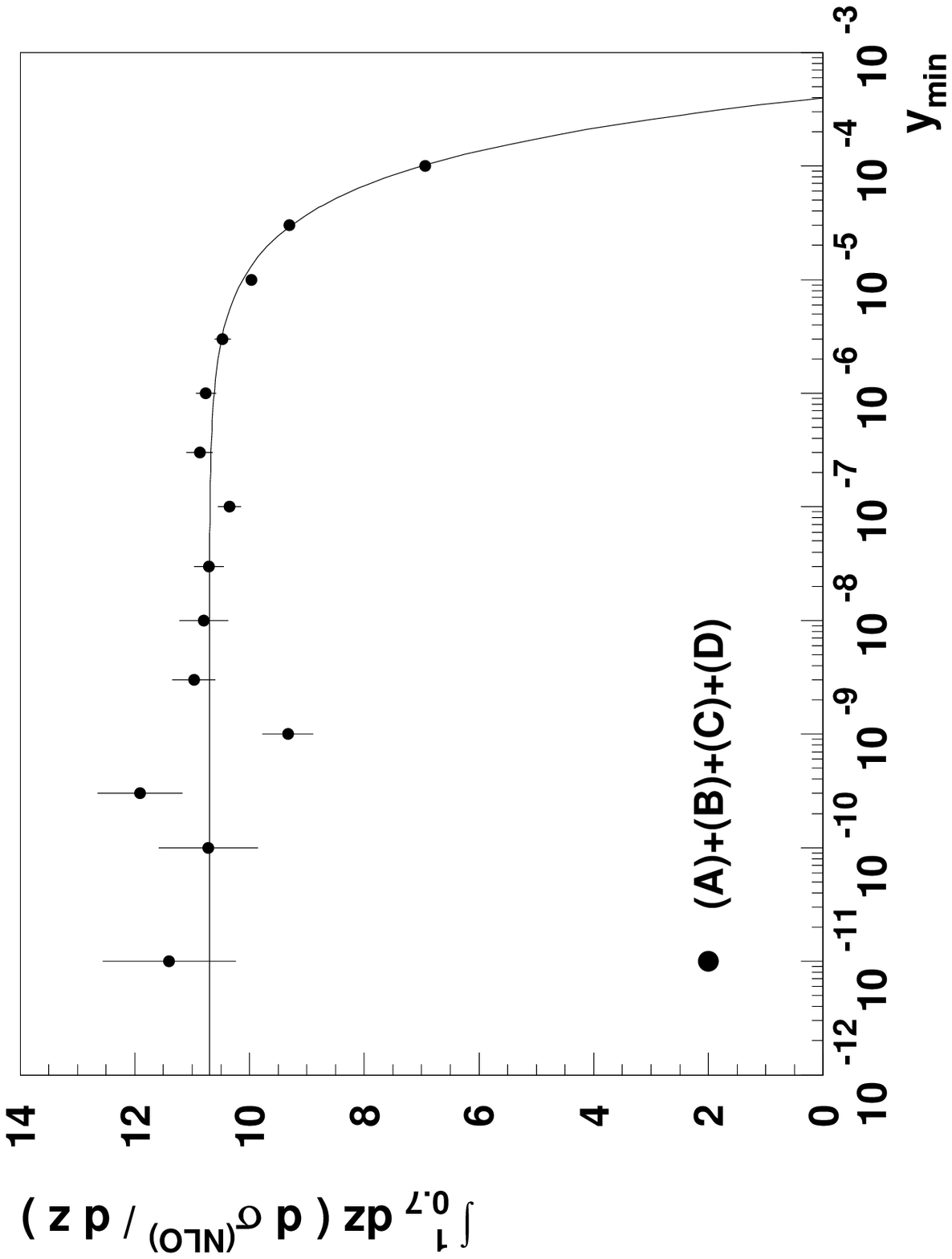}
\caption{The sum of all next-to-leading order contributions 
to the total cross section as function of $\ymin$
for $\ycut=0.1$ and $\zcut=0.7$.
Only the next-to-leading order corrections are shown and
we take  $\alpha e_{q}^2=2\pi$ and $\alpha_s C_F =2\pi$. 
The solid line is a fit of the form 
$a+ b \ymin \ln^2 \ymin$.}
\label{fig:ymindep2}
\end{center}
\end{figure}

The final integrated cross section, which is the sum of 
all theoretically resolved and unresolved contributions, {\em must} of course 
be independent of the artificial parameter $\ymin$.
Each individual term has a very strong dependence on $\ymin$, but, as can be seen in Fig.~\ref{fig:ymindep2}, the sum of all resolved and 
unresolved contributions is clearly $\ymin$ independent (within the 
numerical errors of the calculation) provided $\ymin$ is 
taken small enough.
In practice, this means
values of $\ymin$ ranging between
$10^{-5}\; {\rm to}\; 10^{-9}$ for the chosen value of the experimental jet 
resolution parameter $\ycut=0.1$.
For $\ymin = 10^{-8}$, the magnitude of the individual terms is 
${\cal O}(5000)$, while the final result (after enormous cancellations) is 
${\cal O}(10)$.
This figure demonstrates the consistency 
of our approach -- there is a region of parameter space where the 
choice of the unphysical parameter $\ymin$ does not affect the physically
observable cross section.
Note that
for large values of $\ymin$ the cross section deviates 
from the $\ymin$--independent value. This is because 
for large 
$\ymin$ values the approximations used in the analytic calculation 
become less accurate.
In particular, terms of ${\cal O}(\ymin\,\ln^2(\ymin))$,  which have been neglected, 
become sizeable.
On the other hand,
for values of $\ymin$ below $10^{-9}$ the errors 
on the result become important due to the necessity of cancelling large 
logarithms numerically. The overall result 
becomes therefore less stable numerically for such small values of $\ymin$.
A reasonable choice of $\ymin$, which does not lead to problems of
numerical instability 
is therefore $\ymin/\ycut=10^{-5}$. We shall use this value 
when comparing our results with the ALEPH data in the next section.

\section{Results} 
\setcounter{equation}{0}
\label{sec:results}
In the previous section, we have demonstrated the consistency of our approach 
to evaluate the photon +~1 jet rate at ${\cal O}(\alpha \alpha_{s})$ by showing
that the results   
were $\ymin$ independent.
It is therefore possible now to use these results to determine 
the non-perturbative quark-to-photon fragmentation function 
$D_{q \to \gamma}(z, \mu_{F})$
up to this order
from a comparison with the experimental 
data~\cite{aleph} from the ALEPH Collaboration at CERN.
As a first application we will use 
this newly determined fragmentation function 
to evaluate the isolated photon +~1 jet rate 
in the democratic clustering approach as a function of the jet 
resolution parameter $\ycut$. These will also be compared to experimental 
data~\cite{aleph}. 

\subsection{A determination of $D_{q \to \gamma}(z, \mu_{F})$ at 
next-to-leading order}
\label{subsec:deter}

The quark-to-photon fragmentation function 
$D_{q \to \gamma}(z, \mu_{F})$ which appears explicitly
in the expression of the photon +~1 jet rate given in eq.~(\ref{eq:sumjet3})
has been given as the exact solution  
of the perturbative ${\cal O}(\alpha \alpha_{s})$ evolution equation  by eq.~(\ref{eq:Dnlo}).
As mentioned before,
all unknown non-perturbative contributions to this function 
are contained in its initial value,    
$D_{q \to \gamma}^{np}(z,\mu_{0})$ which needs to be extracted from the data.
We perform a three parameter fit ($\mu_0$ is fitted as well)
to the ALEPH `photon' +~1 jet
data \cite{aleph} for the $z$ distribution, 
$\frac{1}{\sigma_{0}}\frac{d\sigma}{dz}$,  
at a jet resolution parameter $y_{{\rm cut}}=0.06$ 
and $\alpha_s(M_Z) = 0.124$ and obtain \cite{letter}, 
\begin{equation}
\Dq^{np(NLO)}(z,\mu_{0})=\aqed 
\left(-\Pqpzero(z)\ln(1-z)^2 \;+\,20.8\,(1-z)-11.07\right),
\label{eq:fitnlo1}
\end{equation}
where $\mu_{0}=0.64$~GeV.
For reference, the lowest order fit obtained 
by the ALEPH Collaboration \cite{aleph} is,
\begin{equation}
\Dq^{np(LO)}(z,\mu_{0})=\aqed 
\left(-\Pqpzero(z)\ln(1-z)^2 - 13.26\right),
\label{eq:fitlo}
\end{equation}
with $\mu_{0}=0.14$~GeV.
In both cases, the logarithmic term proportional to $P^{(0)}_{q \to \gamma}(z)$
is introduced to ensure that the predicted $z$ distribution is 
well behaved as $z \to 1$ \cite{andrew}. 

\begin{figure}[t]
\vspace{10cm}
\begin{center} 
~ \includegraphics{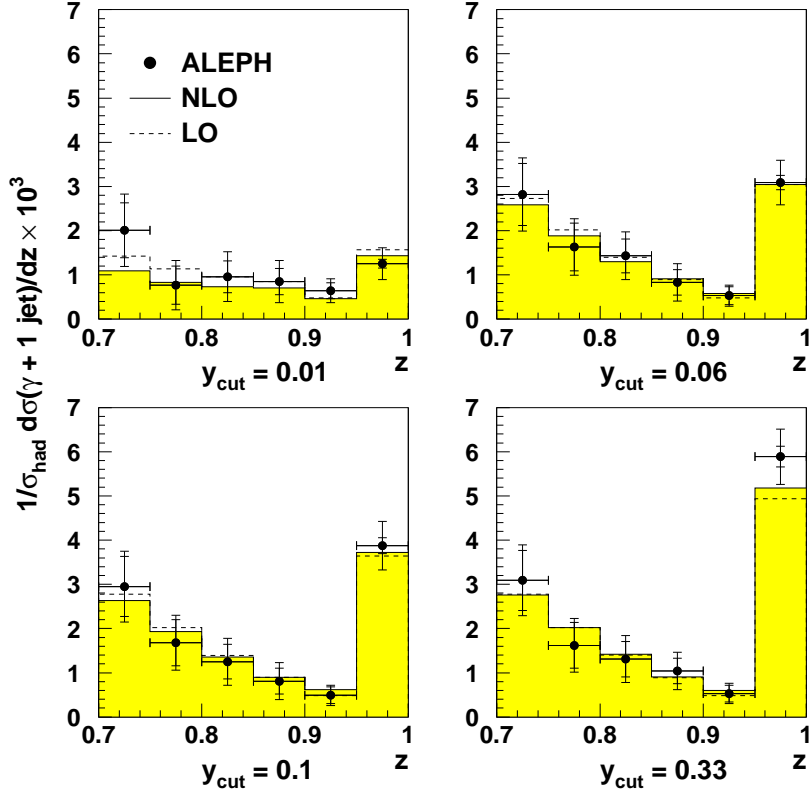}
\caption{Comparison of the photon + 1~jet rate at leading and next-to-leading 
order with the ALEPH data. The non-perturbative quark-to-photon fragmentation 
function is fitted to the data for \protect$\ycut=0.06$ only. The jet 
rates for the other values of $\ycut$ are then predictions from the 
leading order and next-to-leading order calculations.}
\label{fig:aleph}
\end{center}
\end{figure}

The results of the ${\cal O}(\alpha \alpha_s)$ 
calculation using this fitted next-to-leading order 
fragmentation function and the ALEPH data are shown in 
Fig.~\ref{fig:aleph}. Furthermore, the results of this fit 
can be used to evaluate the photon +~1 jet rate
for different values of $\ycut$ ($\ycut =0.01, 0.1, 0.33 $). 
In each case, the calculated $z$-distributions are compared with the leading 
order predictions and the ALEPH data in Fig.~\ref{fig:aleph}.
A good agreement is found for all values of $\ycut$ studied 
in the experimental measurement, 
reflecting the universality of the quark-to-photon fragmentation function.

Since the present calculation is only lowest order in the strong 
coupling constant, there is a large ambiguity in the choice of 
$\alpha_s$ and all values of $\alpha_s$ are in principle equally valid.  
The value $\alpha_{s}(M_Z)=0.124$ used above was chosen 
such that the observed hadronic cross section can be reproduced 
by the ${\cal O}(\alpha_{s})$ calculation. However, jet studies at LEP 
have indicated that lowest order calculations of jet observables 
can only be matched to experimental data if a larger value of the 
strong coupling constant to compensate for missing higher order 
contributions is used. In particular, the experimental data 
on the isolated photon +~2, 3 jet rates~\cite{aleph} are well reproduced  
by a lowest order calculation~\cite{andrew} if $\alpha_s(M_Z)=0.17$. We  
therefore provide a fit of the quark-to-photon fragmentation function 
for this value of $\alpha_s$ as well. We find
\begin{equation}
\Dq^{np(NLO)}(z,\mu_{0}) = \aqed \left(-\Pqpzero(z)\ln (1-z)^2 + 
32.8 (1-z) - 10.35\right),
\label{eq:fitnlo2}
\end{equation}
with $\mu_0=0.59$~GeV. 

\begin{figure}[t]
\vspace{10cm}
\begin{center} 
~ \includegraphics{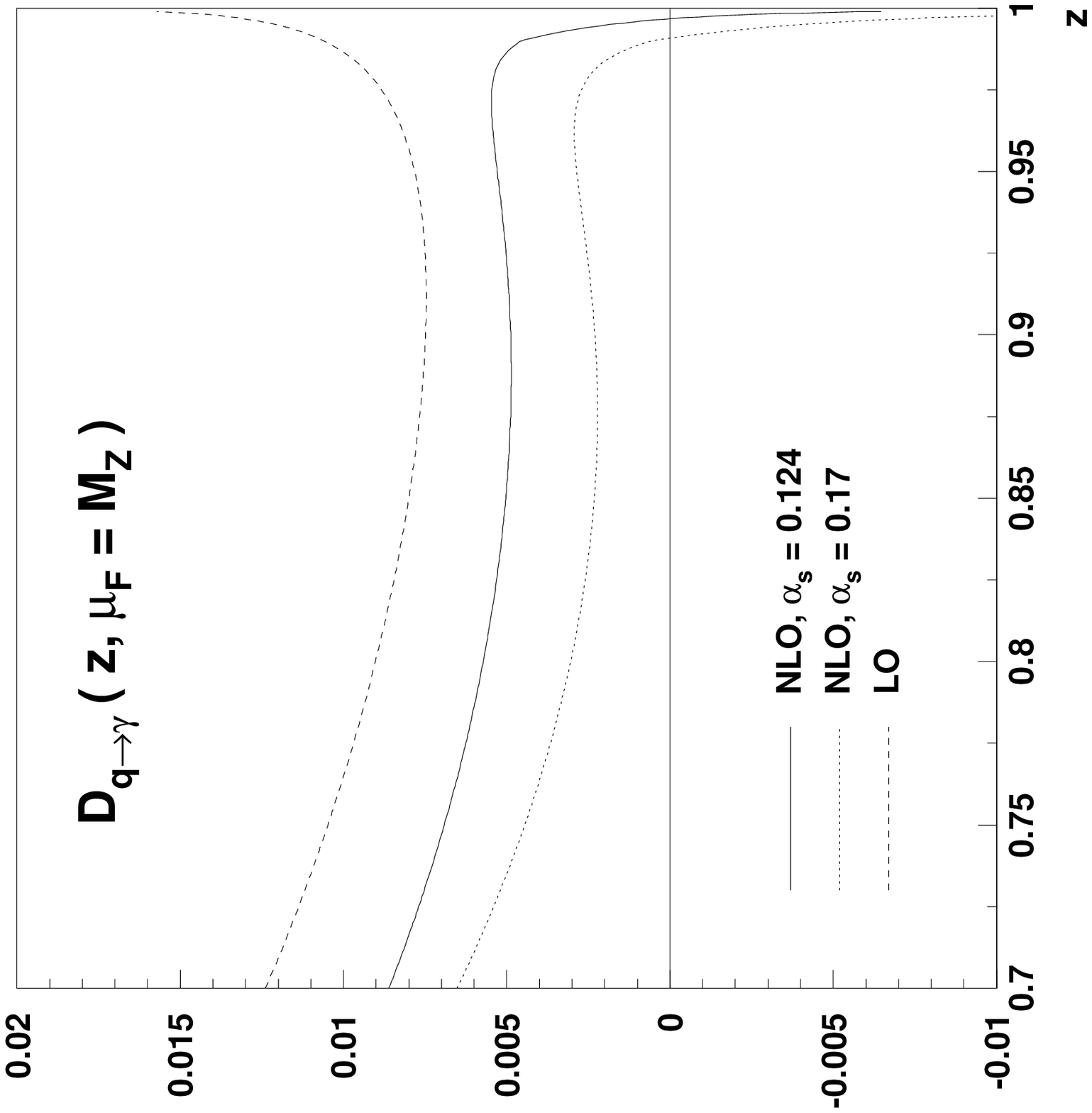}
\caption{The quark-to-photon fragmentation functions at leading and 
next-to-leading order as functions of $z$  for
 a quark of unit charge. 
The factorization scale $\mu_F$ is taken equal to $M_Z$ in both cases.}
\label{fig:dfrag}
\end{center}
\end{figure}

Figure ~\ref{fig:dfrag} displays the fitted ${\cal O}(\alpha \alpha_s)$
quark-to-photon fragmentation functions  
for a quark of unit charge
in the $\overline{{\rm MS}}$--scheme at a factorization scale $\mu_F=M_Z$, 
which is the only hard scale in the process. The corresponding
${\cal O}(\alpha)$
fragmentation function obtained in~\cite{aleph} is shown for comparison.
It is apparent that these fragmentation functions differ largely for 
$z\to 1$, which indicates the need for a resummation of terms proportional 
to $\ln (1-z)$ in this region.

\subsection{The integrated rate for $z>0.95$}
\label{subsec:integ}

If the photon carries a fractional momentum $z$
greater than 0.95 in the photon-jet
it is considered to be {\it isolated} in the democratic clustering appoach
used in the experimental ALEPH analysis. 
In this approach the photon is treated like 
any other hadron when the clustering procedure is applied to form the jets.
The division between isolated and non-isolated photons
suggested by ALEPH 
in~\cite{aleph} is motivated by the fact that hadronization effects 
smear out the isolated photon peak from $z=1$ to slightly smaller values of $z$.

This definition of isolation is in contrast with that 
used in previous theoretical~\cite{gs,ks,kt} and 
experimental~\cite{iso} analyses of isolated photon +~$n$ jet rates, 
where a two-step procedure was used to define an isolated photon.
In these analyses, the photon was isolated from the other hadrons  
using a geometrical cone before these hadrons were clustered into jets. 
After the clustering had taken place 
the photon was required to be well-separated from all of the  
hadronic clusters and was said to be {\it isolated} if it was accompanied only 
by a {\it small} amount of hadronic energy inside the cone.
In these theoretical calculations~\cite{kt,ks,gs},
all of the quark-photon collinear 
and fragmentation contributions considered here 
were assumed to be negligible.
As a result of these studies,
it was found that  
large negative next-to-leading order corrections were needed to provide a reasonable agreement 
between data and theory, precisely because soft gluons were excluded
from the photon cone \cite{softrad}. 
 
\begin{figure}[t]
\vspace{10cm}
\begin{center} 
~ \includegraphics{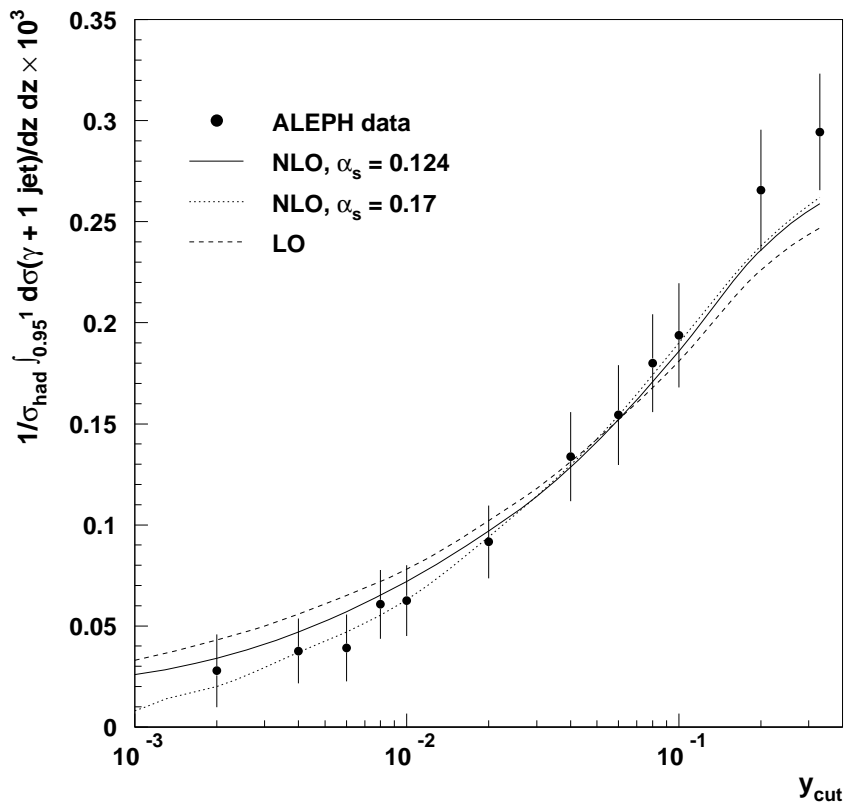}
\caption{The integrated photon +~1 jet rate above $z=0.95$ 
as function of $\ycut$, compared with the full leading-order 
and next-to-leading order calculations including respectively 
the leading-order and next-to-leading order determined 
quark-to-photon fragmentation functions, 
\protect$D_{q \to \gamma}(z,\mu_{F})$. }
\label{fig:ycut}
\end{center}
\end{figure}

Using the fitted fragmentation function, we have evaluated the isolated 
photon +~1 jet rate in the democratic clustering approach up to 
${\cal O}(\alpha \alpha_{s})$.
The results of this calculation for the two different values of $\alpha_{s}$
are compared with the data and the leading order prediction ~\cite{andrew} 
in Fig.~\ref{fig:ycut}.
It appears that 
the leading-order curve provides an adequate description of the data and that 
the next-to-leading order curve improves 
the agreement between theory and experiment over the whole range of $\ycut$. 
Furthermore, it is also apparent that the next-to-leading order 
corrections are moderate thereby demonstrating the perturbative stability 
of the definition of isolation within the democratic approach.

\section{Summary and Conclusions} 
\setcounter{equation}{0}
\label{sec:conc}

In this paper we have presented a complete ${\cal O}(\alpha \alpha_{s})$ 
calculation of the photon +~1 jet rate in $e^+e^-$ annihilation.
Athough this calculation is formally only next-to-leading order 
in perturbation theory, it contains several ingredients appropriate 
to next-to-next-to-leading order calculations of jet observables.
For example, in addition to configurations where one final state particle is unresolved, there are also configurations where two particles are unresolved.
In particular we encountered three {\it double unresolved} factors,
corresponding to the triple collinear, soft/collinear 
and double single collinear limits of the $\gamma^* \to q \bar q g \gamma$
subprocess.
To analytically isolate the singularities associated with these configurations, we employed the resolved parton philosophy 
of \cite{gg} and introduced a theoretical parton resolution criterion $\ymin$.
These previously unknown double unresolved factors 
are universal and are of the type expected to arise 
in any calculation of jet cross 
sections at next-to-next-to-leading order.
Matching the double unresolved parts of phase space with the single unresolved and resolved regions required some thought and was discussed at length 
in section \ref{sec:1jet}.

All these resolved and unresolved contributions 
from the double bremsstrahlung process must be combined
with the single unresolved contributions from one-loop
and fragmentation processes.
Some    
single and double poles in $\e$ remain which
have however exactly the right structure  
to be factorized in the next-to-leading order counterterm 
of the bare quark-to-photon fragmentation function.

All remaining finite contributions can then be numerically evaluated
for arbitrary jet clustering algorithms.
The most stringent test on the consistency of our approach 
is provided 
by Fig.~\ref{fig:ymindep2}, where it is shown that the results 
of the numerical program are independent of the choice of the 
theoretical parameter $\ymin$. 
This relies on the  
numerical cancellation of large logarithms of $\ymin$ taken up 
to the third power.

An important feature of this numerical program is the implementation
of the hybrid subtraction method \cite{eec} which is necessary
to correctly treat the overlapping of two 
single collinear regions in the neighberhood of double unresolved regions 
of the four parton phase space.
This overlapping was shown to be crucial for a correct and complete 
coverage of all singularities in the $\gamma^* \to q \bar{q} \gamma g$ process
in section~\ref{sec:1jet}.  
 
We have presented fits of the non-perturbative part of $\Dq$ for two values of $\alpha_s(M_Z)$ obtained by comparing our results 
with the ALEPH data \cite{aleph}.
In determining the fragmentation function, we have required that $\Dq$ is an exact solution of the ${\cal O}(\alpha \alpha_{s})$ evolution equation.
This solution does not resum logarithms of the factorization scale, but does
ensure that the photon +~1 jet rate 
is $\mu_{F}$ independent.
Furthermore, at large $z$, which corresponds to  the most interesting
`isolated' photon part of the cross section, large logarithms of $(1-z)$ are
present and should be resummed.

As a first application of the fitted next-to-leading order quark-to-photon fragmentation function, 
we have compared the integrated `isolated' photon rate for $z > 0.95$
with the same ALEPH data.
The theoretical next-to-leading order predictions were found 
to describe the data well and to provide a better agreement 
between theory and experiment than the leading 
order calculation.
Furthermore, the relatively small size of the next-to-leading order 
corrections indicates that the isolation definition 
used in this democratic clustering approach is perturbatively stable.

Finally, the next-to-leading order quark-to-photon fragmentation function 
determined here is a universal function and
appears in a variety of processes involving quarks and photons.
The most prominent examples of which are the prompt photon 
cross section at hadron colliders and the photon pair cross section at LHC.
So far the fragmentation contributions to those processes have 
only been evaluated using model 
dependent assumptions for this fragmentation function \cite{grv}.
It is reasonable to expect that the fragmentation function determined directly from LEP data at high $Q^2$ and high $z$ should significantly improve 
the theoretical predictions for these processes,
and, in particular, may help to determine whether a Standard Model Higgs-boson 
of intermediate mass can really be detected via its two photon decay at the LHC.

\section*{Acknowledgements}
We thank Thomas Gehrmann for numerous suggestions and comments throughout
the period of this work.
One of us (A.G.) would like to thank the DESY Theory Group 
for their kind hospitality during the later stages of this work.
A.G. also acknowledges the 
financial support of the University of Durham 
through a Research Studentship (Department of Physics).

\end{document}